\documentclass[runningheads]{svmult}

\usepackage{makeidx}   % allows index generation
\usepackage{graphicx}  % standard LaTeX graphics tool
\usepackage{subeqnar}  % subnumbers individual equations
\usepackage{multicol}  % used for the two-column index
\usepackage{physprbb}  % modified textarea for proceedings,
\makeindex             % used for the subject index
\newcommand{\be}{\begin{equation}}
\newcommand{\ee}{\end{equation}}
\def\bea{\begin{eqnarray}}
\def\eea{\end{eqnarray}}
\def\der{\partial}

\newcommand{\mscr}[1]{\mbox{\scriptsize #1}}
\newcommand{\fscr}[1]{\mbox{\scriptsize \bf #1}}
\newcommand{\ft}[2]{{\textstyle\frac{#1}{#2}}}
\usepackage{amssymb}
\begin{document}
\title*{
Introduction to String Theory
}
\toctitle{
Introduction to String Theory
}
% allows explicit linebreak for the table of content
%
%
\titlerunning{
Introduction to String Theory
}
% allows abbreviation of title, if the full title is too long
% to fit in the running head
%
\author{Thomas Mohaupt}
\authorrunning{Thomas Mohaupt}
% if there are more than two authors,
% please abbreviate author list for running head
%
%
\institute{Friedrich-Schiller Universit\"at Jena,
Max-Wien-Platz 1, D-07743 Jena, Germany}

\maketitle              % typesets the title of the contribution

\begin{abstract}
We give a pedagogical introduction to string theory, D-branes and p-brane
solutions.
\end{abstract}

\section{Introductory remarks}

These notes are based on lectures given at the
271-th WE-Haereus-Seminar \lq Aspects of Quantum Gravity'. Their aim
is to give an introduction to string theory for
students and interested researches.
No previous knowledge of string theory is assumed.
The focus is on gravitational aspects
and we explain in some detail how gravity is described in string
theory in terms of the graviton excitation of the string and
through background gravitational fields. We include Dirichlet 
boundary conditions and D-branes from the beginning and devote
one section to p-brane solutions and their relation to D-branes.
In the final section we briefly indicate how string theory fits
into the larger picture of M-theory and mention some 
of the more recent developments, like brane world scenarios.

The WE-Haereus-Seminar \lq Aspects of Quantum Gravity' covered
both main approaches to quantum gravity: string theory and 
canonical quantum gravity. Both are complementary in many respects.
While the canonical approach stresses background independence
and provides a non-perturbative framework, the cornerstone
of string theory still is perturbation theory in a fixed background
geometry. Another difference is that in the canonical
approach gravity and other interactions are independent
from each other, while string 
theory automatically is
a unified theory of gravity, other interactions and matter.
There is a single dimensionful constant and all couplings are
functions of this constant and of vacuum expectation values of
scalars. The matter content is uniquely fixed by the symmetries
of the underlying string theory. Moreover, when formulating the
theory in Minkowski space, the number of space-time dimensions
is fixed. As we will see, there are 
only five distinct supersymmetric string theories in ten-dimensional
Minkowski space.

The most important feature of string perturbation
theory is the absence of UV divergencies. This allows one to 
compute quantum corrections to scattering amplitudes and to
the effective action, including gravitational effects.
More recently, significant progress has been made
in understanding non-perturbative 
aspects of the theory, through the study of solitons and instantons,
and through string dualities which map the strong coupling behaviour
of one string theory to the weak coupling behaviour of a dual theory. 
Moreover, string dualities relate all five supersymmetric string theories
to one another and lead to the picture of one single underlying theory,
called M-theory. So far, only various limits of this theory are known,
while the problem of finding an intrinsic,
non-perturbative and background-independent definition is unsolved.
One expects that M-theory has an underlying
principle which unifies its various incarnations, presumably
a symmetry principle. One of the obstacles on the way to the
final theory is that it is not clear which 
degrees of freedom are fundamental. Besides strings, also 
higher-dimensional p-branes play an essential role. Moreover, there
is an eleven-dimensional limit, which cannot be described in terms
of strings.

Our presentation of string theory will be systematic rather than
follow the path of historical development. Nevertheless we feel that
a short historical note will be helpful, since many aspects which may seem
somewhat ad hoc (such as the definition of interactions in section 3) become
clearer in their historical context. The story started with the
Veneziano amplitude, which was proposed as an amplitude
for meson scattering in pre-QCD times. The amplitude fitted the
known experimental data very well and had precisely the properties
expected of a good scattering amplitude
on the basis of S-matrix theory, the bootstrap program and
Regge pole theory. In particular it had a very special soft 
UV behaviour. Later work by Y. Nambu, H.B. Nielsen and L.
Susskind showed that the Veneziano amplitude, and various 
generalization thereof  
could be interpreted as describing the scattering of
relativistic strings. But improved experimental data ruled out 
the Veneziano amplitude as a hadronic amplitude: it behaved 
just to softly in order to describe the hard, partonic substructures
of hadrons seen in deep inelastic scattering. J. Scherk and
J. Schwarz reinterpreted string theory as a unified theory of 
gravity and all other fundamental interactions, making
use of the fact that the spectrum of a closed string always contains
a massless symmetric tensor state which couples like a graviton. 
This lead to the development of perturbative string theory, as we
will describe it in sections 2--4 of these lecture notes. More recently
the perspective has changed again, after the role of D-branes, 
p-branes  and string dualities was recognized. This will be 
discussed briefly in sections 5 and 6.

From the historical perspective it appears
that string theory is a theory which is 
\lq discovered' rather than \lq invented'. 
Though it was clear
from the start that one was dealing with an interesting 
generalization of quantum field theory and general relativity, 
the subject has gone through several \lq phase transitions',
and its fundamental principles remain to be made explicit.
This is again complementary to canonical quantum gravity, 
where the approach is more axiomatic, starting from
a set of principles and proceeding to quantize Einstein gravity.

The numerous historical twists, 
our lack of final knowledge about the fundamental principles 
and the resulting diversity of methods and approaches make string theory 
a subject which is not easy to learn (or to teach). 
The 271-th WE-Haereus-Seminar covered a broad variety of topics
in quantum gravity, \lq From Theory to Experimental Search'. 
The audience consisted of two groups: graduate students, mostly without 
prior knowledge of string theory, and researches, 
working on various theoretical and experimental topics in gravity.
The two lectures on string theory were supposed to give a
pedagogical  introduction and to prepare for later lectures on 
branes worlds, large extra dimensions, the AdS-CFT correspondence and 
black holes. These lecture notes mostly follow the lectures, but 
aim to extend them in two ways.
The first is to add more details to the topics I discussed in the lectures.
In particular I want to expand on points which seemed to be either  
difficult or interesting to the audience. The second goal
is to include more material, in order to bring the reader closer
to the areas of current active research. Both goals are 
somewhat contradictory, given that the result is not meant to be
a book, but lecture notes of digestable length. As a compromise I choose 
to explain those things in detail which seemed to be the most important
ones for the participants of the seminar, hoping that they represent 
a reasonable sample of potential readers. On the other side several
other topics are also covered, though in a more scetchy 
way. Besides summarizing advanced topics, which cannot be fully explained
here, I try  to give an overview of 
(almost) all the new developements of the last years and 
to indicate  how they fit into the emerging overall picture of M-theory.

The outline of the lectures is as follows: sections 2--4 are devoted
to perturbative aspects of bosonic and supersymmetric string theories.
They are the core of the lectures. References are given at
the end of the sections. String theory has been a very active field over 
several decades, and the
vast amount of existing literature is difficult to oversee even for 
people working in the field. I will not try to give a complete account
of 
the literature, but only make suggestions for further reading.
The basic references are the 
books \cite{GSW,LT,Pol,Kak1,Kak2}, which contain a huge number
of references to reviews and original papers. The reader interested
in the historical developement of the subject will find 
information in the annotated bibliography of \cite{GSW}.
Section 5 gives an introduction to non-perturbative aspects 
by discussing a particular class of solitons, the
p-brane solutions of type II string theory.  
Section 6 gives an outlook on advanced topics:
while sections 6.1--6.3 scetch how the five supersymmetric
string theories fit into the larger picture of M-theory,
section 6.4 gives an overview of current areas
of research, together with references to lecture notes,
reviews and some original papers.

\section{Free bosonic strings}

We start our study of string theories with the bosonic string.
This theory is a toy-model rather than a realistic theory of
gravity and matter. As indicated by its name
it does not have fermionic states, 
and this disqualifies it as a theory of particle physics. Moreover,
its groundstate in Minkowski space is a tachyon, i.e., a state
of negative mass squared. This signals that the theory is unstable.
Despite these shortcomings, the bosonic string has its virtues as
a pedagogical toy-model: whereas we can postpone to deal with the
additional techniques needed to describe fermions, many features
of the bosonic string carry
over to supersymmetric string theories, which have fermions
but no tachyon.

\subsection{Classical bosonic strings}

We start with a brief overview of classical aspects of 
bosonic strings.

\subsubsection{Setting the stage.}

Let us first fix our notation. We consider a fixed background
Pseudo-Riemannian space-time ${\cal M}$ of dimension $D$,
with coordinates $X=(X^{\mu})$, $\mu = 0,\ldots, D-1$. The
metric is $G_{\mu \nu}(X)$ and we take the signature to
be \lq mostly plus', $(-)(+)^{D-1}$. 

The motion of a relativistic string in ${\cal M}$ is 
described by its generalized worldline, a two-dimensional 
surface $\Sigma$, which is 
called the world-sheet.
For a single non-interacting string the world-sheet has the
form of an infinite strip. We introduce coordinates 
$\sigma =(\sigma^0 , \sigma^1)$ on the world-sheet. The
embedding of the world-sheet into space-time is given by
maps
\be
X: \; \Sigma \longrightarrow {\cal M} \; : \sigma \longrightarrow
X(\sigma) \;.
\ee
The background metric induces a metric on the world-sheet:
\be
G_{\alpha \beta} = \frac{ \der X^{\mu} }{ \der \sigma^{\alpha} }
\frac{ \der X^{\nu} }{ \der \sigma^{\beta} } \; G_{\mu \nu} \;,
\ee
where $\alpha, \beta = 0,1$ are world-sheet indices. 
The induced metric is to be distinguished from an intrinsic
metric $h_{\alpha \beta}$ on $\Sigma$. As we will see below,
an intrinsic metric is used as an auxiliary field in
the Polyakov formulation of the bosonic string.

A useful, but sometimes confusing fact is that the above setting
can be viewed from two perspectives. So far we have taken the 
space-time perspective, interpreting the system as a relativistic
string moving in space-time ${\cal M}$. Alternatively we may view
it as a two-dimensional field theory living on the world-sheet,
with fields $X$ which take values in the target-space ${\cal M}$.
This is the world-sheet perspective, which enables us to use
intuitions and methods of two-dimensional field theory for the
study of strings. 

\subsubsection{Actions.}

The natural action for a relativistic string is its area, 
measured with the induced metric:
\be
S_{\mscr{NG}} = \frac{1}{2 \pi \alpha'} \int_{\Sigma}
d^2 \sigma | \det G_{\alpha \beta} |^{1/2} \;.
\label{NGAct}
\ee
This is the Nambu-Goto action, which is the direct generalization
of the action for a massive relativistic particle. The prefactor
$(2 \pi \alpha')^{-1}$ is the energy per length or tension of the
string, which is the fundamental dimensionful parameter of the
theory. We have expressed the tension in terms of the
so-called Regge slope
$\alpha'$, which has the dimension $(\mbox{length})^2$ in 
natural units, $c=1$, $\hbar =1$. Most of the time we will 
use string units, where in addition we set $\alpha' = \ft12$.

The Nambu-Goto action has a direct geometric meaning, but
is technically inconvenient, due to the square root. Therefore
one prefers to use the Polyakov action, which is equivalent 
to the Nambu-Goto action, but is a standard two-dimensional
field theory action. In this
approach one introduces an intrinsic metric on the world-sheet,
$h_{\alpha \beta}(\sigma)$, as additional datum. The action takes
the form of a non-linear sigma-model on the world-sheet,
\be
S_{\mscr{P}} = \frac{1}{ 4 \pi \alpha'} \int_{\Sigma}
d^2 \sigma \sqrt{h} h^{\alpha \beta} \der_{\alpha} X^{\mu}
\der_{\beta} X^{\nu} G_{\mu \nu}(X) \;,
\label{PolAct}
\ee
where $h = | \det h_{\alpha \beta} |$.

The equation of motion for $h_{\alpha \beta}$ is algebraic.
Thus the intrinsic metric is non-dynamical and can be eliminated,
which brings us back to the Nambu-Goto action. 
Since 
\be
T_{\alpha \beta} := \left( {2 \pi \alpha' \sqrt{h}} \right)^{-1}
\frac{ \delta S_{\mscr{P}}}{\delta h^{\alpha \beta}}
= \der_{\alpha} X^{\mu} \der_{\beta} X_{\mu}
- \frac{1}{2} h_{\alpha \beta} \der_{\gamma} X^{\mu} 
\der^{\gamma} X_{\mu}
\label{2dEMtensor}
\ee
is the energy momentum of the two-dimensional field theory
defined by (\ref{PolAct}), we can interpret the equation of motion of 
$h_{\alpha \beta}$ as the two-dimensional
Einstein equation. The two-dimensional metric is non-dynamical,
because the two-dimensional Einstein-Hilbert action is a topological
invariant, proportional to the Euler number of ${\Sigma}$. Thus its
variation vanishes and the Einstein equation of (\ref{PolAct})
coupled to two-dimensional gravity reduces to $T_{\alpha \beta}=0$.
Note that the energy-momentum tensor (\ref{2dEMtensor})
is traceless, $h^{\alpha \beta} T_{\alpha \beta} = 0$.
This holds before imposing the equations of motion (\lq off shell'). Therefore
$T_{\alpha \beta}$ has only two independent components, which 
vanish for solutions to the equations of motion (\lq on shell').
Since the trace of the energy-momentum tensor is the Noether current
of scale transformations, this shows that the 
two-dimensional field theory (\ref{PolAct}) is scale invariant.
As we will see below, it is in fact  a conformal field theory.

The Polyakov action has three local symmetries. Two are shared
by the Nambu-Goto action, namely reparametrizations of the
world-sheet:
\be
\sigma^{\alpha} \longrightarrow \tilde{\sigma}^{\alpha} (\sigma^0,
\sigma^1) \;.
\ee
The third local symmetry is the multiplication of the metric $h_{\alpha \beta}$
by a local, positive scale factor,
\be
h_{\alpha \beta}(\sigma) \longrightarrow e^{\Lambda(\sigma)} h_{\alpha
\beta}(\sigma) \;.
\ee
This transformation is called a Weyl transformation by physicists,
while mathematicians usually use the term conformal transformation.
The three local symmetries can be used to gauge-fix the metric
$h_{\alpha \beta}$. The standard choice is the conformal gauge,
\be
h_{\alpha \beta}(\sigma) \stackrel{!}{=} \eta_{\alpha \beta} \;,
\;\;\;\mbox{where} \;\;\; (\eta_{\alpha \beta}) = \mbox{Diag}(-1,1) \;.
\label{ConfGauge}
\ee
While this gauge can be imposed globally on the infinite strip 
describing the motion of a single non-interacting string, it can
only be imposed locally on more general world-sheets, which describe
string interactions. We will discuss global aspects of gauge
fixing later. 

The conformal gauge does not provide a complete gauge fixing,
because (\ref{ConfGauge}) is invariant under a residual
symmetry. One can still perform reparametrizations under which
the metric only changes by a local, positive scale factor,
because this factor can be absorbed by a Weyl transformation.
Such conformal reparametrizations are usually called conformal transformations
by physicists. Note that the same term is used for
Weyl transformations by mathematicians. A convenient way
to characterize conformal reparametrizations in terms of coordinates
is to introduce light cone coordinates,
\be
\sigma^{\pm} = \sigma^0 \pm \sigma^1 \;.
\ee
Then conformal reparametrization are precisely those
reparametrizations which do not mix the light cone
coordinates:
\be
\sigma^+ \longrightarrow \tilde{\sigma}^+ ( \sigma^+ ) \;, \;\;\;
\sigma^- \longrightarrow \tilde{\sigma}^- ( \sigma^- ) \;.
\ee
Thus we are left with an infinite-dimensional group of symmetries,
which in particular includes scale transformations. 

\subsubsection{Equations of motion, closed and open strings, and
D-branes.}

In order to proceed we now spezialize to the case of a flat
space-time, $G_{\mu \nu} = \eta_{\mu \nu}$, where
$\eta_{\mu \nu} = \mbox{Diag}(-1,+1,\ldots,+1)$. 
In the conformal gauge the equation of motion for $X$
reduces to a free two-dimensional wave equation,
\be
%replaced \Box here.
\der^2 X^{\mu} = \der^{\alpha} \der_{\alpha} X^{\mu} = 0 \;.
\label{EomX}
\ee
Note that when imposing the conformal gauge on the 
Polyakov action (\ref{PolAct}), the equation of motion
for $h_{\alpha \beta}$, i.e., $T_{\alpha \beta}=0$, becomes
a constraint, which has to be imposed on the solutions
of (\ref{EomX}). 

The general solution of (\ref{EomX}) is a superposition of
left- and right-moving waves,
\be
X^{\mu} (\sigma) = X^{\mu}_L (\sigma^+) + X^{\mu}_R (\sigma^-) \;.
\ee
However, we also have to specify boundary conditions at the ends
of the string. One possible choice are periodic boundary conditions,
\be
X^{\mu} (\sigma^0, \sigma^1 + \pi) = X^{\mu} (\sigma^0, \sigma^1) \;.
\ee
They correspond to closed strings. A convient parametrization of 
the solution is:
\be
X^{\mu}(\sigma) = x^{\mu} + 2 \alpha' p^{\mu} \sigma^0
+ \I \;\sqrt{ 2 \alpha'} \sum_{n \not=0} \ft{\alpha^{\mu}_n}{n}
e^{-2 \I n \sigma^+}  
+ \I \; \sqrt{ 2 \alpha'} \sum_{n \not=0} \ft{\tilde{\alpha}^{\mu}_n}{n}
e^{-2 \I n \sigma^-}  \;.
\ee
Reality of $X^{\mu}$ implies: $(x^{\mu})^{\star} = x^{\mu}$ and
$(p^{\mu})^{\star} = p^{\mu}$ and $(\alpha^{\mu}_m)^{\star} 
= \alpha^{\mu}_{-m}$
and $(\tilde{\alpha}^{\mu}_m)^{\star} = \tilde{\alpha}^{\mu}_{-m}$.
Here $\star$ denotes complex
conjugation.
While $x^{\mu}$ is the position of the center of mass of the string
at time $\sigma^0$, $p^{\mu}$ is its total momentum. Thus, the 
center of mass moves on a straight line in Minkowski space,
like a free relativistic particle. The additional degrees of
freedom are decoupled left- and right-moving waves on the string,
with Fourier components $\alpha^{\mu}_m$ and $\tilde{\alpha}^{\mu}_m$.

When not choosing periodic boundary conditions, the world-sheet
has boundaries and we have open strings.
The variation 
of the world-sheet action yields a boundary term,
$\delta S \simeq \int_{\der \Sigma} d \sigma^0 \der_1 X^{\mu}
\delta X_{\mu}$. The natural choice to make the boundary term
vanish are Neumann boundary conditions,
\be
\left. \der_1 X^{\mu} \right|_{\sigma^1 = 0} = 0 \;, \;\;\;
\left. \der_1 X^{\mu} \right|_{\sigma^1 = \pi} = 0 \;.
\ee
With these boundary conditions, momentum is conserved at the ends
of the string. Left- and right-moving waves are reflected at the
ends and combine into standing waves. The solution takes the form
\be
X^{\mu}(\sigma) = x^{\mu} + (2 \alpha') p^{\mu} \sigma^0
+ \I \; \sqrt{ 2 \alpha'} \sum_{n \not= 0} \ft{\alpha^{\mu}_n}{n}
e^{- \I n \sigma^0} \cos ( n \sigma^1 ) \;.
\ee
There is, however, a second possible choice of boundary conditions
for open strings, namely Dirichlet boundary conditions.  Here
the ends of the string are kept fixed:
\be
\left. X^{\mu} \right|_{\sigma^1=0} = x^{\mu}_{(1)} \;, \;\;\;
\left. X^{\mu} \right|_{\sigma^1=\pi} = x^{\mu}_{(2)} \;.
\ee
With these boundary conditions the solution takes the form
\be
X^{\mu} (\sigma) = x^{\mu}_{(1)} + 
(x^{\mu}_{(2)} - x^{\mu}_{(1)}) \ft{\sigma^1}{\pi}
+ \I \; \sqrt{2 \alpha'} \sum_{n \not=0} \ft{ \alpha^{\mu}_n}{n}
e^{- \I n\sigma^0} \sin ( n \sigma^1 ) \;.
\ee
More generally we can impose Neumann boundary conditions
in the time and in p space directions and Dirichlet
boundary conditions in the other directions. Let us denote
the Neumann directions by $(X^m) = (X^0, X^1, \ldots ,X^p)$ and
the Dirichlet directions by $(X^a) = (X^{p+1} , \ldots, X^{D-1})$.

The most simple choice of Dirichlet boundary conditions is then to require
that all open strings begin and end on a p-dimensional
plane located at an arbitrary position $X^a = x^a_{(1)}$ along
the Dirichlet directions. Such a plane is called a 
p-dimensional Dirichlet-membrane, or D-p-brane, or simply D-brane
for short. While the ends of the strings are fixed in the Dirichlet 
directions, they still can move freely along the Neumann directions.
The world-volume of a D-p-brane is $(\mbox{p}+1)$-dimensional. The Neumann
directions are called the world-volume or the parallel directions,
while the Dirichlet directions are called transverse directions.

An obvious generalization is to introduce 
$N>1$ such D-p-branes, located at positions $x^a_{(i)}$, where
$i=1,\ldots, N$, and to allow strings to begin and end on any of
these. In this setting the mode expansion for a string starting 
on the i-th D-brane and ending on the j-th is:
\bea
X^{m}(\sigma) &=& x^{m} + (2 \alpha') p^{m} \sigma^0
+ \I \; \sqrt{ 2 \alpha'} \sum_{n \not= 0} \ft{\alpha^{m}_n}{n}
e^{- \I n \sigma^0} \cos ( n \sigma^1 ) \;, \nonumber \\
X^{a} (\sigma) &=& x^{a}_{(i)} + 
(x^{a}_{(j)} - x^{a}_{(i)}) \ft{\sigma^1}{\pi}
+ \I \; \sqrt{2 \alpha'} \sum_{n \not=0} \ft{ \alpha^{a}_n}{n}
e^{- \I n\sigma^0} \sin ( n \sigma^1 ) \;.
\eea
(One might also wonder about Dirichlet boundary conditions 
in the time direction. This makes sense, at least  for Euclidean 
space-time signature, and leads to instantons, called D-instantons,
which we will not discuss in these lectures.)

Dirichlet boundary conditions have been neglected for several 
years. The reason is that momentum is not conserved at the
ends of the strings, reflecting that translation invariance
is broken along the Dirichlet directions. Therefore, in a complete
fundamental theory the D-branes must be new dynamical objects, different
from strings. The relevance of such objects was only appreciated
when it became apparent that string theory already includes solitonic
space-time backgrounds, so called ('RR-charged') p-Branes,
which correspond to D-branes. We will return to this point later.

Promoting the D-branes to dynamical objects implies that they
will interact through the exchange of strings. This means that
in general they will repulse or attract, and therefore their
positions become dynamical. But there
exist many static configurations
of D-branes (mainly in supersymmetric string theories), where
the attractive and repulsive forces cancel for arbitrary
distances of the branes.

\subsection{Quantized bosonic strings \label{QBS}}

The definition of a quantum theory of bosonic strings 
proceeds by using standard recipies of quantization.
The two most simple ways to proceed are called
\lq old covariant quantization' and 
\lq light cone quantization'.
As mentioned above imposing the conformal gauge leaves
us with a residual gauge invariance. In light cone
quantization one fixes this residual invariance by
imposing the additional condition
\be
X^+ \stackrel{!}{=} x^+ + p^+ \sigma^+  \;,\;\;\;\mbox{i.e.}\;,\;\;
\alpha^+_m \stackrel{!}{=} 0 \;,
\ee
where $X^{\pm} = \ft1{\sqrt{2}} (X^{0} \pm X^{D-1})$ are 
light cone coordinates in space-time. Then the constraints
$T_{\alpha \beta} = 0$ are solved in the classical theory.
This yields (non-linear) expressions for the oscillators $\alpha^-_n$ in terms
of the transverse oscillators $\alpha^i_n$, $i=1, \ldots D-2$.
In light cone coordinates the world-sheet is embedded into space-time
along the $X^0, X^{D-1}$ directions. The independent 
degrees of freedom are the oscillations transverse
to the world sheet, which are parametrized by the 
$\alpha_n^i$.
One proceeds to quantize these degrees of freedom. In this 
approach unitarity of the theory is manifest, but Lorentz invariance
is not. 

In old covariant quantization one 
imposes the constraints at the quantum level.
Lorentz covariance is manifest, but unitarity is not:
one has to show that there is a positive definite 
space of states and an unitary S-matrix.
This is the approach we will describe in more detail below. 

One might also wonder about \lq new covariant quantization',
which is BRST quantization.
This approach is more involved but also more powerful then
old covariant quantization. When dealing
with advanced technical problems, 
for example the construction of scattering amplitudes
involving fermions in superstring theories, BRST techniques
become mandatory.
But this is beyond the scope of these lectures.

\subsubsection{The Fock space.}

The first step is to impose canonical commutation relations 
on $X^{\mu}(\sigma)$ and its canonical momentum
$\Pi^{\mu}(\sigma) = \der_0 X^{\mu}(\sigma)$. In terms of
modes one gets
\be
[x^{\mu}, p^{\nu}] = i \eta^{\mu \nu} \;, \;\;\,
[\alpha^{\mu}_m, \alpha^{\nu}_n] = m \eta^{\mu \nu} 
\delta_{m+n,0} \;.
\label{ComRel}
\ee
For closed strings there are 
analogous relations for $\tilde{\alpha}^{\mu}_m$.
The reality conditions of the classical theory translate into
hermiticity relations:
\be
(x^{\mu})^+ = x^{\mu} \;,\;\;\,
(p^{\mu})^+ = p^{\mu} \;,\;\;\,
(\alpha^{\mu}_m)^+ = \alpha^{\mu}_{-m} \,.
\label{HermProp}
\ee
While the commutation relations for $x^{\mu}, p^{\nu}$ are those
of a relativistic particle, the $\alpha^{\mu}_m$ satisfy the
relations of creation and annihilation operators of harmonic
oscillators, though with an unconventional normalization.

To proceed, one constructs a Fock space ${\cal F}$ on which the 
commutation relations (\ref{ComRel}) are repesented.
First one chooses momentum eigenstates $|k \rangle$, 
which are annihiliated by half of the oscillators:
\be
p^{\mu} |k \rangle = k^{\mu} | k \rangle \;, \;\;\,
\alpha^{\mu}_m |k \rangle = 0 
= \tilde{\alpha}^{\mu}_m |k \rangle \;, \;\;\; m>0 \;.
\ee
Then a basis ${\cal B}$ of ${\cal F}$ is obtained by acting with
creation operators:
\be
{\cal B} = \{ \alpha^{\mu_1}_{-m_1} \cdots \tilde{\alpha}^{\nu_1}_{-n_1}
\cdots | k \rangle  \;| \; m_l, n_l > 0 \} \;.
\ee
A bilinear form on ${\cal F}$, which is compatible with the hermiticity
properties (\ref{HermProp}), cannot be positive definite.
Consider for example the norm squared of the state
$\alpha^{\mu}_{-m} | k \rangle $:
\be
\langle k | (\alpha^{\mu}_{-m})^+ \alpha^{\mu}_{-m} | k \rangle
\sim \eta^{\mu \mu} = \pm 1 \;.
\ee
However, the Fock space is not the space of physical 
states, because we still have to impose the constraints. 
The real question is whether the subspace of physical 
states contains states of negative norm. 

\subsubsection{The Virasoro Algebra.}

Constraints arise when the canonical momenta of a system are not
independent.
This is quite generic for relativistic
theories. The most simple example is the relativistic particle,
where the constraint is the mass shell condition, 
$p^2 + m^2 = 0$. When quantizing the relativistic particle, physical
states are those annihilated by the constraint, i.e., states 
satisfying the mass shell condition:
\be
(p^2 + m^2) | \Phi \rangle = 0 \;.
\ee
When evaluating this in a basis of formal eigenstates of 
the operator $x^{\mu}$, one obtains the Klein-Gordon equation,
%replaced \Box here
$(\der^2 + m^2) \Phi (x) = 0$, where $\Phi(x) = \langle x | \Phi \rangle$
is interpreted as the state vector in the $x$-basis. This is
a clumsy way to approach the quantum theory of relativistic particles, 
and one usually prefers to use quantum field theory 
(\lq second quantization') rather than quantum mechanics
(\lq first quantization').
But in string theory it turns out that the first 
quantized formulation works nicely for studying the spectrum and
computing amplitudes, whereas string field theory is very
complicated.

Proceeding parallel to the case of a relativistic particle 
one finds that the canonical momentum is $\Pi^{\mu} =
\der_0 X^{\mu}$. The constraints
%, found from either of the
%actions (\ref{NGAct},\ref{PolAct}), 
are
\be
\Pi^{\mu} \der_1 X_{\mu} = 0 \;, \;\;\;
\Pi^{\mu} \Pi_{\mu} + \der_1 X^{\mu} \der_1 X_{\mu} = 0 \;.
\label{Constraints}
\ee
In the Polyakov formulation they are equivalent to $T_{\alpha \beta}=0$.
It is convenient to express the constraints through the
Fourier components of $T_{\alpha \beta}$. Passing to light cone
coordinates, the tracelessness of $T_{\alpha \beta}$, which holds
without using the equation of motion or imposing the constraints,
implies 
\be
T_{+-} = 0 = T_{-+} \;.
\ee
Thus we are left with two independent components, $T_{++}$ and
$T_{--}$, where $T_{\pm \pm} \simeq \der_{\pm} X^{\mu} 
\der_{\pm} X_{\mu}$. For closed strings, 
where $\der_{\pm} X^{\mu}$ are periodic
in $\sigma^1$, 
we expand $T_{\pm \pm}$ in a Fourier series and obtain
Fourier coefficients $L_m, \tilde{L}_m$, $m \in {\bf Z}$.
For open strings, observe that $\sigma^1 \rightarrow - \sigma^1$
exchanges $\der_+ X^{\mu}$ and $\der_- X^{\mu}$. Both fields
can be combined into a single field, which is periodic on
a formally doubled world-sheet with $- \pi \leq \sigma^1 \leq
\pi$. In the same way one can
combine $T_{++}$ with $T_{--}$. By Fourier expansion on the
doubled world-sheet one then obtains one set of Fourier modes
for the energy-momentum tensor, denoted
$L_m$. This reflects that left- and right-moving waves couple
through the boundaries. 

The explicit form for the $L_m$ in terms of oscillators is
\be
L_m = \ft12 \sum_{n = - \infty}^{\infty} \alpha_{m-n} \cdot
\alpha_n  \;,
\label{Lm}
\ee
with an analogous formula for $\tilde{L}_m$ for closed strings.
We have denoted the contraction of Lorentz indices by \lq $\cdot$'
and defined 
$\alpha_0^{\mu} = \ft12 p^{\mu} = \tilde{\alpha}^{\mu}_0$ for
closed strings and $\alpha^{\mu}_0 = p^{\mu}$ for open
strings. In terms of the Fourier modes, the constraints
are $L_m = 0$, and, for closed string, $\tilde{L}_m=0$. 
Translations in $\sigma^0$ are generated by $L_0$ for open and
by $L_0 + \tilde{L}_0$ for closed strings. These functions
are the world-sheet Hamiltonians. The $L_m$ satisfy
the Witt algebra algebra,
\be
\{ L_m, L_n \}_{\mscr{P.B.}} = i (m-n) L_{m+n} \;,
\ee
where $\{ \cdot, \cdot\}_{\mscr{P.B.}}$ is the Poisson
bracket. For closed strings we have two copies of this algebra.
The Witt algebra is the Lie algebra of infinitesimal
conformal transformations. Thus the constraints reflect that we
have a residual gauge symmetry corresponding to conformal
transformations. Since the constraints form
a closed algebra with the Hamiltonian, they are preserved in time.
Such constraints are called first class, and they can be 
imposed on the quantum theory without further modifications
(such as Dirac brackets).

In the quantum theory the $L_m$ are taken to be normal ordered,
i.e., annihilation operators are moved to the right. This
is unambigous, except for $L_0$. We will deal with this ordering
ambiguity below. The hermiticiy properties of the $L_m$ are:
\be
L_m^+ = L_{-m} \;.
\ee

The operators $L_m$ satisfy the Virasora algebra:
\be
[L_m, L_n] = (m-n) L_{m+n} + \frac{c}{12} (m^3 -m) \delta_{m+n,0} \;.
\label{VirAlg}
\ee
The Virasoro
algebra is a central extension of the Witt algebra. On our Fock space
${\cal F}$ the central charge $c$ takes the value
\be
c = \eta^{\mu \nu} \eta_{\mu \nu} = D \;,
\ee 
i.e., each space-time dimension contributes one unit. Since the Poisson
brackets of $L_m$ in the classical theory just give the Witt algebra,
this dependence on the number of dimensions is a new property of the
quantum theory. The extra central term occuring at the quantum level
is related to a normal ordering ambiguity of commutators with 
$m+n=0$. This results in a new \lq anomalous' term in the algebra.
In the context of current algebras such terms are known as
Schwinger terms.

\subsubsection{Imposing the constraints, or, why $D=26$?}

In the classical theory the constraints amount to imposing 
$L_m =0$ on solutions. Imposing this as an operator equation on 
the quantum theory is too strong. In particular it is not compatible
with the algebra (\ref{VirAlg}). What can be imposed consistently
is that matrix elements of the $L_m$ vanish between physical 
states, $\langle \Phi_1 | L_m | \Phi_2 \rangle=0$. Conversely 
this condition singles out the subspace of physical states,
${\cal F}_{\mscr{phys}}  \subset {\cal F}$. Using the hermiticity
properties of the $L_m$, this is equivalent to the statement that
the positive Virasoro modes annihilate physical states,
\bea
L_m | \Phi \rangle &=& 0 \;,\;\;\; m> 0 \;, \nonumber \\
(L_0 -a) | \Phi \rangle &=& 0 \label{Vir-Constr}\;,
\eea
for all $| \Phi \rangle \in {\cal F}_{\mscr{phys}}$. 
Note that we have introduced an undetermined constant $a$ into
the $L_0$-constraint. As mentioned above this operator has an
ordering ambiguity. We take $L_0$ to be normal ordered and
parametrize possible finite ordering effects by the constant $a$.
Since $L_0$ is the Hamiltonian, this might be considered as 
taking into account a non-trivial Casimir effect.
In the case of closed strings there is a second set of 
constraints involving the $\tilde{L}_m$.

The Virasoro operators $L_{-m}$, $m>0$ still act non-trivially 
on physical states and create highest weight representations
of the Virasoro algebra. This corresponds to the fact that
we still have residual gauge symmetries. Therefore it is clear
that ${\cal F}_{\mscr{phys}}$ is not the physical Hilbert space.
${\cal F}_{\mscr{phys}}$ is not positive definite,
but contains null states (states of norm zero) and, depending on
the number of space-time dimensions, also states of negative norm. 
A positive definite space of states can be constructed
if negative norm states are absent, such that ${\cal F}_{\mscr{phys}}$
is positive semi-definite, and if null states are orthogonal to
all physical states. 
Then one can consistently 
identify physical states $|\Phi \rangle$
that differ by null states $| \Psi \rangle $,
\be
| \Phi \rangle \simeq | \Phi \rangle + | \Psi \rangle \;,
\ee
and define the Hilbert space by
\be
{\cal H} = {\cal F}_{\mscr{phys}} / \{ \mbox{Null states} \} \;.
\ee
The working of this construction crucially depends on the values of $D$ and
$a$. This is the  contents of the so-called no-ghost theorem, which 
can be summarized as follows:
\begin{enumerate}
\item
$D=26$ and $a=1$. The construction works as described above.
The resulting theory is known as the critical (bosonic) string
theory, $D=26$ is the critical dimension. Physical states differing
by a null states differ by a residual gauge transformation and represent
the same state in the Hilbert space. We will see explicit examples below.
\item
$D>26$. The physical subspace ${\cal F}_{\mscr{phys}}$ 
always contains states of negative norm and no Hilbert space
${\cal H}$ can be constructed. There is no bosonic string theory
for $D>26$.  
\item
$D \leq 25$. Naively one expects such theories 
to be unitary, because we can just truncate the unitary 
critical string theory and this cannot introduce states
of negative norm. Nevertheless one does not obtain 
a consistent quantum theory by truncation. When 
studying scattering amplitudes at the loop level one
finds poles corresponding to unphysical negative norm states
and there is no unitary S-matrix. Thus truncations of the
critical string do not yield unitary theories.

But there is an alternative to truncation, known as
Liouville string theory or non-critical string theory. This theory
exists in $D<26$, at the price that the quantum theory has a new
degree of freedom, the Liouville mode. (This is most obvious
in a path integral formulation.)
The resulting theory
is much more complicated then the critical string, because 
its world-sheet theory  is interacting even for a flat target
space. For this theory much less is known then about the 
critical string. However, there are arguments indicating
that the non-critical string is 
equivalent to the critical string in a non-trivial background.
\end{enumerate}
We will only consider critical string theories in the
following. Also note that the above analysis applies to strings
in flat space-time, with no background fields. 
When switching on a non-trivial
dilaton background, this can modify the central charge of the
world-sheet conformal field theory, and, hence, the 
dimension of space-time. But this topic is beyond the scope 
of these lectures.

\subsubsection{The spectrum of the bosonic closed string.}

We can now identify the physical states 
by imposing the constraints. 
Let us consider closed strings. We first look at the two
constraints
\be
(L_0 -1 ) | \Phi \rangle = 0\;, \;\;\;
(\tilde{L}_0 -1 ) | \Phi \rangle = 0 \;.
\label{Constr1}
\ee 
The operator $L_0$ can be rewritten as
\be
L_0 = \ft18 p^2 + N \;.
\ee
As mentioned above the operator $L_0$ is the normal 
ordered version of (\ref{Lm}) with $m=0$. The original and
the normal ordered expression formally differ by an infinite
constant. Subtracting this constant introduces a finite 
ambiguity, which was parametrized by $a$. Unitarity then
fixed $a=1$. The oscillator part of $L_0$ is
\be
N = \sum_{n=1}^{\infty} 
\alpha_{-n} \cdot \alpha_n \;.
\label{NumbOp}
\ee
$N$ is called the number operator, because
\be
[N, \alpha^{\mu}_{-m} ] = m \alpha^{\mu}_{-m} \;.
\ee
Since the total momentum is related to the mass of the string 
by $M^2 + p^2 = 0$, the constraints (\ref{Constr1}) determine
the mass of a physical states in terms of the eigenvalues of $N$
and of its right-moving analogue $\tilde{N}$. (We denote
the operators and their eigenvalues by the same symbol.)
We now use the above decomposition of $L_0$, take the sum
and difference of the constraints (\ref{Constr1}) and
reintrodue the Regge slope $\alpha' = \ft12$ by dimensional
analysis:
\bea
\alpha' M^2 &=& 2 ( N + \tilde{N} - 2) \;, \nonumber \\
N  &=& \tilde{N}\;. \label{VirConstr}
\eea
The first equation is the mass formula for string states,
wherea the second equation shows that left- and right-moving
degrees of freedom must constribute equally to the mass.

Let us list the lightest states satisfying these constraints:
\be
\begin{array}{|l|l|l|} \hline
\mbox{Occupation} & \mbox{Mass} & \mbox{State} \\ \hline \hline
N = \tilde{N} = 0 & \alpha'M^2 = -4 & | k \rangle \\ \hline
N = \tilde{N} = 1 & \alpha'M^2 = 0 &  \alpha^{\mu}_{-1} 
\tilde{\alpha}^{\nu}_{-1} | k \rangle \\ \hline
N = \tilde{N} = 2 & \alpha'M^2 = 4 &  \alpha^{\mu}_{-2} 
\tilde{\alpha}^{\nu}_{-2} | k \rangle \\
 & & \alpha^{\mu}_{-2} 
\tilde{\alpha}^{\nu}_{-1} \tilde{\alpha}^{\rho}_{-1} | k \rangle \\
 & & \alpha^{\mu}_{-1} \alpha^{\nu}_{-1} 
\tilde{\alpha}^{\rho}_{-2} | k \rangle \\
 & & \alpha^{\mu}_{-1} \alpha^{\nu}_{-1} 
\tilde{\alpha}^{\rho}_{-1}\tilde{\alpha}^{\sigma}_{-1} 
| k \rangle \\ \hline
\end{array}
\ee

The most obvious and disturbing fact is that the ground state is
a tachyon, i.e., a state of negative mass squared. Since the mass
squared of a scalar corresponds to the curvature of the potential
at the critical point, we seem to be expanding around a maximum
rather then a minimum of the potential.
This signals that the bosonic closed string
quantized in flat Minkowski space is unstable. 
It is a very interesting question whether there is a minimum
of this potential which provides a stable ground state. Since 
the tachyon aquires a vacuum expectation value in this minimum,
this is referred to as tachyon condensation. But since we will be
mostly interested in superstring theories, where tachyons are
absent, we will simply ignore the fact that our toy model has a
tachyon.

The first excited state is massless, and on top of it we find
an infinite tower of states with increasing mass. Since the
mass scale of string theory presumably is very large, we will focus
on the massless states.  So far we only imposed the constraints
(\ref{Constr1}). The other constraints
\be
L_m | \Phi \rangle = 0 \;, \;\;\;
\tilde{L}_m | \Phi \rangle = 0 \;, \;\;\; m>0 \;,
\label{Constr2}
\ee
impose conditions on the polarisations of physical states.   
For the tachyon one gets no condition, while for the first
excited level the constraints with $m=1$ are non-trivial.
Forming a general linear combination of basic states,
\be
\zeta_{\mu \nu} \alpha^{\mu}_{-1} \tilde{\alpha}^{\nu}_{-1}
| k \rangle \;,
\label{StateLevel1}
\ee
the constraints (\ref{Constr2}) imply
\be
k^{\mu} \zeta_{\mu \nu} = 0 = k^{\nu} \zeta_{\mu \nu} \;.
\ee
Since $\zeta_{\mu \nu}$ is the polarization tensor, we see that
only states of transverse polarization are physical. To obtain
the particle content, we have to extract the irreducible 
representations of the D-dimensional Poincar\'e group contained 
in physical $\zeta_{\mu \nu}$. There are three such representations:
the traceless symmetric part describes 
a graviton $G_{\mu \nu}$, the trace part corresponds to a scalar,
the dilaton $\Phi$, and the third representation is an antisymmetric
tensor $B_{\mu \nu}$. In order to disentangle the trace part,
one needs to introduce an auxiliary vector $\overline{k}$, with
the properties:
\be 
\overline{k}\cdot \overline{k} = 0 \;,\;\;\;
k \cdot \overline{k} = -1 \;.
\ee
($k$ is the momentum vector.)
The polarization tensors of the graviton, dilaton and
antisymmmetric tensor are:
\bea
\zeta^{G}_{\mu \nu} &=& \zeta_{(\mu \nu)}
- \ft1{D-2} \zeta^{\rho}_{\rho} ( \eta_{\mu \nu} - k_{\mu}
\overline{k}_{\nu} - k_{\nu} \overline{k}_{\mu}) \;, \nonumber \\
\zeta^{\Phi}_{\mu \nu} &=& \ft1{D-2} 
\zeta^{\rho}_{\rho} ( \eta_{\mu \nu} - k_{\mu}
\overline{k}_{\nu} - k_{\nu} \overline{k}_{\mu}) \;, \nonumber \\
\zeta^{B}_{\mu \nu} &=& \zeta_{[\mu \nu]}  \;,
\eea
where $\zeta_{(\mu \nu)}=\ft12 ( \zeta_{\mu \nu} + \zeta_{\nu \mu})$ 
and $\zeta_{[\mu \nu]} = \ft12 ( \zeta_{\mu \nu} - \zeta_{\nu \mu})$ 
are the symmetric and
antisymmetric parts of $\zeta_{\mu \nu}$. Note that the prefactor
$1/({D-2})$ is needed in order that the trace part is physical.
Using explicit choices for $k,\overline{k}$ one can check
that $\zeta^{G}_{\mu \nu}$ is the polarization tensor
of a plane wave and transforms as a traceless symmetric 
tensor under transverse rotations.

As we discussed above, physical states are only defined up
to the addition of null states, $| \Phi \rangle \sim
| \Phi \rangle + | \Psi \rangle $. In the case at hand 
adding null states corresponds to adding states of longitudinal
polarization, according to:
\bea
\zeta_{(\mu \nu)} &\sim & \zeta_{(\mu \nu)} + k_{\mu} \zeta_{\nu}
+ \zeta_{\mu} k_{\nu} \nonumber \\
\zeta_{[\mu \nu]} &\sim & \zeta_{[\mu \nu]} + k_{\mu} \xi_{\nu}
- \xi_{\mu} k_{\nu} \;.
\eea
$\zeta_{\mu}$ and $\xi_{\mu}$ are arbitrary vectors 
orthogonal to the momentum $k^{\mu}$. Adding null states
can be understood as a residual gauge transformation parametrized
by $\zeta_{\mu}, \xi_{\mu}$. By taking Fourier transforms we see
that these are the standard gauge invariances of a graviton and of
an antisymmetric tensor, repectively:
\bea
G_{\mu \nu} & \sim & G_{\mu \nu} + \der_{\mu} \Lambda_{\nu}
+ \der_{\nu} \Lambda_{\mu} \;, \label{GaugeTGrav} \nonumber \\
B_{\mu \nu} & \sim & B_{\mu \nu} + \der_{\mu} A_{\nu}
- \der_{\nu} A_{\mu} \;.
\eea
A graviton is defined by taking the gravitational action and 
expanding the metric around a flat
background. The gauge transformations are then infinitesimal
reparametrization, which, in a flat background, act according
to (\ref{GaugeTGrav}) on the metric. Note that our
gauge transformations $\Lambda_{\mu}, A_{\mu}$ have a vanishing
divergence, because the corresponding polarization vectors are
orthogonal to the momentum. This reason is that the Virasoro
constraints automatically impose a generalized Lorentz gauge.

Thus far our identification of the symmetric traceless part of 
the state (\ref{StateLevel1}) as a graviton is based on the fact
that this state has the same kinematic properties as
a graviton in Einstein gravity. We will see later, 
after analyzing string interactions, that 
this extends to the dynamical properties.

Finally it is interesting to compare the results of old
covariant quantization to those obtained in light cone quantization.
In light cone quantization unitarity is manifest, but the 
Lorentz algebra of the quantum theory has an anomaly which only
cancels in the critical dimension $D=26$. Moreover the
normal ordering constant must take the value $a=1$. Independently,
the same value of $a$ is obtained when computing the Casimir
energy of the ground state using $\zeta$-function regularization.
One virtue of light cone quantization is that one can write down
immediately all the physical states. A basis is provided by
all states which can be created using transverse oscillators,
\be
\alpha^{i_1}_{-m_1} \cdots \tilde{\alpha}^{j_1}_{-n_1} \cdots | k \rangle \;,
\ee
where $i_1, \ldots, j_1, \ldots = 1, \ldots, D-2$. What remains
is to group these states into representations of the D-dimensional
Poincar\'e group. Massless states are classified by the little group
$SO(D-2)$. Since all states manifestly are tensors with respect
to this subgroup, one immediately sees that the massless states
are a graviton (traceless symmetric tensor), dilaton (trace) 
and antisymmetric tensor. For massive states the little group
is the full rotation subgroup $SO(D-1)$. Using Young tableaux it is
straightforward to obtains these from the given representations 
of $SO(D-2)$.

\subsubsection{Open strings.}

Having treated the closed bosonic string in much detail, we
now describe the results for open
strings. One finds the same critical dimension, $D=26$, and
the same value of the normal ordering constant, $a=1$. 
The constraints read:
\be
(L_0 -1) | \Phi \rangle = 0 \;, \;\;\;
L_m | \Phi \rangle = 0 \;, \;\;m > 0 \;.
\ee
$L_0$ can be decomposed as $L_0 = \ft12 p^2 + N$, where $N$
is the number operator. 
The $L_0$-constraint gives the mass formula:
\be
\alpha' M^2 = N - 1  \;.
\ee
Therefore the lowest states are:
\be
\begin{array}{|l|l|l|} \hline
\mbox{Occupation} & \mbox{Mass} & \mbox{State} \\ \hline \hline
N=0 & \alpha' M^2 = -1 & | k \rangle \\ \hline
N=1 & \alpha' M^2 = 0 & \zeta_{\mu} \alpha^{\mu}_{-1} | k 
\rangle \\ \hline
N=2 & \alpha' M^2 = 1 & \zeta_{\mu \nu} 
\alpha^{\mu}_{-1} {\alpha}^{\nu}_{-1} | k \rangle \\
 & & \zeta_{\mu} \alpha^{\mu}_{-2} | k \rangle \\ \hline
\end{array}
\ee
The other constraints impose restrictions on the polarizations.
Whereas the groundstate is a tachyonic scalar, the massless state
has the kinematic properties of a gauge boson: its polarization
must be transverse,
\be 
\zeta_{\mu} k^{\mu} = 0 \;,
\ee
and polarizations proportional to $k^{\mu}$ correspond to null
states,
\be
\zeta_{\mu} \sim \zeta_{\mu} + \alpha k_{\mu} \;.
\ee
This is the Fourier transform of a $U(1)$ gauge transformation,
\be
A_{\mu} \sim A_{\mu} + \der_{\mu} \chi \;.
\ee
Whereas massless closed string states mediate gravity,
massless open string states mediate gauge interactions.

\subsubsection{Chan-Paton factors.}

Open string theory has a generalization which has
non-abelian gauge interactions. One can assign additional
degress of freedom to the ends of the string, namely 
charges (\lq Chan-Paton factors')
which transform in the fundamental and anti-fundamental
(complex conjugated) 
representation of the group $U(n)$. The massless states 
then take the form
\be
\zeta_{\mu} \alpha^{\mu}_{-1} | k, a, \overline{b} \rangle \;,
\ee
where $a$ is an index transforming in the fundamental
representation $[n]$ of $U(n)$, whereas $\overline{b}$ 
transforms in the anti-fundamental representation $[\overline{n}]$.
Since
\be
[n] \times [\overline{n}] = \mbox{adj} \;U(n) \;,
\ee
the massless states transform in the adjoint of $U(n)$ and
can be interpreted as $U(n)$ gauge bosons. (As for the graviton,
we have only seen the required kinematic properties so far.
But the interpretation is confirmed when studying interactions.)

Note that $U(n)$ is the only compact Lie group where the 
adjoint representation is the product of the fundamental and
anti-fundamental representation. Therefore the construction 
precisely works for these groups.

\subsubsection{Non-oriented strings.}

There is a further modification 
which leads to non-oriented strings.
These are obtained from the theories constructed so far
by a projection. Both closed and open bosonic string 
theories are symmetric under
world-sheet parity, which is defined as a reflection
on the world-sheet:
\be
\Omega: \sigma^1 \longrightarrow \pi - \sigma^1 =
- \sigma^1 \;\;\mbox{modulo} \;\;\pi \;.
\ee
Since $\Omega$ is an involution, $\Omega^2 =1$, the spectrum
can be organized into states with eigenvalues $\pm 1$:
\bea
\Omega | N, k \rangle &=& (-1)^N | N, k \rangle \;, \\
\Omega | N, \tilde{N}, k \rangle &=&
| \tilde{N}, N, k \rangle \;.
\eea
Here $|N,k\rangle$ is an open string state with momentum $k$ and total
occupation number $N$ and $|N,\tilde{N},k\rangle$ is a closed
string state with momentum $k$ and total left and right
occupation numbers $N,\tilde{N}$.

Non-oriented strings are defined by keeping only those states
which are invariant under $\Omega$. The resulting theories
are insensitive to the orientation of the world-sheet. 
Let
us look at the effect of this projection on the lowest states.
For open strings we are left with:
\be
\begin{array}{|l|l|l|} \hline
\mbox{Occupation} & \mbox{Mass} & \mbox{State} \\ \hline \hline
N=0 & \alpha' M^2 = -1 & | k \rangle \\ \hline
N=1 & \alpha' M^2 = 0 &  - \\ \hline 
N=2 & \alpha' M^2 = 1 & \zeta_{\mu \nu} 
\alpha^{\mu}_{-1} {\alpha}^{\nu}_{-1} | k \rangle \\
 & & \zeta_{\mu} \alpha^{\mu}_{-2} | k \rangle \\ \hline
\end{array}
\ee
All states with odd occupation numbers are projected out,
including the gauge boson. For closed strings we obtain:
\be
\begin{array}{|l|l|l|} \hline 
\mbox{Occupation} & \mbox{Mass} & \mbox{State} \\ \hline \hline
N = \tilde{N} = 0 & \alpha'M^2 = -4 & | k \rangle \\ \hline 
N = \tilde{N} = 1 & \alpha'M^2 = 0 &  \zeta_{(\mu \nu)} \alpha^{\mu}_{-1} 
\tilde{\alpha}^{\nu}_{-1} | k \rangle \\ \hline 
N = \tilde{N} = 2 & \alpha'M^2 = 4 &  \zeta_{(\mu \nu)} \alpha^{\mu}_{-2} 
\tilde{\alpha}^{\nu}_{-2} | k \rangle \\
 & & \zeta_{(\mu \rho \nu \sigma)}
\alpha^{\mu}_{-1} \alpha^{\nu}_{-1} 
\tilde{\alpha}^{\rho}_{-1}\tilde{\alpha}^{\sigma}_{-1} 
| k \rangle \\ \hline
\end{array}
\ee
Only states which are left-right symmetric survive. At the massless
level the antisymmetric tensor is projected out, whereas the 
graviton and dilaton are kept.

\subsubsection{Chan-Paton factors for non-oriented strings.}

The above construction can be generalized to open strings with Chan-Paton 
factors.  In this case the two representations
assigned to the ends of the strings must be equivalent. One can
define a generalized involution $\Omega'$, which combines world-sheet
parity with an action on the Chan-Paton indices,
\be
\Omega' | N, a, b \rangle = \varepsilon (-1)^N | N, b, a \rangle \;,
\ee
where $\varepsilon = \pm 1$. The projection is 
$\Omega' | N, a, b \rangle \stackrel{!}{=} | N, a ,b \rangle$.
There are two inequivalent choices of the projection. For $\varepsilon=1$,
the indices
$a,b$ must transform in the fundamental representation of $SO(n)$.
Since the adjoint of $SO(n)$ is the antisymmetric product of the fundamental
representation 
with itself, the massless vector state transforms in the adjoint. More 
generally, states at even (odd) mass level transform as symmetric 
(antisymmetric) tensors.

The other choice is $\varepsilon = -1$. Then $a,b$ transform in the
fundamental of $USp(2n)$ (the compact form of the symplectic group. Our
normalization is such that $USp(2) \simeq SU(2)$.) Since the adjoint
of $USp(2n)$ is the symmetric product of the fundamental representation
with itself,
the massless vector transforms in the adjoint. More generally,
states at even (odd) mass level transform as antisymmetric (symmetric)
tensors.

\subsubsection{D-branes.}

Finally we can consider open strings with Dirichlet boundary
conditions along some directions. Consider first oriented open
strings ending on a D-p-brane located at $x^a_{(1)}$. The ground
state is tachyonic. The massless state of an open string 
with purely Neumann condition is a D-dimensional 
gauge boson $\alpha^{\mu}_{-1}|k \rangle$.
Now we impose Dirichlet boundary conditions
along the directions $a= p+1, \ldots, D-1$, so that 
the string can only
move freely along the Neumann directions
$m=0,1,\ldots,p$. The relevant kinematic
group is now the world-volume Lorentz group $SO(1,\mbox{p})$.
The massless states are a world-volume vector,
\be
\alpha^{m}_{-1} | k  \rangle \;,\;\;\;m=0,1,\ldots, \mbox{p} 
\ee
and
$\mbox{D}-\mbox{p}-1$ scalars,
\be
\alpha^{a}_{-1} | k \rangle \;, \;\;\;a=\mbox{p}+1, \ldots , 
\mbox{D}-1 \;.
\ee
The scalars correspond to transverse oscillations of the brane.
Changing the position of the brane corresponds to changing
the vacuum expectation values of the scalars. The effective
action of the massless modes is given, to leading order in 
$\alpha'$, by the dimensional reduction of the D-dimensional
Maxwell action to $\mbox{p}+1$ dimensions. The full effective action
is of Born-Infeld type.

Next consider $N$ parallel D-p-branes, located at positions
$x^a_{(i)}$. The new feature of this configuration is that
there are strings which start and end on different branes.
For such strings there is an additional term in the mass formula,
which accounts for the stretching:
\be
\alpha' M^2 = N - 1 + \left( 
\frac{| \vec{x}_{(i)} - \vec{x}_{(j)} |}{2 \pi \sqrt{\alpha'}}
\right)^2 \;.
\ee
Here $\vec{x}_{(i)}$ is the position of the i-th brane.
(Remember that the tension of the string is $(2 \pi \alpha')^{-1}$.)
Due to the normal ordering constant, the ground state becomes
tachyonic if two branes come close enough. This signals
an instability of the D-brane configuration. As already 
mentioned this might lead to interesting dynamics (tachyon 
condensation, decay of D-branes), but we will not discuss this
here. Instead, we focus on features shared by D-branes in
supersymmetric string theories. The states at the first 
excited level become massless precisely if the corresponding
D-branes are put on top of each other. Each of the $N$ branes
already carries a $U(1)$ gauge theory: the massless modes of strings 
beginning and ending of the same brane give $N$  
vectors and ${N} \cdot ( \mbox{D}-\mbox{p}-1)$ scalars.
For $N$ coinciding branes we get additional 
$N \cdot (N-1)$ vectors and $N \cdot (N-1) 
\cdot (\mbox{D}-\mbox{p}-1)$ scalars. 
Combining all massless states one gets one vector and
$\mbox{D}-\mbox{p}-1$ scalars  
in the adjoint representation of the non-abelian group $U(N)$.
This suggests that the D-brane system describes 
a $U(N)$ gauge theory with an adjoint Higgs mechanism.
The Higgs mechanism is realized geometrically: Higgs
expectation values correspond to the distances between branes,
and the masses can be understood in terms of stretched 
strings. Again, this interpretation, which is based on 
analyzing the spectrum is confirmed when studying interactions.
Besides Chan-Paton factors, D-branes are a second possibility
to introduce non-abelian gauge groups. In fact Chan-Paton factors
are related to D-branes through T-duality, but we 
will not be able to discuss this in these lectures.

The above construction can be extended to non-oriented 
strings, where other gauge groups occure. There are various
other generalizations, which allow one to construct and study
various gauge theories using strings and D-branes. These 
techniques are known as \lq D-brane engineering' of field theories.
Besides being of interest for the study of field theories
through string methods, D-branes are important for understanding
string theory itself. As we will see later, D-branes are actually
solitons of string theory. Thus we are in the privileged
position of knowing the exact excitation spectrum 
around such solitons in terms of open strings. This
can be used, for example, to compute the entropy and
Hawking radiation of black holes.

Another application of D-branes goes under the name 
of \lq brane worlds' or \lq brane universes' or \lq models with
large extra dimensions'. As we have seen,
D-branes enable one to localize gauge interactions and
matter on a lower-dimension\-al submanifold of space-time.
This leads to models with space-dimensions 
where only gravity (closed strings) but not standard model
matter (open strings) can propagate. Empirical limits on the size 
of the dimensions transverse to the brane only come from gravity, which is
much weaker than all other interactions. Therefore such
dimensions can be quite large, even up to about 1 mm. This is in 
contrast to limits on extra dimensions which are accessible to standard
model interactions. Here the experimental limits are set by the scale 
resolvable in current accelerator experiments.

Brane world
models are nowadays popular in both particle physics and
cosmology. In particular, they can be used to construct models
where the fundamental gravitational scale is of order 1 TeV.
We will come back to these applications of D-branes in section 6.

\subsection{Further reading}

The material covered in this section can be found in all
of the standard textbooks on string theory \cite{GSW,LT,Pol,Kak1,Kak2}.
Dirichlet boundary conditions and D-branes are only covered by
the more recent ones \cite{Pol,Kak1}.

\section{Interacting bosonic strings}

So far we have not specified how strings interact. One might
expect that this can be done by adding interaction terms to
the world-sheet action. However, we have to respect the 
local symmetries of the Polyakov action, which severely restricts
our options. In particular, contact 
interactions, which are frequently used in describing 
non-fundamental string-like objects such as polymers, are not
compatible with Weyl invariance. Admissible interacting
world-sheet actions include marginal deformations 
of the Polyakov action, i.e., deformations which preserve
Weyl invariance. One such deformation replaces the flat
space-time metric by a curved one. As expected intuitively,
such an action does not describe interactions among strings, but
strings moving in a non-trivial background. The same is true
when replacing the Polaykov action by more general conformal 
field theories.

How then do we define interactions? We will give a heuristic
discussion in the next section. The resulting scattering
amplitudes are Lorentz covariant, unitary and UV finite.
They include the 
Veneziano amplitude and its cousins, which historically 
started the subject. 

For definiteness we will focus in the following on closed
oriented strings. The generalization to other string theories
will be indicated briefly.

\subsection{Heuristic discussion}

Intuitively, interactions between strings are described by
world-sheets which connect a given initial configuration of 
strings to a final configuration. One can
draw several such world-sheets, which differ by their topologies. 
Comparing to the similar treatment of point particles by graphs,
we realize that while graphs have vertices,
the world-sheets connecting strings are manifolds without 
distinguished interaction points. This leads to the expectation
that string interactions are less singular then those of point
particles, which is indeed confirmed by the final result of 
the construction. Moreover, it indicates that one does not
have any freedom in defining interactions. For particles, we can
assign couplings to vertices which depend on the species 
of the particles meeting at the vertex. For strings the 
interaction is encoded in the topology of the world-sheet and there
is no such freedom. There is one fundamental interaction,
which couples three closed strings, and all we can do is to
assign a coupling constant $\kappa$ to it. 

Next, we restrict ourselves to finding transition amplitudes 
between asymptotic states in the infinite past and future. 
An asymptoting in- or out-going state is represented by
a semi-infinite cylinder. When mapping this to a punctered disc,
the asymptotic state is represented by the puncture. This leads to the
idea that we can represent the asymptotic state by a local operator
of the world-sheet field theory. Such operators are called 
vertex operators. Note that they do not describe 
interactions. Instead, the vertex operator $V_{\Phi} (\sigma)$ 
describes the creation or annihilation of 
the string state $| \Phi \rangle$ at the position $\sigma$ on the
world-sheet. That is, they allow us to assign a copy of the
space of physical states to every point of the world-sheet.
As we will see below, there is indeed a natural one-to-one map
between physical states $| \Phi \rangle$ and local operators
of the world-sheet field theory.

After replacing the world-sheet punctures by insertions of 
vertex operators we are left with compact closed surfaces.
The topologies of such surfaces are classified by their
genus $g \geq 0$, or equivalently, by their Euler number
$\chi = 2 - 2g$. Here $g=0$ is the two-sphere, and $g=1$ is the
torus. The general genus $g$ surface $\Sigma_g$
is obtained from the sphere by attatching $g$ handles. 
The handles play the role of loops in Feynman
diagrams. When considering
an interaction process on $\Sigma_g$ involving $M$ external states,
we find $M - \chi$ fundamental string interactions and have to  
assign a factor $\kappa^{M-\chi}$.

We now postulate that a scattering amplitude involving
$M$ external states is given by
\be
A(1,\ldots, M) = \sum_{g=0}^{\infty} \kappa^{M-\chi}
A(1,\ldots, M)_g  \;,
\label{Amp1}
\ee
where $A(1,\ldots, M)_g$ is the contribution of $\Sigma_g$.
This is a perturbative expression in the string coupling
$\kappa$. As usual for theories with a single coupling, the 
expansion in the coupling coincides with the expansion in loops,
which in our case is the expansion in the genus $g$.

The genus $g$ contribution is defined to be
\be
A(1,\ldots, M)_g = \langle V_1 \cdots V_M \rangle_g \;,
\label{Amp2}
\ee
where 
\be
V_i = \int_{\Sigma_g} d^2 \sigma_i  \sqrt{h} V_i (\sigma_i)
\label{Amp3}
\ee
are the so-called integrated vertex operators, which are
obtained by integrating the vertex operators $V_i(\sigma_i)$
over the world sheet. (Though our
notation might  suggest otherwise, we do not require that $\Sigma_g$
can be covered by one set of coordinates, which is of course
impossible for compact $\Sigma_g$. We just use a local
representative of the integrand for notational purposes.)
In (\ref{Amp2}) we compute the correlation function 
of the vertex operators $V_i(\sigma_i)$ on $\Sigma_g$ in the world-sheet 
quantum field theory defined by the Polyakov action and 
integrate over the positions of the vertex operators. The result
is interpreted as a scattering amplitude of string states in
space-time, with the in- and out-states represented by the
vertex operators.

Note that it is not possible to introduce arbitrary weight factors
between the contributions of different genera. The reason is that
unitarity requires that scattering amplitudes factorize into
the amplitudes of subprocesses whenever an intermediate state
is on-shell. In fact, in 
the old days of string theory this was used to construct the perturbative
expansion by seewing together tree amplitudes. However, this approach
is more cumbersome then the Polyakov path integral approach that 
we will desribe here.

\subsection{Vertex operators}

We now take a closer look at the vertex operators. Observe 
that the scattering amplitudes defined by (\ref{Amp1},\ref{Amp2},\ref{Amp3})
must be invariant under reparametrizations of the world-sheets.
In particular the local vertex operators $V_i(\sigma_i)$ must
transform such that (\ref{Amp3}) is invariant. When imposing 
the conformal gauge, it still must transform in a specific
way under conformal transformations 
$\sigma^{\pm} \rightarrow \tilde{\sigma}^{\pm} ( \sigma^{\pm} )$.
In conformal field theory fields which transform covariantly
under conformal transformations are called primary conformal fields.
A primary conformal field of weights $(h, \overline{h})$ is an 
object that transforms like a contravariant tensor field of rank
$(h,\overline{h})$:
\be
\tilde{V}( \tilde{\sigma}^+, \tilde{\sigma}^-) = 
\left( \frac{ d \sigma^+ }{ d \tilde{\sigma}^+ } \right)^h
\left( \frac{ d \sigma^- }{ d \tilde{\sigma}^- } \right)^{\overline{h}} 
V(\sigma^+, \sigma^-)\,.
\ee
Invariance of (\ref{Amp3}) implies that vertex operators of
physical states must be primary conformal fields of weights
$(1,1)$. This property is equivalent to imposing 
the Virasoro constraints (\ref{VirConstr}) 
on physical states. States assigned to a point $P$ of
$\Sigma$ are constructed 
from vertex operators by applying them to a ground state $|0\rangle_P$,
\be
| \Phi \rangle = V_{\Phi}(P) | 0 \rangle_P \;.
\ee
To make contact with the space ${\cal F}_{\mscr{phys}}$ constructed
in section \ref{QBS},
one parametrizes $\Sigma$ in the vicinity
of $P$ by a semi-infinity cylinder with $P$ being the asymptotic 
point $\sigma^0 \rightarrow -\infty$. Intuitively
this describes an ingoing state created in the infinite past.
Then,
\be
| \Phi \rangle = \lim_{\sigma^0 \rightarrow - \infty}
V_{\Phi}(\sigma) | 0 \rangle \;,
\ee
where $|0 \rangle := | k =0 \rangle$ is the (unphysical) 
zero-momentum state with occupation numbers
$N=0=\tilde{N}$ in ${\cal F}$.

To indicate how this works in practice, we now specify the vertex operators
for the lowest states. Consider the operator
\be
V(\sigma) = : e^{\I k_{\mu} X^{\mu} } : (\sigma) \;,
\label{TachyVop}
\ee
where $: \cdots : $ indicates normal ordering. Applying this
operator we find
\be
\lim_{\sigma^0 \rightarrow - \infty}  : e^{\I k_{\mu} X^{\mu} } : (\sigma)
| 0 \rangle = e^{\I k_{\mu} x^{\mu} } | 0 \rangle = | k \rangle \;,
\ee
where we used that $e^{\I k_{\mu} x^{\mu} } | 0 \rangle$
is an eigenstate of $p^{\mu}$ with eigenvalues $k^{\mu}$. 
One can show that (\ref{TachyVop}) has weights
$( \ft18 k^2, \ft18 k^2 )$. Thus it has weights $(1,1)$
if $k^2 = 8$, which is the physical state condition
$M^2 = -8$ for the tachyonic ground state of the closed string.
(We have set $\alpha'=\ft12$.)

The vertex operator for the first excited level is
\be
V(\sigma) = : \zeta_{\mu \nu} \der_{+} X^{\mu}
\der_{-} X^{\nu} e^{\I k_{\rho} X^{\rho}} : (\sigma) \;.
\ee
This has weights $(1,1)$ if 
\be
k^2 = 0 \;, \;\; \;k^{\mu} \zeta_{\mu \nu} = 0 = k^{\nu} \zeta_{\mu \nu}\;,
\ee
which is precisely the physical state condition for the state
\be
\zeta_{\mu \nu} \alpha^{\mu}_{-1} \tilde{\alpha}^{\nu}_{-1} 
| k \rangle \;.
\ee
More generally, vertex operators of the form
\be
V(\sigma) = : \zeta_{\mu_1 \cdots \nu_1 \cdots }
\der_+^{m_1} X^{\mu_1} \cdots \der_-^{n_1} X^{\nu_1} \cdots
e^{\I k_{\rho} X^{\rho}} : (\sigma)
\ee
generate states of the form
\be
 \zeta_{\mu_1 \cdots \nu_1 \cdots }
\alpha^{\mu_1}_{-m_1} \cdots \tilde{\alpha}^{\nu_1}_{-n_1} \cdots
| k \rangle \;.
\ee

\subsection{Interactions in the path integral formalism}

The next step is to explain in more detail how the amplitudes 
(\ref{Amp1})--(\ref{Amp3}) are defined and how they are computed
in practice. 
As usual one can use either the path integral (Lagrangian)
or the operator (Hamiltonian) formulation. We will use Polyakov's
path integral formulation. This has the advantage of immediately 
providing explicit formal expressions for correlation functions.
The mathematical complications of defining the interacting 
quantum theory are hidden in the path integral measure. 
We will not discuss this is full detail, but mention and illustrate the most 
important points.

\subsubsection{The path integral.}

We now turn to the Polaykov path integral, which is one way
to give a precise meaning to (\ref{Amp1}). In this approach the
correlation function (\ref{Amp2}) is computed by functional methods.
Intuitively we integrate over all paths that strings can take
in space-time. However, in order to have a well defined path
integral, we need to study the theory in Euclidean signature,
both on the world-sheet and in space-time. A Euclidean formulation
of the world-sheet theory is needed to have a well defined 
functional integral for the world-sheet field theory. In particular,
we want to have well defined world-sheet metrics on general
surfaces $\Sigma_g$, which is not possible for Lorentzian
signature. Second, one also has to work in Euclidean space-time,
in order to have a standard Gaussian integral for the \lq time'
coordinate $X^0$. Wick-rotating $X^0$ can be interpreted as
continuing to unphysical Euclidean momenta and polarizations.
As we have seen in our discussion of vertex operators the
string coordinates $X^{\mu}$ are always contracted with momenta
and polarizations. Physical scattering amplitudes are thus
obtained by computing (\ref{Amp1}) in the Euclidean theory
and evaluating the result for physical momenta and polarizations.
This uses the analycity properties expected to hold for any relativistic
unitary scattering amplitude. For tree-level amplitudes one
can study how the Wick rotation works explicitly, by comparing
to results obtained by operator methods.

Our starting point is the Polyakov action on a
world-sheet $\Sigma$ with positive definite metric $h_{\alpha \beta}$
and local complex coordinate $z$,
\be
S_{\mscr{P}} = \frac{1}{4 \pi \alpha'} \int_{\Sigma} d^2 z
\sqrt{h} h^{\alpha \beta} \der_{\alpha} X^{\mu} \der_{\beta} X_{\mu} \;.
\label{EucPol}
\ee
The quantum theory is now defined by summing over all topologies
of $\Sigma$ and integrating over $X^{\mu}$ and $h_{\alpha \beta}$:
\be
A(1,\ldots, M) = \sum_{g=0}^{\infty} \kappa^{M - \chi} 
N_g \int DX^{\mu} D h_{\alpha \beta} e^{- S_P[X,h]}
V_1 \cdots V_M \;,
\label{PolPI}
\ee
where $V_i$ are the integrated vertex operators of the physical 
states and $N_g$ are normalization factors needed to define the
path integral. The $V_i$ depend on $X^{\mu}$ through the local
vertex operators $V_i(\sigma_i)$, while the world-sheet metric
enters through the integration over $\sigma_i$.

One expects that one can properly define and compute the 
expression (\ref{PolPI}), because the integration over $X^{\mu}$
is Gaussian (in flat space-time) and $h_{\alpha \beta}$ 
is non-dynamical. This turns out to be true, though several 
interesting complications arise. Let us consider the integration
over $h_{\alpha \beta}$. Since we can locally impose the conformal
gauge,
\be
h_{\alpha \beta} = \delta_{\alpha \beta} \;,
\label{ConfGaugeEuc}
\ee
we expect that we can use the Faddeev-Popov method and 
trade the integration over the metric for an integration over
reparametrizations and the Weyl factor. As long as these are symmetries,
the corresponding integration 
factorizes and can be absorbed in the
normalization factor $N_g$. The first obstruction encountered is
that there is a conformal anomaly when the quantum theory
based on (\ref{EucPol}) lives on a curved world-sheet. This has the
consequence that the integration over the Weyl factor does not
factorize in general. One option is to accept it as a new,
purely quantum degree of freedom: this is non-critical string theory,
also called Liouville string theory, because the dynamcis of the
Weyl factor is given by the Liouville action. The other option
is to observe that the anomaly is proportional to $D-26$, and therefore
cancels for $D=26$ space-time dimensions. This is the 
critical string theory we study in these lectures.

\subsubsection{Moduli and modular transformations.}

The next point is that the gauge (\ref{ConfGaugeEuc}) cannot be imposed
globally. All that can be achieved is to map $h_{\alpha \beta}$
to a metric of constant curvature,
\be
h_{\alpha \beta} \stackrel{!}{=} \hat{h}_{\alpha \beta} [\vec{\tau}] \;.
\label{GlobalGauge}
\ee
As indicated, $\Sigma_{g}$ in general posesses a continuous family
of such metrics, para\-me\-trized by moduli 
$\vec{\tau}=(\tau_1, \ldots )$. The space of constant curvature metrics
on a two-dimensional closed compact surface is isomorphic to the
space of complex structures. By reparametrizations and Weyl
transformations we cannot change the complex structure of the
metric but we can map it to the unique representative (\ref{GlobalGauge}) 
of the complex structure class which
has constant curvature. Then the path integral over all metrics
reduces to a finite-dimensional integral over the
space ${\cal M}_g$ of complex structures. The dimension of this
space is known from a Riemann-Roch theorem. For $g=0$ the 
complex structure is unique, and every metric can be mapped to 
the standard round metric on the sphere. For $g>1$ there is
a non-trivial moduli space,
\bea
\dim_{\bf C} {\cal M}_g &=& 1 \;, \;\;\;\mbox{for}\;\;g=1 \;,\nonumber \\
\dim_{\bf C} {\cal M}_g &=& 3g-3 \;, \;\;\;\mbox{for}\;\;g>1  \;.
\eea
After carrying out the integration over the metric,
amplitudes take the form
\be
A(1,\ldots, M) = \sum_{g=0}^{\infty} \kappa^{M - \chi}
N'_g \int_{{\cal M}_g} d\mu(\vec{\tau}) \int D X^{\mu} e^{-S_P[X, \hat{h}]}
J( \hat{h} ) V_1 \ldots V_M \;.
\label{PolPI2}
\ee
$N_g'$ are normalization factors needed to deal with the
$X^{\mu}$-integration and $J(\hat{h})$ is the Faddeev-Popov determinant,
which one can rewrite as a functional integral over 
Faddeev-Popov ghost fields. As indicated the $X^{\mu}$-integral
depends on the moduli through the world-sheet metric
$\hat{h}_{\alpha \beta} = \hat{h}_{\alpha \beta}(\vec{\tau})$.
One finds that the measure $d\mu(\vec{\tau})$ for the moduli is
the natural measure on the space of complex structures, the 
so-called Weil-Petersson measure.

The precise characterization of the moduli space has further
interesting details. We examplify this with the two-torus.
We can represent a torus as a parallelogram in the complex plane
with opposite sides identified. 
Since the complex structure does not depend on the overall
volume, we can restrict ourselves to parallelograms
with edges $0,1,\tau,\tau+1$, where
$\mbox{Im}(\tau) > 0$. In one complex dimension holomorphic
maps are conformal maps,
and vice versa. Thus the complex structure is 
varied by moving $\tau$ in the upper half-plane,
\be
{\cal H} = \{ \tau \in {\bf C} \; |\; \mbox{Im}(\tau) > 0 \} \;.
\ee
This is the modulus we are looking for. 
${\cal H}$ has a metric of constant negative curvature,
the Poincar\'e metric,
\be
d\mu(\tau) = \frac{d^2\tau}{(\mbox{Im}(\tau))^2} \;.
\ee
With this $SL(2,{\bf R})$--invariant metric,  
${\cal H}$ is the symmetric space
$Sl(2, {\bf R})/SO(2)$. However, ${\cal H}$ is not our moduli
space, because it overcounts complex structures. On ${\cal H}$
the group $Sl(2,{\bf R})$ acts from the right. Taking $\tau$ as
coordinate, the operation is 
\be
\tau  \longrightarrow \frac{a \tau + b}{ c \tau + d} \;,\;\;\;
\mbox{where} \;\;\;
\left( \begin{array}{ll} a & b \\ c & d \\ \end{array} \right)
\in Sl(2, {\bf R}) \;.
\ee
The subgroup
$Sl(2,{\bf Z})$ maps parallelograms to parallelograms which
define the same torus, because they form basic cells of the same
lattice in ${\cal H}$. Such transformations are called modular 
transformations. Their action on the torus is given by 
cutting the torus along a non-contractible loop, twisting
and regluing. This corresponds to a large reparametrization which cannot
be continously connected to the identity. Clearly, we have to
require that string amplitudes are invariant under such large
reparametrizations. This implies a consistency condition,
known as modular invariance: the 
$\vec{\tau}$-integral in (\ref{PolPI2}) must be invariant
under modular transformations. This condition becomes
non-trivial when considering more general background geometries or
string theories with fermions.

The moduli space is obtained by restricting
to a fundamental domain ${\cal F}$ of the action of $Sl(2,{\bf Z})$ on
${\cal H}$. By modular invariance we can consistently restrict
the $\tau$-integration to such an ${\cal F}$. The standard choice
is found by looking at the action of the two generators of 
$Sl(2,{\bf Z})$,
\be
\tau \rightarrow \tau +1 \;,\;\;\;
\tau \rightarrow - \frac{1}{\tau} \;.
\ee
Therefore the  most convenient choice is 
\be
{\cal F} = \{ \tau \in {\cal H} | - \ft12 \leq \mbox{Im}(\tau) < \ft12 \tau 
\;,\;\;\; | \tau | \geq 1 \}
\ee
(with certain identifications along the boundary).

Modular invariance has deep consequences for the short distance 
behaviour of string theory. In fact, modular invariance is 
what makes closed string theories UV finite. To illustrate how
this works, note that a one-loop amplitude in closed string theory
takes the form
\be
A^{\mscr{String}}_{\mscr{1-loop}} \sim \int_{\cal F} 
\frac{d^2 \tau}{ (\mbox{Im}(\tau))^2 } F(\tau) \;.
\ee
An analogous expression for one loop amplitudes in quantum field
theory is given by Schwinger's proper time parametrization,
\be
A^{\mscr{QFT}}_{\mscr{1-loop}} \sim 
\int^{\infty}_{\varepsilon} \frac{dt}{t} f(t) \;,
\ee
where $t$ is the proper time and $\varepsilon$ is an UV cutoff.
In this formulation UV divergencies occure at short times $t\rightarrow 0$.
In string theory $\mbox{Im}(\tau)$ plays the role of proper time,
and potential UV divergencies occure for $\mbox{Im}(\tau) \rightarrow 0$.
However, by restricting to the fundamental domain we have cut out
the whole dangerous region of small times and high momenta. 
This confirms the intuitive idea that strings should have a 
particularly soft  UV behaviour, because the
theory has a minimal length scale, which works like a physical
UV cutoff. Note that one still has 
IR divergencies. In bosonic string theory one has divergencies
related to the tachyon, which show that the theory is unstable
in Minkowski space. This problem is absent in supersymmetric string
theories. In addition one can have IR divergencies related to
massless states. 
Since there is only a finite number of massless string states,
this problem has the same character as in field theory.

Also note that the modular transformation 
$\tau \rightarrow - 1/ \tau$ maps the UV region of 
${\cal H}$ to its IR region. Thus, modular transformations map
UV divergencies to IR divergencies and enable us to 
reinterpret them in terms of low energy physics (namely,
intermediate massless states which go on-shell). 

For higher genus surfaces $\Sigma_g$ with $g>1$ the story is 
similar, but more complicated. There is an analogue of the upper
half plane, which is called Siegels upper half plane and has
complex dimension $\ft{g(g+1)}2$. Since there are only 
$3g -3$ complex moduli, this space contains more parameters then
needed for $g\geq 4$. The Teichm\"uller space is embedded in
a complicated way into Siegel's upper half plane. On top of this
there is a modular group which has to be divided out.

\subsubsection{Global conformal transformations.}

The integration over complex structure moduli in
(\ref{PolPI2}) reflects that surfaces with $g>0$ 
have metrics that cannot be related by reparametrizations.
Therefore there is a finite left-over integration when
replacing the integral over metrics by an integral over 
reparametrizations. For $g<2$ one has in addition the 
reciprocal phenomenon: these surfaces have global conformal
isometries. This means that there are reparametrizations
which do not change the metric, implying an overcounting
of equivalent contributions in (\ref{PolPI2}). Formally this is
taken care of by the normalization factors $N_0', N_1'$.
The overcounting yields a multiplicative factor, which is the 
volume of the group of conformal isometries. This has to be
cancelled by the normalization factors. For $g=0$ the
conformal group is $Sl(2,{\bf C})$ and has infinite volume.
Thus one has to formally divide out an infinite constant.
For $g=1$ the conformal group is $U(1)^2$, and has a volume
which depends on the complex structure modulus $\tau$ of the 
world-sheet. This factor is crucial for world-sheet modular
invariance. 

The systematic approach is to treat the global conformal
isometries as a residual gauge invariance and to apply the
Faddeev-Popov technique.%\footnote{This is nicely explained
%in \cite{Weinberg}.} 
Then the volumes of residual gauge
groups are properly taken care of. 
So far we have been sloppy about how and 
when to carry out the integration over the positions of the
vertex operators.
The proper order 
is as follows: one first carries out the $X^{\mu}$-integration
to obtain a correlation function on a world-sheet of given
topology and complex structure:
\be
\langle V_1(z_1, \overline{z}_1) \cdots \rangle_{g,\tau}
= N'_g \int DX e^{-S_P[X,\hat{h}(\vec{\tau})]} J(\hat{h}({\vec{\tau}}))
V_1 (z_1, \overline{z}_1) \cdots \;.
\ee
Next one integrates over the positions of the vertex operators.
For $g<2$ one treats the global conformal isometries by the 
Faddeev Popov method. The result is
\be
\langle  V_1 \cdots \rangle_{g, \vec{\tau}} 
= \int d \mu (z_1, \overline{z}_1, \ldots) 
\langle V_1(z_1, \overline{z}_1) \cdots \rangle_{g,\vec{\tau}}  \;,
\ee
where $d\mu(z_1, \overline{z}_1, \ldots )$ for $g<2$ is a measure
invariant under the global isometries. 

For $g=0$ the
measure vanishes if less than three vertex operators are present.
This reflects the infinite volume of the global conformal group:
by $Sl(2,{\bf C})$ transformations one can map three points on
the sphere to three arbitrary prescribed points. Thus, the $Sl(2,{\bf C})$
symmetry can be used to by keep three vertex operators at fixed 
positions. In other words the first three integrations over 
vertex operators compensate the infinite volume of the 
global conformal group that one has to divide out. For less then
three vertex operators one cannot compensate this infinite
normalization factor and the result is zero. Thus, the integrated
zero-, one- and two-point functions vanish. This implies that at
string tree level the cosmological constant and all tadpoles 
diagrams vanish.

The final step in evaluating (\ref{PolPI2}) is to integrate
over complex structures and to sum over topologies:
\be
A(1,\ldots, M) = \sum_{g=0}^{\infty} \kappa^{M-\chi(g)} \int_{{\cal M}_g} 
d \mu (\vec{\tau}) \langle V_1 \ldots \rangle_{g, \vec{\tau}} \;.
\ee
Through the vertex operators, $A(1,\ldots, M)$ is a function
of the momenta $k^{\mu}_i$ and polarization tensors
$\zeta^{\mu_1 \cdots}_i$ of the external states.

\subsubsection{Graviton scattering.}

Though we cannot go through the details of a calculation here, we 
would like to discuss the properties of string scattering amplitudes
in a particular example. Our main interest being gravity, we choose
the scattering of two massless closed string states.
The corresponding external states are
\be
\zeta_{\mu \nu}^{(i)} \alpha^{\mu}_{-1} \tilde{\alpha}^{\nu}_{-1}
| k^{(i)} \rangle \;,
\ee
with $i=1,2,3,4$. The resulting amplitude takes the
following form:
\be
A_4^{\mscr{String}} = \kappa^2 
\frac{ \Gamma ( - \ft{\alpha'}{4} s) 
\Gamma ( - \ft{\alpha'}{4} t) 
\Gamma ( - \ft{\alpha'}{4} u) }{
\Gamma ( 1+  \ft{\alpha'}{4} s) 
\Gamma ( 1+  \ft{\alpha'}{4} t) 
\Gamma ( 1+  \ft{\alpha'}{4} u) }
\cdot K(\zeta^{(i)}, k^{(i)}) \;.
\label{VirAmpl}
\ee
Here $s,t,u$ are the Mandelstam variables
\be
s = - (k^{(1)} + k^{(2)})^2 \;, \;\;
t = - (k^{(2)} + k^{(3)})^2 \;, \;\;
u = - (k^{(1)} + k^{(3)})^2 
\ee
and $K(\zeta^{(i)}, k^{(i)})$ is the kinematic factor,
a complicated function of momenta and polarizations that we 
do not display.

Scattering amplitudes have poles whenever an intermediate
states can be produced as a real physical state. 
Unitarity 
requires that the residue of the pole describing such a resonance
is the product 
of the amplitudes of the subprocesses through which the
intermediate state is produced and decays. In this way 
the pole structure of amplitudes is related to the particle spectrum
of the theory.

The amplitude (\ref{VirAmpl}) has poles when the argument
of one of the $\Gamma$-functions in the numerator takes a non-positive
integer value,
\be
- \ft{\alpha'}{4} x = 0,-1,-2,\ldots \;,
\;\;\; \mbox{where} \;\;x = s,t,u \;.
\ee
Comparing to the
mass formula of closed strings,
\be
\alpha' M^2 = 2 ( N + \tilde{N} -2) \;,
\ee
we see that the poles precisely correspond to massless and 
massive string states with $N= \tilde{N}=1,2,3, \ldots$.
There is no pole corresponding to the tachyon ($N=0$)
in this amplitude, because the tachyon cannot be produced as
a resonance for kinematic reasons. 
When computing the amplitude for 
tachyon scattering instead, one also finds
a tachyon pole. 

The particular pole structure of (\ref{VirAmpl}) and of
related string amplitudes was
observed before the interpretation of the amplitudes in terms of
strings was known. In the late 1960s it was observed experimentally
that hadronic resonances obey a linear relation between the
spin and the square of the mass, called Regge behaviour. 
This behaviour was correctly captured by the Veneziano amplitude,
which has a structure similar to (\ref{VirAmpl}) and describes
the scattering of two open string tachyons. The Regge behaviour
was the clue for the interpretation of the Veneziano amplitudes
and its cousins in terms of strings.

To see that string states show Regge behaviour,
consider the truncation of  string theory to four space-time
dimension (which is consistent at tree level). Closed string states with 
level $N=\tilde{N}$ have spins $J \leq 2N$, because the 
spin $J$ representation of the four-dimensional Lorentz
group is the traceless symmetric tensor of rank $J$. 
Open string states have spins $J \leq N$.
The states lie on lines in the $(M^2, J)$--plane,
which are called Regge trajectories. The closed string
Regge trajectories are given by
\be
\alpha_{\mscr{closed}}(M^2) = \alpha'_{\mscr{closed}} M^2 
+ \alpha_{\mscr{closed}}(0) \;,
\ee
where 
\be
\alpha'_{\mscr{closed}} = \frac{\alpha'}{4} \;,\;\;\;
\alpha_{\mscr{closed}}(0) = 1,0,-1,\ldots \;.
\ee
String states correspond to those points on the Regge trajectories
where \newline
$\alpha'_{\mscr{closed}}(M^2) = N + \tilde{N}$. States with
the maximal possible spin $J = 2 N = N + \tilde{N}$ for a  given mass
lie on the leading Regge trajectory $\alpha_{\mscr{closed}}(0) = 1$.
Since $\alpha'$ determines the slope of the trajectories, it is called
the Regge slope. The corresponding expressions for open
strings are:
\be
\alpha_{\mscr{open}}(M^2) = \alpha'_{\mscr{open}} M^2 
+ \alpha_{\mscr{open}}(0) \;,
\ee
where 
\be
\alpha'_{\mscr{open}} = {\alpha'} \;,\;\;\;
\alpha_{\mscr{open}}(0) = 1,0,-1,\ldots \;.
\ee
The resonances found in open string scattering lie on the 
corresponding Regge trajectories.

When computing scattering amplitudes in terms of Feynman diagrams
in field theory, individual diagrams only have poles in one
particular kinematic channel, i.e., in the s-channel or t-channel
or u-channel. The full scattering amplitude, which has poles
in all channels, is obtained by summing up all Feynman diagrams.
In (closed oriented) string theory there is only one diagram in
each order of perturbation theory, which simultanously has poles
in all channels. 
The total amplitude can be written as a sum over resonances
in one particular channel, say the s-channel. This is consistent
with the existence of poles in the other channels, because there is
an infinite set of resonances. 
When instead writing the amplitude  in the form (\ref{VirAmpl}),
it is manifestly symmetric under permutations of
the kinematic variables $s,t,u$. This special property was called
'duality' in the old days of string theory 
(a term that nowadays is used for a variety of other, unrelated phenomena
as well).

Another important property of (\ref{VirAmpl}) and other string
amplitudes is that they fall off exponentially for large $s$, 
which means that the behaviour for large external momenta is much
softer then in any field theory. This is again due to the
presence of an infinite tower of excitations. Since loop 
amplitudes can be constructed by sewing tree amplitudes,
this implies that the UV behaviour of loop diagrams
is much softer than in field theory. This
lead to the expectation that string loop amplitudes are UV
finite, which was confirmed in the subsequent development
of string perturbation theory.

Though we did not explicitly display the kinematic factor
$K(\zeta^{(i)}, k^{(i)})$ we need to emphasize one of its
properties: it vanishes whenever one of the external states is
a null state. As we learned above, null states have
polarizations of the form
\be
\zeta^{(i)}_{\mu \nu} = k^{(i)}_{\mu} \xi^{(i)}_{\nu} +
k^{(i)}_{\nu} \zeta^{(i)}_{\mu}
\ee
and are gauge degrees of freedom. They have to decouple from
physical scattering amplitudes, as it happens in the above 
example. This property is called \lq on shell gauge invariance',
because it is the manifestation of local gauge invariance at
the level of scattering amplitudes. It can be proven to hold for general
scattering amplitudes.

If we take the polarization tensors of the external states
to be symmetric and traceless, then (\ref{VirAmpl})
describes graviton--graviton scattering. 
So far our identification of this string state with the graviton
was based on its kinematic properties. Since Einstein gravity
is the only known consistent interaction
for a second rank, traceless symmetric tensor field 
(\lq massless spin--2--field'), we 
expect that this also holds dynamically. 
We will now check this explicitly.
In the field theoretical perturbative approach to quantum gravity
one starts from the Einstein-Hilbert action,
\be
S = \frac{1}{2 \kappa^2} \int d^D x \sqrt{g} R
\label{EH}
\ee
and expands the metric around flat space
\be
g_{\mu \nu}(x) = \eta_{\mu \nu} + \kappa \psi_{\mu \nu}(x) \;.
\ee
The field $\psi_{\mu \nu}(x)$ is the graviton field. Expanding
(\ref{EH}) in $\kappa$ one gets a complicated non-polynomial
action for $\psi$ that one quantizes perturbatively. 
The resulting theory is non-renormalizable, but tree diagrams
can be consistently defined and computed. In particular one can
compute graviton--graviton scattering at tree level and compare
it to the string amplitude (\ref{VirAmpl}). Denoting the
field theory amplitude by $A_4^{\mscr{FTh}}$, the relation 
is
\be
A^{\mscr{String}}_4 = \frac{
\Gamma( 1 - \ft{\alpha'}{4} s )
\Gamma( 1 - \ft{\alpha'}{4} t )
\Gamma( 1 - \ft{\alpha'}{4} u )}{
\Gamma( 1 + \ft{\alpha'}{4} s )
\Gamma( 1 + \ft{\alpha'}{4} t )
\Gamma( 1 + \ft{\alpha'}{4} u )}
A^{\mscr{FTh}}_4 \;.
\label{AmplStrField}
\ee
In the limit $\alpha' \rightarrow 0$, which corresponds to sending
the string mass scale to infinity, the string amplitude reduces to
the field theory amplitude:
\be
\lim_{\alpha' \rightarrow 0} 
A^{\mscr{String}}_4
= 
A^{\mscr{FTh}}_4 \;.
\ee
At finite $\alpha'$ string theory deviates from field theory.
The correction factor in (\ref{AmplStrField}) contains precisely
all the poles corresponding to massive string states, whereas
the massless poles are captured by the field theory amplitude.
One can construct an effective action which reproduces the 
string amplitude order by order in $\alpha'$. At order $\alpha'$
one obtains four-derivative terms, in particular terms quadratic
in the curvature tensor,
\be
S_{\mscr{eff}} = \frac{1}{ 2 \kappa^2} \int d^D x \sqrt{g}
( R + \alpha' c_1 R_{\mu \nu \rho \sigma}R^{\mu \nu \rho \sigma}
+ \cdots + {\cal O}( (\alpha')^2 ) )  \;,
\label{EffAlphaPrime}
\ee
where $c_1$ is a numerical constant. The $\alpha'$-expansion
of the effective action is an expansion in derivatives. It is valid
at low energies, i.e., at energies lower than the scale set 
by $\alpha'$, where corrections due to massive string scales are small.

Obviously, it is very cumbersome to construct the effective 
action
by matching field theory amplitudes with string amplitudes.
In practice one uses symmetries to constrain the
form of the effective action. This is particularly efficient
for supersymmetric actions, which only depend on a few
independent parameters or functions, which can be extracted from
a small number of string amplitudes. A different technique,
which often is even more efficient, 
is to study strings in curved backgrounds, and,
more generally, in background fields.

\subsection{Strings in curved backgrounds}

So far we only discussed strings in flat backgrounds. Let us
now consider the case of a curved background with Riemannian
metric $G_{\mu \nu}(X)$. Then the Polyakov action takes the form
of a non-linear sigma-model
\be
S_P = \frac{1}{4 \pi \alpha'} \int d^2 \sigma \sqrt{h}
h^{\alpha \beta} \der_{ \alpha} X^{\mu} \der_{\beta}
X^{\nu} G_{\mu \nu}(X) \;.
\label{NLSM}
\ee
As emphasized above, the local Weyl invariance 
\be
h_{\alpha \beta} \rightarrow e^{\Lambda(\sigma)} h_{\alpha \beta}
\ee
is crucial for the consistency of string theory, since the
construction of states, vertex operators and amplitudes
is based on having a conformal field theory on the world-sheet.
If the space-time metric is curved, then the Weyl invariance 
of the classical action (\ref{NLSM}) is still manifest. 
But at the quantum level it becomes non-trivial and imposes
restrictions on $G_{\mu \nu}(X)$. In the non-linear sigma-model
defined by (\ref{NLSM}) one can define a modified beta function
$\overline{\beta}$, which measures the violation of 
local Weyl invariance. In order to have local Weyl invariance 
this function must vanish,
\be
\overline{\beta} =0 \;.
\ee
Since $G_{\mu \nu}(X)$ are the field-dependent couplings 
of the non-linear sigma-model, the beta function $\overline{\beta}$
is a functional of $G_{\mu \nu}(X)$. It can be computed perturbatively,
order by order in $\alpha'$. The dimensionless expansion parameter
is the curvature scale of the target space
(i.e., space-time) measured in units of the string length $\sqrt{\alpha'}$.

The leading term in this expansion is:
\be
\overline{\beta}^G_{\mu \nu} = - \frac{1}{2 \pi} R_{\mu \nu} \;.
\label{Beta1}
\ee
Thus the space-time background has to be Ricci-flat, i.e., it satisfies
the vacuum Einstein equation. The condition imposed on the background
field by local Weyl invariance on the world-sheet is its space-time
equation of motion. This relation between world-sheet and space-time
properties holds for other background fields as well and can
be used as an efficient method to construct effective actions.
One can also compute the $\alpha'$-corrections to (\ref{Beta1}):
\be
\overline{\beta}^G_{\mu \nu} = - \frac{1}{2 \pi} \left( 
R_{\mu \nu} +
\frac{\alpha'}{2} R_{\mu \alpha \beta \gamma} R_{\nu}^{\;\;\alpha
\beta \gamma} \right) \;.
\ee
The corresponding $\alpha'$-corrections to the Einstein-Hilbert
action take the form (\ref{EffAlphaPrime}).

At this point we need to reflect a little bit on how gravity
is described in string theory. So far we have seen that it enters in 
two ways: first, there is a graviton state $\zeta_{(\mu \nu)}
\alpha^{\mu}_{-1} \tilde{\alpha}^{\nu}_{-1} | k \rangle$
in the string spectrum.
Second, there is a background metric $G_{\mu \nu}(X)$.
If gravity is described consistently, then these two objects
must be related. To explore this we expand $G_{\mu \nu}(X)$
around flat space,
\be
G_{\mu \nu}(X) = \eta_{\mu \nu} + \kappa \psi_{\mu \nu}(X) \;,
\ee
and observe that the action (\ref{NLSM}) is related to the
Polyakov action in flat space by
\be
S_P[G_{\mu \nu}] = S_P [\eta_{\mu \nu}]
+ \kappa V[\psi_{\mu \nu}] \;, 
\label{RelateActions}
\ee
where
\be
V[\psi_{\mu \nu}] = \frac{1}{4 \pi \alpha'}
\int d^2 \sigma \sqrt{h} h^{\alpha \beta} \psi_{\mu \nu}(X)
\der_{\alpha} X^{\mu} \der_{\beta} X^{\nu} \;.
\label{Vpsi}
\ee
Taking the Fourier transform of $\psi_{\mu \nu}(X)$ we obtain
\be
V[\psi_{\mu \nu}] = \frac{1}{4 \pi \alpha'} \int d^D k
\int d^2 \sigma \sqrt{h} V_\psi(k,\tilde{\psi}(k))
\ee
where
\be
V_\psi(k,\tilde{\psi}(k)) = \tilde{\psi}_{\mu \nu}(k) 
\der_{\alpha} X^{\mu} \der^{\alpha} X^{\nu} e^{ \I k_{\rho} X^{\rho}}
\label{GravVOp2}
\ee
is the graviton vertex operator with polarisation tensor
$\tilde{\psi}_{\mu \nu}(k)$.

Thus the curved space action $S_P[G_{\mu \nu}]$ is obtained by deforming 
the flat space action $S_P[\eta_{\mu \nu}]$ by the graviton
vertex operator. 
Since both actions must be conformal, $V[\psi]$ must
be a so-called exactly marginal operator of the world-sheet 
field theory. These are the operators which generate 
deformations of the action while preserving conformal invariance.
A necessary condition is that $V[\psi]$ must be a
marginal operator, which means it has weights $(0,0)$ with respect
to the original action. Such operators have the correct weight for
being added to the action and generate infinitesimal deformations
which preserve conformal invariance.
Note that it is not guaranteed that a marginal
operator is still marginal in the infinitesimally deformed theory.
Only those marginal operators which stay marginal under deformation
generate finite deformations of a conformal field theory and
are called exactly marginal (or truly marginal).

If the integrated vertex operator $V[\psi]$ has weights
$(0,0)$, then the vertex operator (\ref{GravVOp2}) must have 
weights $(1,1)$. This is the condition for a vertex
operator to create a physical state. The resulting conditions
on momenta and polarization are
\be
k^2 = 0 \;, \;\;\;k_{\mu} \tilde{\psi}^{(\mu \nu)} = 0 \;,
\ee
which we now recognize as the Fourier transforms of the 
linearized Einstein equation. This the free part of the
equations of motion for the graviton and characterizes its 
mass and spin.

Marginal operators are not necessarily exactly marginal.
The flat space action defines a free field theory on the world-sheet,
which is conformally invariant at the quantum level. Thus $V[\psi]$
is exactly marginal if and only if the curved space action $S_P[G_{\mu \nu}]$ 
is conformally invariant. By the beta--function analysis, this 
is equivalent to the full vacuum Einstein equation for the metric
$G_{\mu \nu}$, plus corrections in $\alpha'$. This is the full,
non-linear equation of motion for the graviton string state.

In order to understand the relation between the graviton string state
and the background metric even better we use (\ref{RelateActions})
to relate amplitudes computed using the curved space action
$S_P[G_{\mu \nu}]$ and the flat space action $S_P[\eta_{\mu \nu}]$:
\be
\langle V_1 \cdots V_M \rangle_{G} = 
\langle V_1 \cdots V_M e^{V[\psi]} \rangle_{\eta}
\label{insertion}
\ee
The operator $e^{V[\psi]}$ generates a coherent state of gravitons
in flat space. This can be seen as follows:
in quantum mechanics (think of the harmonic oscillator)
coherent states are defined
as states with minimal Heisenberg uncertainty. They are eigenstates
of annihilation operators and can be constructed by exponentiating
creation operators. The resulting states are not eigenstates 
of the number operator but are  superpositions of states with all possible
occupation numbers. In (\ref{insertion}) the role of the 
creation operator is played by the graviton vertex operator.

In quantum field theory,
coherent states are the states corresponding to classical fields.
For example, in quantum electrodynamics a classical electrodynamic
field can be represented as a coherent state of photons. 
Similarly, in gravity
a curved metric (modulo global properties)
can be described as a coherent state of gravitons
in the Minkowski vacuum. This is realized in the above formula,
where the
amplitudes  in the curved background can be 
computed equivalently by inserting the vertex operator for a coherent
state of gravitons into the flat space amplitude.
This is a manifestation of background independence:
though we need to pick a particular background to define
our theory, other backgrounds are different states in the 
same theory. Since consistent backgrounds must satisfy the
equations of motion, one also calls them solutions of string theory.
In this terminology different background geometries are 
different solutions of the single underlying string theory.

\subsection{Effective actions}

In the last section we have seen that the equation of motion
of the metric/graviton can be obtained from an effective action.
Such effective actions are very convenient, because they allow us
to describe string states in terms of D-dimensional field theory.
Effective actions are obtained in an expansion in $\alpha'$ and
therefore their use is limited to scales below the string scale.
But given that the string scale probably is very large, they
are extremely useful to extract particle physics or gravitational
physics from string theory. Therefore they 
play a mayor role in string theory.
We have also seen that there are two methods for deriving
effective actions: the matching of string theory amplitudes with
field theory amplitudes and solving the conditions for
Weyl invariance $\overline{\beta}=0$ in a non-trivial background.

So far we found that the Einstein-Hilbert action is the leading
part of the effective action for the graviton. We have seen that
the closed string has two further massless modes, the dilaton
$\Phi$ and the antisymmetric tensor field $B_{\mu \nu}$. 
We can now switch on the corresponding non-trivial background fields.
The total world-sheet action is:
\be
S = S_P[G] + S[B] + S[\Phi] \;.
\label{TotAct}
\ee 
Here $S_P[G]$ is the action (\ref{NLSM}),
\be
S[B] = \frac{1}{4 \pi \alpha'} \int d^2 \sigma \varepsilon^{\alpha \beta}
\der_{\alpha} X^{\mu} \der_{\beta} X^{\nu} B_{\mu \nu}(X)
\ee
and 
\be 
S[\Phi] = \frac{1}{4 \pi} \int d^2 \sigma \sqrt{h} R^{(2)}(h) \Phi(X) \;.
\label{DilEffAct}
\ee
Here $\varepsilon^{\alpha \beta}$ is the totally antisymmetric 
world-sheet tensor density and $R^{(2)}(h)$ is the Ricci scalar of
the world-sheet metric. Note that the dilaton action is higher order
in $\alpha'$. 

The beta-function for the dilaton starts with a 
term proportional to $(D-26)$ and has $\alpha'$-correction proportional
to derivatives of $\Phi$. The leading term of the beta-function
corresponds to a cosmological constant in the effective action. 
When considering 
string theory around backgrounds with constant dilaton, the
only solution to the dilaton beta-function equation is to work
in the critical dimension $D=26$. We will only consider such backgrounds
here, and therefore the cosmological term in the effective
action vanishes. But let us note
that there are known exact solutions to the beta-function equations
with non-constant dilaton. These describe exact string backgrounds with
$D\not=26$. 

Let us now return to dilaton term of the world-sheet action. When evaluated
for constant dilaton, (\ref{DilEffAct}) is proportional to 
the Euler number of the world-sheet. For a Euclidean closed string 
world-sheet of genus $g$ we have:
\be
\chi = \frac{1}{4 \pi} \int_{\Sigma_g} d^2 z \sqrt{h} R^{(2)}(h)
= 2 - 2 g \;.
\ee
Therefore shifting the dilaton by a constant $a$, 
\be
\Phi(X) \rightarrow \Phi(X) + a
\ee
has the effect of shifting the total action (\ref{TotAct}) by a
constant proportional to the Euler number:
\be
S \rightarrow S + a \chi(g) \;.
\ee
For the corresponding partition function this is equivalent to
rescaling the coupling by $e^a$:
\be
Z = \sum_{g=0}^{\infty} \kappa^{-\chi(g)} \int DX Dh e^{-S}
\longrightarrow 
\sum_{g=0}^{\infty} ( \kappa e^a )^{-\chi(g)} \int DX Dh e^{-S}  \;.
\label{PartFctCplg}
\ee
This shows that the coupling constant $\kappa$ and vacuum expectation
value $\langle \Phi \rangle$ of the dilaton are not independent.
To clarify the physical meaning of both quantities, we now investigate
the effective action of the massless modes. The conditions for 
Weyl invariance of (\ref{TotAct}) are the Euler-Lagrange equation
of the following effective action:
\be
S^{\mscr{StrFr}}_{\mscr{tree}} = \frac{1}{2 \kappa^2} 
\int d^D x \sqrt{G} e^{-2 \Phi} \left( R(G) - \ft1{12} H_{\mu \nu \rho}
H^{\mu \nu \rho} + 4 \der_{\mu} \Phi \der^{\mu} \Phi
+ {\cal O}(\alpha') \right) \;.
\label{StrFrAct}
\ee
This way of parametrizing the effective action is called the string-frame.
The string frame metric $G_{\mu \nu}$ is the metric appearing in the
world-sheet action (\ref{NLSM}). The field strength of the antisymmetric
tensor field is
\be
H_{\mu \nu \rho} = 3! \; \der_{[\mu} B_{\nu \rho]}  \;,
\ee
where $[\mu \nu \rho]$ denotes antisymmetrization.

Concerning the dilaton we note that its vacuum expectation value
is not fixed by the equations of motion. Like in the partition function
(\ref{PartFctCplg}), shifting the dilaton by a constant is equivalent
to rescaling the coupling. In order to determine the relation of
the string coupling constant $\kappa$ to the physical gravitational
coupling $\kappa_{\mscr{phys}}$ one has to perform a field redefinition
that transforms the gravitational term in (\ref{StrFrAct})
into the standard Einstein-Hilbert action. The coefficient in front of
this term is the physical gravitational coupling. 

The transformation which achieves this is the following 
Weyl rescaling of the
metric:
\be
g_{\mu \nu} := G_{\mu \nu} e^{ - \frac{4}{D-2} ( \Phi - \langle \Phi \rangle)
} \;.
\label{Frames}
\ee
Expressing everything in terms of the Einstein frame
metric $g_{\mu \nu}$ one obtains:
\be
S^{\mscr{EinstFr}}_{\mscr{tree}} = \frac{1}{2 \kappa_{\mscr{phys}}^2}
\int
\sqrt{g} 
\left( R(g)  - \ft1{12} e^{ - 8 \frac{ \Phi - \langle \Phi \rangle }{
D-2}} H_{\mu \nu \rho} H^{\mu \nu \rho} - \ft{4}{D-2} \der_{\mu}
\Phi \der^{\mu} \Phi + {\cal O}(\alpha') \right) \;.
\label{EinFrAct}
\ee
The physical gravitational coupling is
\be
\kappa_{\mscr{phys}} = \kappa e^{\langle \Phi \rangle } \;.
\ee
Since the coupling $\kappa$ can be rescaled by shifting $\Phi$, it can
be set to an arbitrary value. This is used to fix $\kappa$:
\be
\kappa \stackrel{!}{=} (\alpha')^{\frac{D-2}{4}} \;.
\label{kappa-phys}
\ee
(Note that the $D$-dimensional gravitational couplings
$\kappa, \kappa_{\mscr{phys}}$ have dimension
$(\mbox{length})^{{D-2}/{2}}$.) Since $\kappa_{\mscr{phys}}$ and
$\alpha'$ are related by the vacuum expectation value of the dilaton
we see that there is only one fundamental dimensionful parameter in
string theory, which we can take to be either the gravitational coupling
$\kappa_{\mscr{phys}}$ or the string scale set by $\alpha'$.
They are related by the vacuum expectation value of the dilaton, which 
classically is a free parameter labeling different ground states 
in one theory. Defining the dimensionless string coupling constant 
by
\be
g_S = e^{\langle \Phi \rangle} \;,
\ee
we have the relation
\be
\kappa_{\mscr{phys}} = (\alpha')^{\frac{D-2}{4}} g_S \;.
\label{kapppa-phys2}
\ee

The effective actions (\ref{StrFrAct},\ref{EinFrAct})
have been constructed to leading order in $\alpha'$ and at tree level in the
string coupling $g_S$. Loop corrections in $g_S$ can
be obtained, either by considering loop amplitudes or from the contribution
of higher genus world-sheets to the Weyl anomaly (Fischler-Susskind
mechanism). One might expect that loop corrections generate a potential
for the dilaton and lift the vacuum degeneracy. But for the bosonic
string one does not know the stable ground state, because of the 
tachyon. In supersymmetric string  
theories tachyons are absent, but no dilaton potential is created
at any loop level. Thus the value of the string coupling remains a
free parameter. This is (part of) the problem of vacuum degeneracy
of superstring theories. Since the flatness of the dilaton potential
is a consequence of supersymmetry, the solution of the vacuum
degeneracy problem is related to understanding supersymmetry breaking.

For practical applications, both the string frame effective action
and the Einstein frame effective action (and their higher-loop
generalizations) are needed. The string frame action is adapted
to string perturbation theory and has a universal dependence
on the dilaton and on the string coupling:
\be
S^{\mscr{StrFr}}_{\mscr{g-loop}} \sim {g_S^{-2+2g}} \;.
\ee
The Einstein frame
action is needed when analyzing gravitational physics, in particular
for solutions of the effective action that describe black holes and
other space-time geometries. Note 
that concepts such as 
the ADM mass of an asymptotically flat space-time are tied to 
the gravitational action written in the Einstein frame. The
relation between the Einstein frame metric and the string frame metric
is non-trivial, because it
involves the dilaton, which in general is a space-time dependent
field. Therefore various quantities, most importantly the metric
itself, can take a very different form in the two frames. For example
one metric might be singular wheras the other is not. In order to
decide whether a field configuration is singular or not, one has
of course to look at all the fields, not just at the metric. 
If the metric is singular in one frame but not in the other, then
the dilaton must be singular.

\subsection{Interacting open and non-oriented strings}

We now indicate how the above results 
extend to open and non-oriented strings.

\subsubsection{Open strings.}

The world-sheets describing the interactions of open strings have two kinds of
boundaries: those corresponding to the initial and final strings
and those corresponding to the motions of string endpoints.
Boundaries corresponding to external strings can be 
mapped to punctures and are then replaced by
vertex operators. The boundaries corresponding to the motions of
string endpoints remain. They are the new feature compared to 
closed strings.
Perturbation theory for open strings can then be 
developed along the same lines as for closed strings. 
Instead of closed oriented surfaces it involves oriented surfaces
with boundaries, and the vertex operators for open string states
are inserted at on the boundaries.

Again there is one
fundamental interaction, which couples three open strings, and we
assign to it a coupling constant $\kappa_o$. 
The most simple world-sheet, analogous to the sphere for closed strings,
is the disc. It is leading in the expansion in $\kappa_o$ and
describes scattering at tree level. 
The computation of tree level scattering 
amplitudes confirms the interpretation of the massless state
as a gauge boson. The resulting effective action, to leading order
in $\alpha'$, is the Maxwell or, with Chan-Paton factors, the
Yang-Mills action. It receives higher order corrections in
$\alpha'$ and one can show that the resulting actions are
of Born-Infeld type. 

Higher order diagrams in open string perturbation theory correspond
to surfaces with more than one boundary component. They 
are obtained from the disc by removing discs from the interior. 
Each removal of a disc corresponds to an open string loop.
The one loop diagram is the the annulus.

One can also introduce a coupling of two open strings to one closed string
with coupling $\kappa_{oc}$ and consider theories of open and closed
strings. Unitarity then implies that the three couplings $\kappa_{o},
\kappa_{oc}, \kappa$ are not independent. To see this consider first
a disc diagram with two open string vertex operators at the boundaries
and two closed string vertex operators in the interior. This amplitude
can be
factorized with an intermediate closed string. Looking at 
string interactions we see that one has one interaction between three
closed strings and one between one closed and two open strings.
Therefore the amplitude is proportional to $\kappa \kappa_{oc}$.
The amplitude can also be factorized with an intermediate open
string. This time one sees two interactions involving two open
and one closed string. Therefore the amplitude is proportional
to $\kappa_{oc}^2$. Comparing both forms of the amplitude we deduce
\be
\kappa \simeq \kappa_{oc}
\ee
(the numerical factor has to be  determined by explicit computation).

Next consider the open string one loop diagram, the annulus.
Putting two vertex operators on each boundary one can again
factorize it with either a closed or an open intermediate state.
This way one finds
\be
\kappa \simeq \kappa_o^2 \;.
\label{closed-open}
\ee
Note that the above amplitude does not involve external 
closed string states. This indicates an important property
of open string theories: the coupling to closed strings is not
optional, but mandatory. When computing open string loop diagrams,
one finds that they have poles which correspond to 
closed string states. Therefore consistency of open string
theories at the quantum level requires the inclusion of closed
strings. This means in particular that every consistent quantum
string theory has to include gravity. The
relation between open and closed strings becomes obvious when one
realizes that the 
annulus is topologically equivalent to the cylinder. While the
annulus intuitively is the open string one loop diagram, the
cylinder is the closed string propagator. This is reflected by
the properties of the corresponding string amplitudes, which can
be written either as a sum over poles corresponding to open
strings (open string channel) or as a sum over poles corresponding
to closed strings (closed string channel).

The UV finiteness of closed string theories is due to modular
invariance. Open string world-sheets do not have a modular group.
The role of modular invariance is played by another property,
called tadpole cancellation. The underlying observation is that
the cancellation of divergencies between different 
diagrams is equivalent to the vanishing of the dilaton tadpole.
It turns out that tadpole cancellation
cannot be realized in a theory of  oriented open and closed strings.
In theories of non-oriented open and closed strings tadpole
cancellation  
fixes the gauge group to be
$SO(2^{D/2})$. For bosonic strings the critical dimension
is $D=26$ an the gauge group must be $SO(8192)$. 
Since the primary problem of bosonic strings is the tachyon, it is
not clear whether tadpole cancellation plays a fundamental role there.
But for type I superstrings this is the condition which makes the theory
finite.

Since we only discussed orientable world-sheets so far, we 
next collect some properties of the world-sheets of non-oriented strings.

\subsubsection{Non-oriented strings.}

Theories of non-oriented strings are obtained by keeping only states
which are invariant under world-sheet parity. Since such  
theories are insensitive to the orientation of the world--sheet
one now has to include non-orientable world-sheets. 
Theories of closed non-oriented strings
involve orientable and non-orientable surfaces without boundaries,
whereas theories of open and closed non-oriented strings involve
in addition orientable and non-orientable world-sheets with 
boundaries. Let us summarize which types of world-sheets
occure in string theory, 
depending on boundary conditions and orientability of the
world-sheet:
\be
\begin{array}{|c|c|c|c||c|c|c|c|} \hline
\multicolumn{4}{|c|}{\mbox{Strings}} & 
\multicolumn{4}{|c|}{\mbox{Surfaces}} \\ \hline
 & & & & \multicolumn{2}{|c|}{\mbox{boundaries}} &
\multicolumn{2}{|c|}{\mbox{orientable}} \\ 
\mbox{open} & \mbox{closed} & \mbox{oriented} & \mbox{non-oriented} & 
\mbox{without}  &
\mbox{with}  & \mbox{yes}  &  \mbox{no}  \\ \hline \hline
\mbox{x} & - & \mbox{x} & - & \mbox{x} & - & \mbox{x} & - \\ \hline
\mbox{x} & \mbox{x} & \mbox{x} & - & \mbox{x} 
& \mbox{x} & \mbox{x} & - \\ \hline
\mbox{x} & - & - & \mbox{x} & \mbox{x} & - & \mbox{x} & \mbox{x} \\ \hline
\mbox{x} & \mbox{x} & - & \mbox{x} & \mbox{x} 
& \mbox{x} & \mbox{x} & \mbox{x} \\ \hline
\end{array}
\ee

The simplest example of a non-orientable surface without boundary 
is the real projective
plane  ${\bf RP}^2$, which is obtained from ${\bf R}^2$ by adding a circle
at infinity, such that every line through the origin in ${\bf R}^2$ 
intersects the circle in one point. Equivalently, 
${\bf RP}^2$ is obtained from the disc by identifying antipodal
points on its boundary. Thus, ${\bf RP}^2$ is a closed, but non-orientable
surface, and it is a world-sheet occuring in theories of closed 
non-oriented strings. It is useful to note that ${\bf RP}^2$ 
can be obtained from the sphere, which is the tree-level world-sheet
already familiar from oriented closed strings, by the following 
procedure: start with the sphere, remove a disc, (realize that the 
result is a disc itself,) 
then identify antipodal points on the resulting boundary. This
operation is called \lq adding a crosscap'. By iterating this process we get
an infinite series of new non-orientable surfaces. For example,
by adding a second crosscap we get the Klein bottle. As we
discussed above, there is a similar operation that generates all
orientable closed surfaces from the sphere: adding a handle. 
By adding both handles and crosscaps we can generate all closed
surfaces, orientable and non-orientable. In fact, it is sufficient
to either add handles (generating all orientable surfaces) or
to add crosscaps (generating all non-orientable surfaces). 
The reason is that adding a crosscap and a handle is equivalent 
to adding three crosscaps. 

When considering theories of non-oriented open strings 
one has to add world-sheets with boundaries. These are obtained
from the world-sheets of closed strings by removing discs. For example,
removing one disc from ${\bf RP}^2$ gives the M\"obius strip.
As we discussed in the last section, the couplings between open strings,
$\kappa_{o}$, and between open and closed strings, $\kappa_{og}$, 
are related to the closed string coupling $\kappa$ by unitarity. 
The order of a given world-sheet in string perturbation theory is
$\kappa^{-\chi(g,b,c)}$, where the Euler number is now determined
by the number $g$ of handles, the number $b$ of boundary components 
and the number $c$ of crosscaps:
\be
\chi(g,b,c) = 2 - 2 g - b - c \;.
\ee
Let us write down explicitly the first few world-sheets:
\be
\begin{array}{|l|l|l|l||l||l|} \hline
g & b & c & \chi & \mbox{Surface} & \mbox{Coupling} \\ \hline \hline
0 & 0 & 0 & 2 & \mbox{Sphere} & \kappa^{-2} \\ \hline
0 & 1 & 0 & 1 & \mbox{Disc} & \kappa^{-1} \\ \hline
0 & 0 & 1 & 1 & \mbox{Real projective plane}  & \kappa^{-1} \\ \hline
1 & 0 & 0 & 0 & \mbox{Torus} & \kappa^0 \\ \hline
0 & 0 & 2 & 0 & \mbox{Klein bottle} & \kappa^0 \\ \hline
0 & 2 & 0 & 0 & \mbox{Annulus}= \mbox{cylinder} & \kappa^0 \\ \hline
0 & 1 & 1 & 0 & \mbox{M\"obius strip} & \kappa^0 \\ \hline
\end{array}
\ee

\subsection{Further reading}

Vertex operators and the Polyakov path integral are discussed
in all the standard textbooks \cite{GSW,LT,Pol,Kak1}. A very nice
introduction to the use of conformal field theory in string theory
is provided by \cite{Ginsparg}. A more detailed introduction to the
Polyakov path integral can be found in \cite{Weinberg}. For an
extensive review of this subject, see \cite{DHoPho}. A pedagogical
treatment of the mathematical ingredients needed to treat higher
genus surfaces can be found in \cite{Schlich}, whereas 
\cite{Jost} discusses the Polyakov path integral from the mathematicians
point of view.

\section{Supersymmetric strings}

The bosonic string does not have fermionic states and therefore
it cannot be used as a unified theory of particle physics and
gravity. One way to introduce fermionic states is an extension
known as the Ramond-Neveu-Schwarz string (RNS string). In this 
model one introduces new dynamical fields $\psi^{\mu} = (\psi^{\mu}_A)$
on the world-sheets, which are vectors with respect to space-time
but spinors with respect to the world-sheet. We will suppress
the world-sheet spinor indices $A=1,2$ most of the time.
Surprisingly, the presence of such fields, when combined with a 
certain choice of boundary conditions, yields states which are spinors
with respect to space-time, as we will see below.

The RNS model contains space-time bosons and fermions, but 
still has a tachyonic ground state. One then observes that
there are projections of the spectrum which simultanously
remove the tachyon and make the theories 
space-time supersymmetric. A closer inspection shows that these
projections are not optional, but required by consistency at the
quantum level. This way one obtains three consistent 
supersymmetric strings theories, called type I, type IIA and type IIB.
Finally there are also two so-called heterotic string theories, which are 
the result of a hybrid construction, combining type II and bosonic
strings. This makes a total of five supersymmetric string theories.

\subsection{The RNS model}

We now discuss the classical and quantum properties of the RNS 
string, proceeding along the same lines as we did for the 
bosonic string.

\subsubsection{The RNS action.}

The action of the RNS model is obtained from the Polyakov action
by extending it to an action with supersymmetry on the world-sheet.
Note that that world-sheet supersymmetry 
is different from, and does not imply, 
supersymmetry in space-time. The action of the RNS model
is constructed by extending the Polyakov action (\ref{PolAct})
to an action with local world-sheet supersymmetry. 
This action also  has local Weyl symmetry, and further
local fermionic symmetries which make it locally superconformal.
We will not need its explicit form here.
The analogue of the conformal
gauge is called superconformal gauge. In this gauge the action
reduces to 
\be
S_{\mscr{RNS}} = \frac{1}{4 \pi \alpha'} \int_{\Sigma} d^2 \sigma
\left(
\der_{\alpha} X^{\mu} \der^{\alpha} X_{\mu} + \I \; \overline{\psi}^{\mu}
\rho^{\alpha} \der_{\alpha} \psi_{\mu} \right)\;.
\label{RNS}
\ee
The fields $\psi^{\mu} = (\psi^{\mu}_A)$ are Majorana spinors
with respect to the world-sheet and vectors with respect to
space-time, while $\rho^{\alpha} = (\rho^{\alpha}_{AB})$ 
are the two-dimensional spin matrices. 
We will usually suppress the world-sheet spinor index
$A,B=1,2$. 
The action (\ref{RNS})
is invariant under global world-sheet supersymmetry
transformations:
\be
\delta X^{\mu} = \overline{\varepsilon} \psi^{\mu} \;, \;\;\;
\delta \psi^{\mu} = - \I\; \rho^{\alpha} \varepsilon \der_{\alpha}
X^{\mu} \;.
\ee
The equations of motion are: 
\be
\der^2 X^{\mu} = 0 \;, \;\;\;
\rho^{\alpha} \der_{\alpha} \psi^{\mu} = 0 \;.
\ee
To these one has to add the constraints, which arise from
the locally superconformal action. In this action the supersymmetric
partner of the world-sheet metric is a vector-spinor, the 
gravitino. This field is non-dynamical in two dimensions and
is set to zero in the superconformal gauge.
The equation of motion for the metric implies that the 
energy-momentum tensor vanishes on shell:
\be
T_{\alpha \beta} = \der_{\alpha} X^{\mu} \der_{\beta} X_{\mu}
+ \ft{i}2 \overline{\psi}^{\mu} \rho_{(\alpha} \der_{\beta)} \psi_{\mu}
- \mbox{   Trace   }
= 0 \;.
\ee
The equation of motion for the gravitino implies that the
world-sheet supercurrent $J_{\alpha}$ vanishes on shell:
\be
J_{\alpha} = \ft12 \rho^{\beta} \rho_{\alpha} \psi^{\mu} \der_{\beta}
X_{\mu} = 0 \;.
\ee

In order to solve the equation of motion for $\psi^{\mu}$ it is 
convenient to choose the following spin matrices:
\be
\rho^0 = \left( \begin{array}{ll} 0 &\I \\ \I & 0 \\ \end{array} \right)
\;, \;\;\;
\rho^1 = \left( \begin{array}{ll} 0 & -\I \\ \I & 0 \\ \end{array} 
\right)  \;.
\label{rho-matrices}
\ee
Using the chirality matrix $\overline{\rho} = \rho^0 \rho^1$ we see
that 
the components $\psi^{\mu}_{\pm}$ of $\psi^{\mu}$, defined by
\be
\psi^{\mu} = \left( \begin{array}{l} \psi_- \\ \psi_+ \\ \end{array}
\right)
\ee
with respect to the basis (\ref{rho-matrices})
are Majorana-Weyl spinors. The equations of motion decouple,
\be
\der_- \psi_+^{\mu} = 0 \;,\;\;\;
\der_+ \psi^{\mu}_- = 0 \;,
\label{PsiEom}
\ee
and have the general solution
\be
\psi_+^{\mu} = \psi^{\mu}_+ ( \sigma^+ ) \;,\;\;\;
\psi_-^{\mu} = \psi^{\mu}_- ( \sigma^- ) \;.
\ee
Next we have to specify the boundary conditions. Requiring the
vanishing of the boundary terms when varying the action implies:
\be
\left. (\psi^{\mu}_- \delta \psi_{\mu - }  - \psi^{\mu}_+ \delta \psi_{\mu +}) 
\right|_{\sigma^1 = 0}
=
\left. (\psi^{\mu}_- \delta \psi^{\mu}_-  - \psi^{\mu}_+ \delta \psi_{\mu +}) 
\right|_{\sigma^1 = \pi} \;.
\ee

For open strings we take
\bea
\psi^{\mu}_+ (\sigma^0, \sigma^1 = 0) &=& \psi^{\mu}_- (\sigma^0,
\sigma^1 = 0) \label{Psi} \\
\psi^{\mu}_+ (\sigma^0, \sigma^1 = \pi) &=& \pm \psi^{\mu}_- (\sigma^0,
\sigma^1 = \pi) \;. \label{PsiBC}
\eea
This couples $\psi^{\mu}_+$ and $\psi^{\mu}_-$ at the boundaries.
Depending on the choice of sign in (\ref{PsiBC})
one gets Ramond boundary conditions (\lq +' sign) or Neveu-Schwarz
boundary conditions (\lq $-$' sign). 
One can use the same doubling trick that we used to obtain
the Fourier expansion for bosonic 
open strings. Setting
\be
\psi^{\mu} ( \sigma^0, \sigma^1) := \left\{
\begin{array}{ll}
\psi_-^{\mu} ( \sigma^0, - \sigma^1 )  & \mbox{if} 
- \pi \leq \sigma^1 \leq 0 \;, \\
\psi_+^{\mu} ( \sigma^0,  \sigma^1 )  & \mbox{if} 
\;\;0 \leq \sigma^1 \leq \pi \;, \\
\end{array} \right. 
\label{DoublePsi}
\ee
we find that $\psi$ is periodic for R(amond)-boundary conditions
and antiperiodic for N(eveu-)S(chwarz)-boundary conditions on
the doubled world-sheet. Consistency at the loop level requires
that both types of boundary conditions have to be included.
The Hilbert space has both an NS-sector and an R-sector.

For closed strings we can make $\psi_+$ and $\psi_-$
either periodic (R-boundary conditions)
or antiperiodic (NS-boundary conditions):
\bea
\psi^{\mu}_+ ( \sigma^0, \sigma^1 = \pi)
&=& 
\pm \psi^{\mu}_+ ( \sigma^0, \sigma^1 = 0) \;,\\
\psi^{\mu}_- ( \sigma^0, \sigma^1 = \pi)
&=& 
\pm \psi^{\mu}_- ( \sigma^0, \sigma^1 = 0) \;.
\eea
Since $\psi^{\mu}_+$ and $\psi^{\mu}_-$ are 
independent, one has four different choices of fermionic
boundary conditions: R-R, NS-R, R-NS, NS-NS. Again 
considerations at the loop level require that all
four sectors have to be included.

We can now write down solutions of (\ref{PsiEom}) subject to
the boundary conditions that we admit. For open strings we use the
doubling trick and Fourier expand (\ref{DoublePsi}). For
R-boundary conditions one obtains,
\be
\psi_{\mp}^{\mu} = \frac{1}{\sqrt{2}} \sum_{n \in \fscr{Z}}
d^{\mu}_n e^{- \I n \sigma^{\mp}} \;,
\ee
while for NS-boundary conditions the result is: 
\be
\psi_{\mp}^{\mu} = \frac{1}{\sqrt{2}} \sum_{r \in \fscr{Z} + \ft12}
b^{\mu}_r e^{- \I r \sigma^{\mp} } \;.
\ee
For closed strings R-boundary conditions in the right-moving sector
we get:
\be
\psi^{\mu}_- = \sum_{n \in  \fscr{Z}} d^{\mu}_n
e^{-2 \I n \sigma^-}  \;,
\ee
while with NS-boundary conditions this becomes
\be
\psi^{\mu}_- = \sum_{r \in \fscr{Z}+\ft12} b^{\mu}_r
e^{-2 \I r \sigma^-} \;. 
\ee
The Fourier coefficients of the left-moving fields are
denoted $\tilde{d}^{\mu}_n$ and $\tilde{b}^{\mu}_r$, respectively.

Likewise, one obtains Fourier coefficients of the 
energy momentum tensor $T_{\alpha \beta}$ and of the
supercurrent $J_{\alpha}$. For open strings the Fourier
coefficients of $J_+$, $J_-$ (in the doubled intervall)
are denoted $F_m$ in the R-sector and $G_r$ in the NS-sector.
For closed strings the Fourier modes of $J_+$ are denoted
$F_m,G_r$, while those of $J_-$ are $\tilde{F}_m$ and $\tilde{G}_r$.
The Fourier components of $T_{++}$ and $T_{--}$ are denotes as
before.

\subsubsection{Covariant quantization of the RNS model.}

The covariant quantization of the RNS model proceeds along the
lines of the bosonic string. We will consider open strings for
definiteness.
The canonical commutation relations
of the $\alpha^{\mu}_m$ are as before. The fermionic modes 
satisfy the canonical anticommutation  relations
\be
\{ b^{\mu}_r, b^{\nu}_s \} = \eta^{\mu \nu} \delta_{r+s,0}
\ee 
in the NS-sector and
\be
\{ d^{\mu}_m, d^{\nu}_n \} = \eta^{\mu \nu} \delta_{m+n,0}
\ee 
in the R-sector. (For closed strings there are analogous
relations for the second set of of modes.)

The Virasoro generators get contributions from both the
bosonic and the fermionic oscillators, $L_m = 
L_m^{(\alpha)} + L_m^{(NS)/(R)}$.
The bosonic part $L_m^{(\alpha)}$ is given by (\ref{Lm}), while 
the  contributions from the fermionic oscillators 
in the respective sectors are:
\bea
L_m^{(NS)} &=& \ft12 \sum_{r = - \infty}^{\infty} (r + \ft12 m) \;
b_r \cdot  b_{m+r}  \;, \\
L_m^{(R)} &=& \ft12 \sum_{n = - \infty}^{\infty} (n + \ft12 m) \;
d_n \cdot  d_{m+n} \;.
\eea
The explicit formulae for the modes of the supercurrent are:
\bea
G_r &=& \sum_{n=-\infty}^{\infty} \alpha_{-n}\cdot b_{r+n} \;,\\
F_m &=& \sum_{n=-\infty}^{\infty} \alpha_{-n} \cdot d_{m+n}
\;. 
\eea

The modes of $T_{\alpha \beta}$ and $J_{\alpha}$ generate a
supersymmetric extension of the Virasoro algebra. In the NS
sector this algebra takes the form
\bea
[L_m, L_n] &=& (m-n) L_{m+n} + \frac{D}{8} (m^3 -m) \delta_{m+n,0} \;,\\
{}[L_m, G_r]  &=& ( \ft{1}{2} m -r) G_{m+r} \;,\\
\{ G_r, G_s \} &=& 2 L_{r+s} + \frac{D}{2} ( r^2 - \ft14) 
\delta_{r+s,0} \;,
\eea
while in the R-sector one finds
\bea
[L_m, L_n] &=& (m-n) L_{m+n} + \frac{D}{8} m^3 \delta_{m+n,0}\;, \\
{}[L_m, F_n] &=& ( \ft12 m -n) F_{m+n} \;,\\
\{ F_m, F_n \} &=& 2 L_{m+n} + \frac{D}{2}  m^2  \delta_{m+n,0} \;.
\eea

The subspace of physical states ${\cal F}_{\mscr{phys}} \subset {\cal F}$
is found by imposing the corresponding super Virasoro constraints.
In the NS-sector the constraints are:
\bea
L_n | \Phi \rangle &=& 0 \;,\;\;\;n> 0 \;, \nonumber \\
(L_0 -a) | \Phi \rangle &=& 0 \;,\nonumber \\
G_r | \Phi \rangle &=& 0 \;, \;\;\;r > 0 \;.
\eea 
Absence of negative norm states is achieved for
\be
D=10 \mbox{   and   } a= \ft12 \;.
\ee
(Like for bosonic strings there is the option to have
a non-critical string theory with $D<10$, which we will
not discuss here.) Thus the critical dimension has been 
reduced to 10.

In the R-sector the constraints are:
\bea
L_n | \Phi \rangle &=& 0 \;,\;\;\;n> 0 \;,\nonumber \\
(L_0 -a) | \Phi \rangle &=& 0 \;,\nonumber \\
F_n | \Phi \rangle &=& 0 \;, \;\;\;n \geq 0 \;.
\eea 
Note that there is no normal ordering ambiguity in $F_0$.
Since $F_0^2 = L_0$ we conclude $a=0$. The 
critical dimension is 10, as in the NS-sector:
\be
D=10 \mbox{   and   } a= 0 \;.
\ee

Let us construct explicitly the lowest states of the open
string in both sectors.
In the NS-sector the basic momentum eigenstates satisfy
\bea
\alpha^{\mu}_m | k \rangle &=& 0 \;, \;\;\;m>0\;, \\ 
b^{\mu}_r | k \rangle &=& 0\;,\;\;\; r > 0
\eea
and the constraint $(L_0 - \ft12) | \Phi \rangle =0$ provides
the mass formula:
\be
\alpha' M^2 = N - \ft12 \;,
\ee
where we reinstated $\alpha'$. The number operator
gets an additional term $N^{(b)}$ compared to (\ref{NumbOp}),
which counts fermionic oscillations:
\bea
N^{(d)} &=& \sum_{r=1/2}^{\infty} r \; b_{-r} \cdot b_r \;,\\
{}[N, b^{\mu}_{-r}] &=& r\; b^{\mu}_{-r} \;.
\eea
Now we can list the states:
\be
\begin{array}{|l|l|l|} \hline
\mbox{Occupation} & \mbox{Mass} & \mbox{State} \\ \hline \hline
N = 0 & \alpha' M^2 = - \ft12 & |k \rangle \\ \hline
N = \ft12 & \alpha' M^2 = 0 & b^{\mu}_{-1/2} |k \rangle \\ \hline
N = 1 & \alpha' M^2 = \ft12 & b^{\mu}_{-1/2} b^{\nu}_{-1/2} 
|k \rangle \\
 & & \alpha^{\mu}_{-1} | k \rangle \\ \hline
N = \ft32 & \alpha' M^2 = 1 & b^{\mu}_{-1/2} b^{\nu}_{-1/2} 
b^{\rho}_{-1/2} |k \rangle \\
 & & \alpha^{\mu}_{-1} b^{\nu}_{-1/2} | k \rangle \\
 & & b^{\mu}_{-3/2} | k \rangle \\ \hline
\end{array}
\ee
Thus the NS-sector of the open string consists of space-time bosons and has a 
tachyonic ground state. The massless state is a gauge boson.

The basic momentum eigenstates in the R-sector defined by:
\bea
\alpha^{\mu}_m | k \rangle &=& 0 \;, \;\;\;m>0 \;,\\ 
d^{\mu}_m | k \rangle &=& 0\;,\;\;\; m > 0 \;.
\eea
The constraint $L_0 | \Phi \rangle =0$ yields the mass
formula 
\be
\alpha' M^2 = N \;.
\ee
The number operator gets an additional fermonic contribution
\be
N^{(d)} = \sum_{m=1}^{\infty} m \; d_{-m} \cdot d_m \;.
\ee
The zero modes $d_0^{\mu}$ of the fermionic fields play a
distinguished role. Their algebra is, up to normalization,
the Clifford algebra $Cl(1,9)$:
\be
\{ d_0^{\mu}, d_0^{\nu} \} = \eta^{\mu \nu} \;.
\ee
The unique irreducible representation of this algebra is
the spinor representation of the Lorentz group
$SO(1,9)$. Introducing standard Clifford generators 
$\gamma^{\mu} = \sqrt{2} d_0^{\mu}$, the generators
of the spinor representation are 
$\sigma^{\mu \nu} = \ft14 [ \gamma^{\mu}, \gamma^{\nu} ]$.
Since the $d_0^{\mu}$ are real, this representation is
the 32-dimensional Majorana representation, denoted $[32]$.

The zero modes $d_0^{\mu}$ commute with the number operator.
Therefore the states in the R-sector organize themselves
into spinor representations of the Lorentz group. This is
how space-time spinors are described in the RNS model.
To construct the states, we denote the ground state of
the R-sector by
\be
| a \rangle \;,\;\;\; a = 1 , \ldots , 32 = 2^{D/2}\;,
\ee
where $a$ transforms in the $[32]$ representation. Then
the first states are:
\be
\begin{array}{|l|l|l|} \hline
\mbox{Occupation} & \mbox{Mass} & \mbox{State} \\ \hline \hline
N = 0 & \alpha' M^2 = 0 & |a  \rangle \\ \hline
N = 1 & \alpha' M^2 = 1 & d^{\mu}_{-1} |a \rangle \\ \
 & & \alpha^{\mu}_{-1} |a \rangle \\ \hline
\end{array}
\ee
The constraints $L_n | \Phi \rangle =0$ ($n>0$) and the new 
constraints $F_n | \Phi \rangle =0 $ ($n\geq 0$) 
impose restrictions on the polarization. For example,
$F_0 | a \rangle =0$ is easily seen to be the Fourier transform
of the massless Dirac equation and reduces the number of
independent components by a factor $\ft12$. Excited states
are obtained by acting with creation operators $\alpha^{\mu}_{-m}$,
$d^{\mu}_{-m}$ on the gound state. Since the product of a 
tensor representation of the Lorentz group with a spinor representation 
always gives spinor representations, we see that all states in 
the R-sector are space-time spinors.

\subsubsection{The GSO projection for open strings.}

The RNS model solves the problem of describing space-time
fermions but still has a tachyon. Gliozzi, Scherk and Olive
observed that one can make a projection of the spectrum,
which removes the tachyon. Moreover the resulting spectrum
is supersymmetric in the space-time sense. This 
so-called GSO projection is optional at the classical 
level, but it becomes mandatory at the quantum level, as we will
discuss below. 

The GSO projector in the NS-sector is defined as follows:
\be
P^{(NS)}_{GSO} = - (-1)^{ \sum_{r=1/2}^{\infty} b_{-r} \cdot b_r } \;.
\ee
Imposing $P^{(NS)}_{GSO} \stackrel{!}{=} 1$, one projects out
all the states which contain an even number of $b^{\mu}_{-r}$
creation operators. This in particular removes the tachyon.
The GSO projector in the R-sector is
\be
P^{(R)}_{GSO} = \overline{\gamma} (-1)^{ \sum_{m=1}^{\infty} 
d_{-m} \cdot d_m} \;,
\label{GSO-R}
\ee
where $\overline{\gamma}$ is the ten-dimensional chirality
operator. On the ground state $|a \rangle$ of the R-sector
the projection $P^{(R)}_{GSO} | \Phi \rangle \stackrel{!}{=} 1$
removes one chirality of the spinor. This is consistent, because
in ten space-time dimensions the irreducible spinor representations are
Majorana-Weyl spinors. The $[32]$ representation decomposes
according to
\be
[32] = [16]_+ + [16]_- \;.
\ee
With the GSO projection one only keeps one chirality (which we
have taken to be the $[16]_+$, for definiteness):
\be
| a \rangle = | a_+ \rangle + | a_- \rangle
\longrightarrow |a _+ \rangle \;,
\ee
where $a_+ = 1,\ldots, 16$ is a Majorana-Weyl index.

At the massive
level just projecting out one chirality would not be consistent,
as massive particles cannot be chiral. The projection with
(\ref{GSO-R}) keeps states which either have 
\lq $+$' chirality and an even
number of $d^{\mu}_{-m}$ creation operators or 
\lq $-$' chirality and an odd number of 
$d^{\mu}_{-m}$ creation operators. 

By writing down the first few states one can easily verify
that after the projection the NS-sector and R-sector have an equal number
of states, and that the  
massive states in the R-sector combine into full (non-chiral)
massive Lorentz representations.

Checking the equality of states at every mass level is done
by computing the one-loop partition function. Moreover one
can construct explicitly the representation of the ten-dimensional
super Poincar\'e algebra on the physical states. 
This is done using BRST techniques and 
lies  beyond the scope of these lectures. Here we 
restrict ourselves to noting that the ground state of the
open string, after GSO projection, is a ten-dimensional
vector supermultiplet:
\be
\{ b^{\mu}_{-1/2} | k \rangle \;, \;\;\;
|a_+\rangle \} \;.
\ee

\subsubsection{Spectrum and GSO projection for closed strings.}

Let us next study the spectrum of closed RNS strings.
The masses of states are determined by
\bea
\alpha' M^2 = 2 ( N - a_x + \tilde{N} - \tilde{a}_x)\;, & &  \nonumber \\
N - a_x = \tilde{N} - \tilde{a}_x \;,& & \label{MassII}
\eea
with normal ordering constants
$a_R = 0 = \tilde{a}_R$ and $a_{NS} = \ft12 = \tilde{a}_{NS}$.

We start by listing the first states in the NS-NS sector:
\be
\begin{array}{|l|l|l|} \hline
\mbox{Occupation} & \mbox{Mass} & \mbox{States} \\ \hline \hline
N = \tilde{N} = 0 & \alpha' M^2 = -2 & | k \rangle \\ \hline
N = \tilde{N} = \ft12 & \alpha' M^2 = 0 & b^{\mu}_{-1/2}
\tilde{b}^{\nu}_{-1/2} | k \rangle \\ \hline
N = \tilde{N} = 1 & \alpha' M^2 = 2 & 
\alpha^{\mu}_{-1} \tilde{\alpha}^{\nu}_{-1} | k \rangle \\ 
 & & 
\alpha^{\mu}_{-1} \tilde{b}^{\nu}_{-1/2} \tilde{b}^{\rho}_{-1/2} 
| k \rangle \\ 
 & & 
b^{\mu}_{-1/2} {b}^{\nu}_{-1/2} \tilde{\alpha}^{\rho}_{-1} 
| k \rangle \\ 
 & & 
b^{\mu}_{-1/2} {b}^{\nu}_{-1/2} \tilde{b}^{\rho}_{-1/2} 
\tilde{b}^{\sigma}_{-1/2} 
| k \rangle \\ \hline
\end{array}
\ee
All these states are bosons, and at the massless level
we recognize the graviton, the dilaton and the antisymmetric
tensor.

In the R-R sector, the ground state transforms in the 
$[32] \times [32]$ representation and is denoted
$|a, \tilde{a} \rangle$. The first states are
\be
\begin{array}{|l|l|l|}\hline
\mbox{Occupation} & \mbox{Mass} & \mbox{State} \\ \hline \hline
N = \tilde{N} = 0 & \alpha' M^2 = 0 & | a, \tilde{a} \rangle \\ \hline
N = \tilde{N} = 1 & \alpha' M^2 = 2 & 
\alpha^{\mu}_{-1} \tilde{\alpha}^{\nu}_{-1} | a, \tilde{a} \rangle \\ 
 &  & 
d^{\mu}_{-1} \tilde{\alpha}^{\nu}_{-1} | a, \tilde{a} \rangle \\ 
& & 
\alpha^{\mu}_{-1} \tilde{d}^{\nu}_{-1} | a, \tilde{a} \rangle \\ 
& & 
d^{\mu}_{-1} \tilde{d}^{\nu}_{-1} | a, \tilde{a} \rangle \\ \hline
\end{array}
\ee
The product of two spinor representations is a vector-like
representations. Therefore the states in the R-R sector
are bosons. In more detail, the $[32] \times [32]$ representation
is the direct sum of all the antisymmetric tensor representations
of rank zero to ten. Using the ten-dimensional $\Gamma$-matrices
we can decompose a general massless state into irreducible
representations:
\be
| \Phi_{RR} \rangle = 
( F \delta_{a \tilde{a}} 
+ F_{\mu} \Gamma^{\mu}_{a \tilde{a}} +
F_{\mu \nu} \Gamma^{\mu \nu}_{a \tilde{a}} + \cdots ) | a,  \tilde{a} \rangle
\;.
\ee
By evaluating the remaining constraints 
$F_0 | \Phi_{RR} \rangle = 0 = \tilde{F}_0 | \Phi_{RR} \rangle$
one obtains the 
conditions 
\be
k^{\mu_1} F_{\mu_1 \mu_2 \ldots \mu_n} = 0 
\mbox{   and   }
k_{[\mu_0} F_{\mu_1 \mu_2 \ldots \mu_n]} = 0\;,
\ee
which are the Fourier transforms of the equation of motion
and Bianchi identity of an $n$-form field strength:
\be
d \star F_{(n)} =0 \mbox{   and   }
d F_{(n)} = 0 \;.
\ee
The physical fields are antisymmetric tensor gauge fields
or rank $n-1$. Note that in contrast to the antisymmetric 
NS-NS field, the states in the R-R sector (and the corresponding
vertex operators) describe the field strength and not the
gauge potential. When analyzing interactions one finds that
there are no minimal gauge couplings but only momentum
couplings of these fields (i.e. couplings involving the field
strength). In other words the perturbative
spectrum does not contain states which are charged under these
gauge fields. This is surprising, but a closer analysis shows
that the theory has solitonic solutions which carry R-R charge.
These so called R-R charged p-branes turn out to be
an alternative description of D-branes.

Now we turn to the NS-R sector. The first states are:
\be
\begin{array}{|l|l|l|} \hline
\mbox{Occupation} & \mbox{Mass} & \mbox{State} \\ \hline \hline
N= \ft12 \;, \tilde{N} =0& \alpha' M^2 = 0 &
b^{\mu}_{-1/2} | \tilde{a} \rangle \\ \hline
N= \ft32 \;, \tilde{N} =1& \alpha' M^2 = 4 &
\alpha^{\mu}_{-1} {b}^{\nu}_{-1/2} \tilde{\alpha}^{\rho}_{-1}
| \tilde{a} \rangle \\ 
 & & b^{\mu}_{-1/2} {b}^{\nu}_{-1/2} b^{\rho}_{-1/2}
\tilde{\alpha}^{\sigma}_{-1}
| \tilde{a} \rangle \\ 
 & & \alpha^{\mu}_{-1} {b}^{\nu}_{-1/2} \tilde{d}^{\rho}_{-1}
| \tilde{a} \rangle \\ 
 & & b^{\mu}_{-1/2} {b}^{\nu}_{-1/2} b^{\rho}_{-1/2} \tilde{d}^{\mu}_{-1}
| \tilde{a} \rangle \\ \hline
\end{array}
\ee
The massless state is a product of a vector $[D]$ and a 
spinor $[2^{D/2}]$. It decomposes into a vector-spinor and a 
spinor:
\be
[D] \times [2^{D/2}] = [ (D-1) 2^{D/2}] + 2^{D/2} \;.
\ee
Therefore this state and all other states in the NS-R sector
are space-time fermions. The spectrum of the R-NS sector
is obtained by exchanging left- and right-moving fermions.

We observe that the massless states contains two vector-spinors.
The only known consistent interaction for such fields
is supergravity. There these fields are called gravitini. They
sit in the same supermultiplet as the graviton, they are the
gauge fields of local supertransformations and couple 
to the conserved supercurrent. The spectrum of the closed RNS
model is obviously not supersymmetric. This suggests that we
have to make a projection in order to obtain consistent 
interactions. This brings us to the GSO projection for 
closed strings, which makes the spectrum supersymmetric and removes
the tachyon. The GSO projection is applied both in the
left-moving and in the right-moving sector. In the R-sectors
one has to decide which chirality one keeps. There are two
inequivalent projections of the total spectrum: one either 
takes opposite chiralities of the R-groundstates (type A)
or the same chiralities (type B). The resulting theories
are the type IIA and type IIB superstring. Let us 
look at their massless states.
The NS-NS sectors of both theories are identical. The states
\be
b^{\mu}_{-1/2} \tilde{b}^{\nu}_{-1/2} | k \rangle
\ee
are the graviton $G_{\mu \nu}$ the dilaton $\Phi$ and
the antisymmetric tensor $B_{\mu \nu}$. The number of
on-shell states is $8 \cdot 8 = 64$. 
The ground states of the R-R sectors are:
\bea
| a_+ , \tilde{a}_- \rangle & & \mbox{(type A)} \;, \\
| a_+ , \tilde{a}_+ \rangle & & \mbox{(type B)} \;.
\eea
In both cases we have $8 \cdot 8 = 64$ on-shell states.
Again we can decompose these representations into irreducible
antisymmetric tensors. For type IIA we get a two-form and
a four-form field strength, corresponding to a one-form
and a three-form potential:
\be
\mbox{IIA}: \;\;A_{\mu} \;,\;\;A_{\mu \nu \rho} \;.
\ee
There is also a zero-form field strength which has no
local dynamics. It corresponds to the so-called massive
deformation of IIA supergravity, which is almost but 
not quite a cosmological constant. (In the effective action
the corresponding term is a dimensionful constant multiplied by
the dilaton. This is as close as one can get to a cosmological
constant in ten-dimensional supergravity.)

In the IIB theory one has a one-form, a three-form and 
a selfdual five-form field strength. The corresponding 
potentials are:
\be
\mbox{IIB}: \;\; A \;, \;\;A_{\mu \nu} \;, \;\;
A_{\mu \nu \rho \sigma} \;.
\ee

The massless states in the NS-R sector and R-NS sector are:
\bea
\mbox{IIA}: \; b^{\mu}_{-1/2} | \tilde{a}_- \rangle & & 
\tilde{b}^{\mu}_{-1/2} | a_+ \rangle \;, \\
\mbox{IIB}: \; b^{\mu}_{-1/2} | \tilde{a}_+ \rangle & & 
\tilde{b}^{\mu}_{-1/2} | a_+ \rangle \;, 
\eea
The total number of fermionic states is $128$ in both
cases. The decomposition into irreducible representations 
gives two vector-spinors, the gravitini, and two spinors,
called dilatini. For type IIA they have opposite chiralities,
whereas for type IIB they have the same chiralities.
The corresponding space-time fields are:
\bea
\mbox{IIA}: \; & & \psi^{\mu}_+ \;, \;\; \psi^{\mu}_- \;,
\;\;\psi_+ \;, \;\; \psi_- \;,   \nonumber \\
\mbox{IIB}: \; & & \psi^{\mu}_{+(1)} \;, \;\; \psi^{\mu}_{+(2)} \;,
\;\;\psi_{+(1)} \;, \;\; \psi_{+(2)} \;.  
\eea

All together we get the field content of the type IIA/B
supergravity multiplet with 128 bosonic and 128 fermionic
on shell states. The IIA theory is non-chiral whereas the
IIB theory is chiral. The massive spectra are of course 
non-chiral, and, moreover, they are identical.

\subsection{Type I and type II superstrings}

We will now begin to list all consistent supersymmetric string
theories. A priori, we have the following choices:
strings can be (i) open or closed, (ii) oriented or non-oriented,
(iii) one can make the GSO projection, with two inequivalent choices
(type A and B) for closed strings and (iv) one can choose 
gauge groups for open strings: $U(n)$ for oriented and
$SO(n)$ or $Usp(2n)$ for non-oriented strings. 

We have already seen that not all combinations of these choices
are consistent at the quantum level. Since theories of 
open strings have closed string poles in loop diagrams,
we can either have closed or closed and open strings. 
The next restriction comes from modular invariance. 
On the higher genus world-sheets of closed oriented strings,
one has to specify boundary conditions around every handle.
Since modular invariance maps one set of boundary conditions
to others, these choices are not independent. It turns out
that one has to include both 
NS- and R-boundary conditions around every handle,
but one has the freedom of choosing relative signs between
different orbits of action of the modular group on the set of
boundary conditions. There are four possible
choices. Two of them correspond to the IIA and IIB superstrings.
The other two choices are non-supersymmetric theories without
fermions, known as type 0A and 0B, which we will not 
discuss here.

Type IIA and IIB are theories of oriented closed strings. Can we 
construct supersymmetric string theories with oriented
closed and open strings? 
The states of the oriented open string 
fall into representations of 
the minimal $N=1$ supersymmetry algebra in $D=10$. This algebra
has 16 supercharges, which transform as a Majorana-Weyl 
spinor under the Lorentz group.
In ten dimensions there are two further 
supersymmetry algebras, called $N=2A$ and $N=2B$. They have
32 supercharges which either combine into two Majorana-Weyl
spinors of opposite chirality (A) or into two Majorana-Weyl
spinors of the same chirality (B).
The states of the oriented closed string form multiplets
of the $N=2A$ or $N=2B$ supersymmetry algebra.
In particular one has two gravitini, which must couple
to two independent supercurrents.
Therefore oriented open and closed strings cannot be coupled in
a supersymmetric way. One can also show that any such theory
has divergencies, due to the presence of dilaton tadpoles. 

Next we have to consider non-oriented strings. A theory of non-oriented 
closed strings can be obtained by projecting the 
type IIB theory onto states invariant under world-sheet parity.
(IIA is not invariant under world-sheet parity, because the
R-groundstates have opposite chirality.)
This theory has divergencies, which are related to the non-vanishing
of dilaton tadpole diagrams. One can also see from the space-time
point of view that this theory is inconsistent: the massless states
form the $N=1$ supergravity multiplet, which is chiral. Pure
$N=1$ supergravity 
has a gravitational anomaly, which can only be cancelled by
adding precisely 496 vector multiplets.

Therefore we have to look at theories with non-oriented  closed and
open strings. Tadpole cancellation precisely occurs if the
gauge group is chosen to be $SO(2^{D/2}) = SO(32)$. 
This is one of the gauge groups for which gravitational 
anomalies cancel. The other anomaly-free gauge groups are
$E_8 \times E_8$, $E_8 \times U(1)^{248}$ and $U(1)^{496}$, which,
however, cannot be realized through Chan-Paton factors.
Thus there is one supersymmetric string theory with
non-oriented closed and open strings and gauge group $SO(32)$.
This is the type I superstring.

Let us construct the massless spectrum of this theory.
The closed string sector is obtained by projecting
the IIB theory onto states invariant under world-sheet parity.
Parity acts by exchanging left- and right-moving quantities:
\be
\alpha^{\mu}_m \leftrightarrow
\tilde{\alpha}^{\mu}_m \;, \;\;\;
b^{\mu}_r \leftrightarrow
\tilde{b}^{\mu}_r \;, \;\;\;
d^{\mu}_m \leftrightarrow
\tilde{d}^{\mu}_m \;, \;\;\;
|a_+ \rangle \leftrightarrow | \tilde{a}_+ \rangle \;.
\ee
The action on the R-R ground state is:
\be
| a_+, \tilde{a}_+ \rangle \leftrightarrow 
- | \tilde{a}_+, a_+ \rangle \;.
\ee
The \lq $-$' sign reflects that one exchanges two fermionic states.
(To make this precise one needs to construct the so-called
spin fields $S^a$, $S^{\tilde{a}}$ which generate the R-groundstates
from the NS-groundstate. This can be done in the framework
of BRST quantization, which we did not introduce here.)

We can now write down the massless states of the type IIB string
which are invariant under world-sheet parity and survive the
projection. In the NS-NS sector we find
\be
\mbox{NS-NS} \;: \;\;\; \ft12 \left (
b^{\mu}_{-1/2} \tilde{b}^{\nu}_{-1/2} 
+ b^{\nu}_{-1/2} \tilde{b}^{\mu}_{-1/2} \right) 
| k \rangle \;.
\ee
Therefore the $B_{\mu \nu}$ field is projected out and we are left 
with the graviton $G_{\mu \nu}$ and dilaton $\Phi$.
In the R-R sector the invariant massless state is:
\be
\mbox{R-R} \;: \;\;\; \ft12 \left( |a_+, \tilde{a}_+ \rangle 
- | \tilde{a}_+ , a_+ \rangle \right) \;.
\ee
Thus only the antisymmetric part of the tensor product of the
two Majorana-Weyl spinors survives the projection. This  corresponds
to the three-form field strength $F_{\mu \nu \rho}$, as is most
easily seen by computing the dimensions of the representations.
Thus the two-form R-R gauge field $A_{\mu \nu}$ survives the
projection. 

In the NS-R and R-NS one finds the following invariant
state:
\be
\mbox{R-NS/NS-R} \;: \;\; 
\ft12 \left(
b^{\mu}_{-1/2} | \tilde{a}_+ \rangle +
\tilde{b}^{\mu}_{-1/2} | a_+ \rangle \right) \;.
 \ee
Therefore one gravitino $\psi^{\mu}_+$ and one dilatino
$\psi^{\mu}$ are kept.

In the NS-sector of the open string we get massless vectors
$A^i_{\mu}$, which transform in the adjoint representations
of $SO(32)$: $i = 1,\ldots, \dim ( \mbox{adj} SO(32) ) = 496$.
The R-sector contains massless spinors $\psi^i$ which 
combine with the vectors to form vector supermultiplets.

Combining the massless states of the closed and open string sector
we get the field content of $N=1$ supergravity coupled to 
Super-Yang-Mills theory with gauge group $SO(32)$.

\subsection{Heterotic strings}

There is yet another construction of supersymmetric
string theories. It is a hybrid construction, which combines the  
bosonic string with the type II superstring and is called the 
heterotic string. The right-moving sector is taken from
the type II superstring, whereas the left-moving 
sector is taken from the bosonic string. To get a modular
invariant theory, the sixteen extra left-moving coordinates
have to be identified periodically,
\be
X^I \simeq X^I + w^I_{(i)} \;,\;\;\;I=1,\ldots,16 \;. 
\ee
The vectors
$\vec{w}_{(i)}= (w^I_{(i)})$, $i=1,\ldots,16$ 
generate a sixteen dimensional lattice $\Gamma_{16}$.
Modular invariance requires that $\Gamma_{16}$ is an even 
self-dual lattice. Modulo rotations, there are only two
such lattices, the root lattice of $E_8 \times E_8$ and
the lattice generated by the roots and the weights of one
of the Majorana-Weyl spinor representations of $SO(32)$.
Thus, there are two different heterotic string theories.

The bosonic massless states come from the NS-sector and take
the form
\bea
\alpha^{\mu}_{-1} \tilde{b}^{\nu}_{-1/2} | k\rangle & & \\
\alpha^I_{-1} \tilde{b}^{\nu}_{-1/2} | k\rangle & & \\
e^{\I {k}^{(i)}_I x^I_L} \tilde{b}^{\nu}_{-1/2} | k\rangle & & 
\eea
Here $\alpha^I_{-1}$ are the oscillators corresponding
to the sixteen extra left-moving directions. 
The vectors $k^{(i)}= (k^{(i)}_I)$ are discrete momentum vectors in 
the extra dimensions. The above states are massless 
if the vectors $k^{(i)}$ have norm-squared two.
The two lattices $\Gamma_{16}$ have 480 such vectors,
corresponding to the roots of $E_8 \times E_8$ and
$SO(32)$, respectively. Together with the states generated
by the internal oscillators one gets bosons in the adjoint
representations of theses two groups.
The massless fermionic states are obtained by replacing
$\tilde{b}^{\nu}_{-1/2} |k \rangle$ by the R-ground state
$| a_+ \rangle$. In total one gets the $N=1$ supergravity
multiplet plus vector multiples in the adjoint representation
of $E_8 \times E_8$ or $SO(32)$. 

The massless sectors of the five supersymmetric string theories 
correspond to four different supergravity theories.
The type I and the heterotic string with gauge group $SO(32)$ have
the same massless spectrum, but their massive spectra and interactions
are different.

Let us summarize the essential properties of the five
supersymmetric string theories:
\be
\begin{array}{|c|c|c|c|c|c|} \hline
\mbox{Type} & \mbox{open/closed?} & \mbox{oriented?} 
& \mbox{chiral?} 
& \mbox{supersymmetry} 
& \mbox{gauge group} \\ \hline \hline 
\mbox{I} & \mbox{both} & \mbox{no} &
\mbox{yes} & N=1 & SO(32) \\ \hline
\mbox{II A} & \mbox{closed} & \mbox{yes} & \mbox{no} & N=2A & - \\ \hline
\mbox{II B} & \mbox{closed} & \mbox{yes} & \mbox{yes} & N=2B & - \\ \hline
\mbox{Heterotic} & \mbox{closed} & \mbox{yes} & \mbox{yes} & N=1 & 
E_8 \times E_8 \\ \hline
\mbox{Heterotic} & \mbox{closed} & \mbox{yes} & \mbox{yes} & N=1 & 
SO(32) \\ \hline
\end{array}
\ee

\subsection{Further reading}

Supersymmetric string theories are discussed in all of the
standard textbooks \cite{GSW,LT,Pol,Kak1,Kak2}. To prove 
the necessity of the GSO projection and the consistency of the
heterotic string as a perturbative quantum theory one needs
properties of the multiloop path integral \cite{DHoPho}. 
A paedagogical treatment of the relation between the GSO projection
and boundary conditions in the path integral can be found in
\cite{Ginsparg}.

\section{p-Branes in type II string theories}

In this section we discuss a class of solitons of the type II 
string theories, which turn out to be alternative descriptions
of the D-branes introduced earlier.

\subsection{Effective actions of type II string theories}

The effective actions for the massless states of type IIA/B
superstring theory are the corresponding type IIA/B supergravity
actions. Since we will be interested in bosonic solutions
of the field equations, we will only display the bosonic parts.
The effective action for the fields in the NS-NS sector is
the same for both theories. Moreover it is identical 
to the effective action of the bosonic string:
\be
S_{\mscr{NS-NS}} = \frac{1}{2 \kappa^2} \int d^{10} x \sqrt{-G}
e^{-2 \Phi} \left( R + 4 \der_{\mu} \Phi \der^{\mu} \Phi
- \ft1{12} H_{\mu \nu \rho} H^{\mu \nu \rho} \right)\;.
\ee
The R-R sectors consist of antisymmetric tensor gauge fields.
For an $(n-1)$ form gauge potential $A_{(n-1)}$ 
with field strength $F_{(n)} = d A_{(n-1)}$ the generalized
Maxwell action is
\be
S \simeq  - \ft12 \int F_{(n)} \wedge \star F_{(n)} = 
- \ft12 \int d^D x \sqrt{-G} | F_{(n)} |^2 \;,
\label{Maxwell}
\ee
where
\be
|F_{(n)}|^2 := \frac{1}{n!} F_{\mu_1 \cdots \mu_n}
F^{\mu_1 \cdots \mu_n} \;.
\ee
In the effective R-R actions one has in addition Chern-Simons terms.

In the IIA theory the R-R fields are $A_{(1)}$ and $A_{(3)}$. It is convenient
to define a modified field strength
\be
\tilde{F}_{(4)} = d A_{(3)} - A_{(1)} \wedge H_{(3)} \;,
\ee 
where $H_{(3)} = d B_{(2)}$ is the field strength of the 
antisymmetric NS-NS tensor field. Then the R-R action is the sum
of a Maxwell and a Chern-Simons term:
\bea
S^{\mscr{IIA}}_{\mscr{R-R}} &=& - \frac{1}{4 \kappa^2}
\int d^{10} x \sqrt{-G} \left( | F_{(2)} |^2 +
| \tilde{F}_{(4)} |^2 \right)  \nonumber \\
 & & - \frac{1}{4 \kappa^2} \int B_{(2)} \wedge F_{(4)}
\wedge F_{(4)}  \;.
\eea

The massless R-R fields of IIB string theory are $A_{(0)}$, $A_{(2)}$ and
$A_{(4)}$. Again it is useful to define modified field strengths
\bea
\tilde{F}_{(3)} &=& F_{(3)} - A_{(0)} \wedge H_{(3)} \;,\nonumber \\
\tilde{F}_{(5)} &=& F_{(5)} - \ft12 A_{(2)} \wedge H_{(3)} 
+ \ft12 B_{(2)} \wedge F_{(3)}  \;.
\eea
Since $\tilde{F}_{(5)}$ must be selfdual, the kinetic
term (\ref{Maxwell}) vanishes and does not give a field equation.
The simplest way out is to impose the selfduality condition only
at the level of the equation of motion. Then one can use the
action
\bea
S^{\mscr{IIB}}_{\mscr{R-R}} & = &  - \frac{1}{4 \kappa^2} \int d^{10}x
 \sqrt{-G} \left( | F_{(1)} |^2 + | \tilde{F}_{(3)} |^2 
+ \ft12 | \tilde{F}_{(5)} |^2 \right)\nonumber \\
 & & - \frac{1}{4 \kappa^2} \int A_{(4)} \wedge H_{(3)} \wedge F_{(3)} \;.
\eea
The correct covariant equations of motion result when 
varying the action and imposing selfduality of $\tilde{F}_{(5)}$
afterwards.

\subsection{R-R charged p-brane solutions}

The type II effective actions have static solutions which are charged under
the R-R gauge fields. The solution charged under $A_{(p+1)}$ has
$p$ translational isometries. From far it looks like a p-dimensional
membrane and therefore one calls it a p-brane solution or just
a p-brane.

For $0 \leq p \leq 2$ the solution has the following form:
\bea
ds^2_{\mscr{Str}} &=& H^{-1/2}(r) \left( - dt^2 + (dx^1)^2 + \cdots
+ (dx^p)^2 \right) \nonumber \\
 & & + H^{1/2}(r) \left( (dx^{p+1})^2 + \cdots
+ (dx^9)^2 \right) \;,\nonumber \\
F_{(p+2)} &=& d H^{-1}(r) \wedge dt \wedge dx^1 \wedge \cdots
\wedge dx^p \;, \nonumber \\
e^{- 2 \Phi} &=& H^{(p-3)/2}(r) \;, \label{RR-p}
\eea 
where
\be
r^2 = (x^{p+1})^2 + \cdots + (x^{9})^2 
\ee
and $H(r)$ is a harmonic function of the transverse coordinates
$(x^{p+1}, \ldots, x^9)$:
\be
\Delta^{\perp} H = \sum_{i=p+1}^9 \der_i \der_i H = 0 \;.
\ee
We require that the solution becomes asymptotically flat
at transverse infinity and normalize
the metric such that it approaches the standard
Minkowski metric. This fixes 
\be
H(r) = 1 + \frac{Q_p}{r^{7-p}} \;.
\ee
$Q_p$ measures the flux of the R-R field strength at transverse infinity.
A convenient way to parametrize it is:
\be
Q_p = N_p c_p \;,\;\;\;
c_p = \frac{ (2 \pi)^{7-p} }{(7-p) \omega_{8-p} } ( \alpha' )^{\frac{7-p}{2}}
g_S \;.
\label{c-p}
\ee
$N_p$ is a constant, which a 
priori is real, but will turn out later to be an integer.
Therefore $c_p$ is the fundamental quantum of R-R
p-brane charge. $\omega_n$ is the volume of the n-dimensional
unit sphere,
\be
\omega_n = \frac{2 \pi^{(n+1)/2} }{\Gamma \left( \frac{n+1}{2} \right) } \;.
\ee
Besides geometrical factors, $Q_p$ contains the appropriate
power of $\alpha'$ to give it the correct dimension. $g_S$ is the
dimensionless string coupling. Note that in the above solution for the dilaton 
we have subtracted the dilaton vacuum expectation value from $\Phi$.

The metric used in this solution is the string frame metric, as indicated
by the subscript. (The effective action was also given in the string
frame.) Using (\ref{Frames})  we can find the corresponding Einstein 
frame metric:
\bea
ds^2_{\mscr{Einst}} &=& - H^{\frac{p-7}{8}}(r) 
\left( - dt^2 + (dx^1)^2 + \cdots
+ (dx^p)^2 \right) \nonumber \\
 & & + H^{\frac{p+1}{8}}(r) \left( (dx^{p+1})^2 + \cdots
+ (dx^9)^2 \right)  \;.
\eea

The above solution is most easily understood as a generalization
of the extreme Reissner-Nordstrom solution of four-dimensional 
Einstein-Maxwell theory. Let us review its properties. 

The isometry directions $t,x^1, \ldots, x^p$ are called 
longitudinal or world-volume directions, the others transverse
directions.
Since the
solution has translational invariance, it has infinite mass, as long
as one does not compactify the world-volume directions. However, the
tension $T_p$ (the energy per world volume) is finite. Since the solution
becomes asymptotically flat in the transverse directions, the tension
can be defined by a generalization of the ADM construction of general
relativity. Concretely, the tension of a p-brane
can be extracted from the Einstein frame 
metric by looking
at the leading deviation from flatness:
\be
g_{00} = - 1 + \frac{16 \pi G_N^{(D)} T_p}{(D-2) \omega_{D-2-p} r^{D-3-p}}
+ \cdots 
= - 1 + \frac{16 \pi G_N^{(10)} T_p}{8 \omega_{8-p} r^{7-p}} + \cdots
\label{ADM-Tension}
\ee
The Schwarzschild radius $r_S$ of the brane is:
\be
r_S^{D-3-p}  = \frac{16 \pi G_N^{(D)} T_p}{(D-2) \omega_{D-2-p} } \;.
\label{SchwSch}
\ee

Since there is only one independent dimensionful constant,
which we take to be $\alpha'$, we can express the ten-dimensional
Newton constant $G_N^{(10)}$ in terms of $\alpha'$ and the
dimensionless string coupling $g_S$:
\be
G_N^{(10)} = 8 \pi^6 (\alpha')^4 g_S^2 \;.
\ee
Since Newton's constant is related to the physical gravitational
coupling by
\be
8 \pi G_N^{(D)} =  \kappa_{(D),\mscr{phys}}^2
\ee
in any dimension, this corresponds to replacing the conventional
choices (\ref{kappa-phys},
\ref{kapppa-phys2}) by
$\kappa^2 \stackrel{!}{=} 64 \pi^7 (\alpha')^4$ and 
$\kappa_{\mscr{phys}}^2 = 64 \pi^7 (\alpha')^4 g_S^2$.

Using (\ref{ADM-Tension}) we can compute the tension
of the p-brane solution (\ref{RR-p}):
\be
T_p = \frac{N_p}{g_S ( \alpha')^{\frac{p+1}{2}} ( 2 \pi)^p }  \;.
\label{T-p} 
\ee

For $r\rightarrow 0$ the solution (\ref{RR-p}) has a null singularity, that is
a curvature singularity which is lightlike and coincides with an
event horizon. The p-brane (\ref{RR-p}) is the extremal limit of 
a more general black p-brane solution, which has a time-like singularity
along a p-dimensional surface and a regular event horizon. In the extremal
limit, the singularity and the even horizon coincide.
This behaviour is similar to the Reissner-Nordstrom black hole.
The behaviour of the black p-brane in the extremal limit is slightly 
more singular, because for the extremal Reissner-Nordstrom black hole
singularity and horizon do not coinicide. But since the singularity
of the p-brane solution is not naked,
we can think about it as describing an extended charged black hole.
The charge (density) carried under the gauge field $A_{(p+1)}$ can be read off
from the asymptotic behaviour of the field strength,
\be
F_{01 \ldots p} \simeq \frac{Q_p}{r^{8-p}} \;.
\ee
Instead of $Q_p$ we can define use a redefined charge, which 
has the dimension of a tension:
\be
\hat{Q}_p =  \frac{1}{2 \kappa^2} \oint_{S_{8-p}}
\star F_{(p+2)} \;,
\ee
which gives
\be
\hat{Q}_p = N_p \frac{\mu_p}{g_S} \;, \;\;\;
\mu_p = \frac{1}{ (2 \pi)^p ( \alpha')^{\frac{ p+1 }{2} }} \;.
\label{mu-p}
\ee
We now observe that tension and charge are equal:
\be
T_p = \hat{Q}_p \;.
\ee

More generally, 
black p-brane solutions satisfy the Bogomol'nyi 
bound 
\be
T_p \geq \hat{Q}_p \;.
\label{BPS-bound}
\ee
This inequality guarantees the existence of an event horizon,
just as for charged black holes.

A feature that distinguishes our solutions from 
Reissner-Nordstrom type black holes is that
one also has a non-trivial scalar, the dilaton.

The extremal solution has a multicentered generalization. When
replacing $H(r)$ by
\be
H(\vec{x}_{\perp}) =  1  + \sum_{i=1}^N 
\frac{ |Q_p^{(i)}| }{ | \vec{x}_{\perp} - \vec{x}_{\perp}^{(i)} |^{7-p} } \;,
\ee
one still has a static solution, provided that all the charges
$Q_p^{(i)}$ have the same sign. Here $\vec{x}_{\perp} = (x^{p+1},
\ldots, x^{9})$ and $\vec{x}_{\perp}^{(i)}$ is the position of
(the horizon of) the i-th p-brane. It is remarkable that the solution
is static for arbitrary positions $\vec{x}_{\perp}^{(i)}$, because
this implies that the gravitational attraction and the 
\lq electrostatic' repulsion cancel for arbitrary distances. 
(If one flips the sign of one charge, one has to flip the corresponding
tension, which makes the solution unphysical.) 
Systems of extremal Reissner-Nordstrom black holes have the
same properties. The corresponding multi-centered solutions
are known as Majumdar-Papapetrou solutions.

The remarkable properties of these (and other related) solutions
can be understood in terms of supersymmetry. The solution (\ref{RR-p}) is
a supersymmetric solution, i.e., it has Killing spinors.
Killing spinors are the supersymmetric analogues of 
Killing vectors $v(x)$, which satisfy
\be
{\cal L}_{v(x)} \left .\Psi(x)  \right|_{\Psi_0(x)} = 0 \;,
\ee
where ${\cal L}$ is the Lie derivative. Here $\Psi(x)$ 
collectively denotes all the fields, and $\Psi_0(x)$ is
the particular field configuration, which is invariant under
under the transformation generated by the vector field $v(x)$. 
In supergravity theories one can look for field configurations
$\Psi_0(x)$ which are invariant under supersymmetry transformations. 
From the action
one knows the supersymmetry variations of all the fields,
$\delta_{\varepsilon(x)} \Psi(x)$, where the spinor (field)
$\varepsilon(x)$ is the transformation parameter. Then one can
plugg in a given field configuration $\Psi_0(x)$ and check 
whether the variation vanishes for a specific choice 
of $\varepsilon(x)$:
\be
\delta_{\varepsilon(x)} \left. \Psi(x) \right|_{\Psi_0(x)} = 0 \;.
\label{KillingSpinor}
\ee
Since the supersymmetry transformations involve derivatives
of $\varepsilon(x)$, this is a system of first order differential equation
for $\varepsilon(x)$. Solutions of (\ref{KillingSpinor}) are called
Killing spinors.

The type II superalgebras have 32 independent real transformation parameters, 
which organize themselves into two Majorana-Weyl spinors 
$\varepsilon_{i}(x)$. The equation (\ref{KillingSpinor})
fixes the space-time dependence of the $\varepsilon_{i}(x)$. 
For the p-brane one finds 
\be
\varepsilon_{i}(x) = g_{tt}^{1/4}(x) \varepsilon^{(0)}_i \;,
\ee
where the constant Majorana-Weyl spinors $\varepsilon^{(0)}_i$, $i=1,2$
are related by
\be
\varepsilon^{(0)}_2 = \Gamma^0 \cdots \Gamma^p \varepsilon^{(0)}_1 \;.
\label{KSRR}
\ee
Since half of the components of the $\varepsilon_{i}^{(0)}$ is fixed
in terms of the other half,  we see that we have 16 independent
solutions, i.e., 16 Killing spinors. The maximal number
of Killing spinors equals the number of sypersymmetry transformation
parameters, which is 32 in type II theory. Solutions with the maximal number
of Killing spinors are
invariant under all supersymmetry transformations. They are the analogues
of maximally symmetric spaces in Riemannian geometry, which by definition
have as many isometries as flat space.
One example of a maximally supersymmetric solution of type II theory
is flat ten-dimensional Minkowski space. Here the Killing spinor equation is
solved by all constant spinors. 
The p-brane solution (\ref{RR-p}) has 16 Killing spinors,
and only is invariant under half of the supersymmetry transformations.
Solutions with residual supersymmetry are called BPS solutions, and 
solutions which preserve half of the supersymmetry are called
\lq $\ft12$ BPS solutions'. 

The
Bogomol'nyi bound (\ref{BPS-bound}) can be shown to follow 
from supersymmetry. In this context it is then also called the BPS bound.
In theories where the supersymmetry algebra contains central charges,
(\ref{BPS-bound}) is a relation between the mass or tension of a state
and its central charge. In our case the charges carried under the
R-R gauge fields are such central charges. The representations
of the supersymmetry algebra fall into distinct classes, depending on
whether they saturate the bound or not. Representations which saturate
the bound are called short representations or BPS representations. Since
BPS states have the minimal tension possible for their charge they are
absolutely stable. This minimization of energy also accounts for the
existence of static multicentered solutions.

So far we restricted ourselves to p-branes solutions with 
$0\leq p \leq 2$. There is a second class, where the solution
(\ref{RR-p}) and the other formulae take the form form, 
but with $p$ replaced by $\tilde{p}$ with 
$4 \leq \tilde{p} \leq 6$. The field strength 
$F_{\tilde{p}+2}$
in equation (\ref{RR-p}) is the $\star$--dual of $F_{p+2} = d A_{p+1}$. 
Since $F_{\tilde{p} + 2} = \star F_{p+2}$ implies 
$p+ \tilde{p} + 4 = D =10$, each of the so-called electric
solutions with $p=0,1,2$ has a dual magnetic solution with 
$\tilde{p}=6,5,4$.

There is also a solution with $p=3$. 
The five-form gauge field is selfdual, and the solution for $F_5$ is
from (\ref{RR-p}) by adding the $\star$-dual of the right hand side
of the equation. The solutions for the metric and for the dilaton 
are not modified. 
Note that for $p=3$  the dilaton is constant. The three-brane solution is not
singular at $r=0$. Instead one has a regular horizon, and the geometry
is asymptotic to $AdS^5 \times S^5$. This geometry has 32 Killing spinors
and is fully supersymmetric. The interior of this geometry is isometric
to the exterior, in particular it is non-singular. Since the field
strength is selfdual, the three-brane carries an equal amount of electric
and magnetic charge (it is not only dyonic, carrying both electric and
magnetic charge, but selfdual).

Electric and magnetic charges are subject to a generalized Dirac 
quantization condition, which can be found by generalizing either the
Dirac string or the Wu-Yang construction known from four-dimensional
magnetic monopoles. In our conventions
the condition is:
\be
( 2\pi)^7 g_S^2 (\alpha')^4 \;
\hat{Q}_p \hat{Q}_{\tilde{p}} 
\in  2 \pi {\bf Z} \;.
\ee
This fixes the possible magnetic charges in terms of the electric
charges. Using T-duality and S-duality one can fix the electric and magnetic
charge units. T-duality is a symmetry that can be proven to  hold in string
perturbation theory. It acts on our solutions by transforming 
p-branes into $(\mbox{p} \pm 1)$-branes. 
In this way one can relate the tensions and
charges of all R-R charged 
p-branes. S-duality is a conjectured non-perturbative
symmetry of IIB string theory. It relates the R-R one-brane to 
a solution which describes the fundamental
IIB string. This way one relates the fundamental unit of R-R one-brane
charge to the charge carried by a fundamental IIB string under the
NS-NS B-field. The resulting R-R p-brane charge units 
are given by $\mu_p$ (\ref{mu-p}) and satisfy Dirac quantization in a minimal
way: $\mu_p \mu_{\tilde{p}} (2 \pi)^7 (\alpha')^4 = 2 \pi$.
Thus $N_p$ in (\ref{mu-p}) is an integer which counts
multiples of the fundamental R-R charge. When using $Q_p$ instead
of $\hat{Q}_p$ to measure charges then $c_p$ as defined in (\ref{c-p})
is the unit charge.

We now summarize the R-R charged p-brane solutions of type II string theories:
\be
\begin{array}{|l|l|l|l|}\hline
\mbox{Theory} & \mbox{R-R potential} & \mbox{electric sol.} & 
\mbox{magnetic sol.} \\ \hline \hline
\mbox{IIA} & A_{(1)} & p=0 & p = 6 \\ \hline
\mbox{IIB} & A_{(2)} & p=1 & p= 5 \\ \hline
\mbox{IIA} & A_{(3)} & p=2 & p=4 \\ \hline
\mbox{IIB} & A_{(4)} & \multicolumn{2}{l|}{ p=3 \mbox{   (selfdual)}} \\ \hline
\end{array}
\ee

The R-R p-brane solutions have properties which qualify them as solitons:
They are static, stable (BPS bound), regular (no naked singularities)
solutions of the field equations and have finite tension.
The three-brane has an additional property familiar from two-dimensional
solitons: it interpolates between two vacua, Minkowski space at infinity
and $AdS^5 \times S^5$ at the event horizon. (We call $AdS^5 \times S^5$
a vacuum, because it is maximally supersymmetric.)
For solitons one
expects that the tension depends on the coupling as 
$T \sim \ft1{g^2}$. This is, for example, what one finds for monopoles
in Yang-Mills-Higgs theories. In this respect the R-R p-branes show
an unusal behaviour as their tension is proportional to the inverse
coupling, $T_p \sim \ft1{g_S}$, see (\ref{T-p}). This behaviour is in
between the one expected for a soliton $T \sim \ft1{g^2_S}$
and the one of a fundamental
string, $ T \sim 1$, which is independent of the coupling.

One clue to this unexpected behaviour is that the fundamental coupling of 
three closed strings is -- up to a constant -- the square of the coupling
of three open strings, see (\ref{closed-open}). 
Thus a R-R p-brane has the coupling dependence 
expected for a soliton in a theory of open strings.
The type II string theories,  as defined so far, are theories of 
oriented closed strings. Consider now an extension where one adds
to the theory open strings with Dirichlet boundary conditions 
along p directions. If we manage to identify
the corresponding D-p-branes with the R-R p-brane solutions, this
provides a description of type II string theory in these solitonic
backgrounds.

\subsection{p-branes and D-branes}

Surprising as it may be, the identification of R-R p-branes and D-branes
can be supported by convincing arguments. Let us compare the known
properties of these objects. R-R p-branes preserve half of
supersymmetry and can be located at arbitrary positions in transverse space.
The same is true for D-branes with 
$\mbox{p}=0,2,4,6$ in type IIA and $\mbox{p}=1,3,5$ 
in type IIB string theory. The corresponding Killing spinors are constant and 
are given by (\ref{KSRR}). The translational symmetries trivially agree.
These D-p-branes are BPS states and since the central charge associated
with a BPS state with Killing spinors (\ref{KSRR}) is precisely 
the R-R charge, they must carry R-R charge. A crucial quantitative test
is to compute the R-R charge carried by a single D-p-brane. To do so
one has to compute the force due to exchange of R-R gauge fields
between to D-p-branes. 

One first  computes an annulus 
diagram with Dirichlet boundary conditions on both boundaries.
This diagram can be factorized in two ways: either as a sum over
intermediate open strings, or a as a sum over intermediate closed strings.
In the closed string channel the diagram can be visualized as a cylinder
(closed string propagator)
ending on the two D-branes. In this picture it is obvious that one
measures the total force between the D-branes resulting from the exchange
of arbitrary closed string states. This amplitude vanishes, which tells us
that the total force vanishes, as expected for a BPS state. 
To extract the long range part of the force one takes the two D-branes
to be far apart and expands the amplitudes in the masses of the 
closed string states. Then the exchange of massless states
dominates. In detail one finds an attractive force due to 
graviton and dilaton exchange which is cancelled exactly by 
a repulsive force due to exchange of rank $(\mbox{p}+1)$ tensor gauge fields.
The static R-R forces correspond to a generalized Coulomb potential,
\be
V_{\mscr{R-R}} = \frac{Q_{p}}{r^{D-p-3}} = \frac{Q_{p}}{r^{7-p}} \;.
\ee
It turns out that one D-p-brane carries precisely one unit of
R-R p-brane charge,
\be
Q_{p} = c_p = \frac{ (2 \pi)^{7-p} }{ (7-p) \omega_{8-p} } 
( \alpha' )^{ \frac{7-p}{2} } g_S \;.
\ee
This shows that one should identify a R-R p-brane of
charge $N_p c_p$ with a system of $N_p$ D-p-branes. 
People also have computed various other quantities, including
the  low energy scattering, absorption and emission (encoded in 
the so-called greybody factors)
of various strings states on R-R p-branes and D-p-branes,
and the low velocity interactions between p-branes and D-branes.
All these test have been successful.

Since p-branes are extended supergravity solutions with non-trivial
space-time metric, whereas D-branes are defects in flat space-time,
we should of course be more precise in what we mean by identification.
We have seen that both kinds of objects have the 
same charges, tensions and 
low energy dynamics. They have the same space-time and 
supersymmetries and saturate the same BPS bound. Thus they seem
to represent the same BPS state of the theory, but in different
regions of the parameter space. A description in terms of $N_p$
D-branes works within string perturbation theory. In presence of 
D-branes the effective string loop counting parameter is 
$N_p g_S$ instead of $g_S$. The reason is as follows: as we have
seen in section 3 each boundary component gives rise to a factor $g_S$
in scattering 
amplitudes. In a background with D-branes every boundary component
can end on each of the $N_p$ D-branes and therefore 
$g_S$ always occures multiplied
with $N_p$. Since we are inerested in describing macroscopic objects with
large $N_p$, we need to impose that $N_p g_S$ is small in order to 
apply perturbation theory.

Thus we are in the perturbative regime if
\be
N_p g_S \ll 1 \;.
\label{D1}
\ee
Using the Schwarzschild radius (\ref{SchwSch}) we see that this
equivalent to 
\be
r_S \ll \sqrt{\alpha'} \;,
\label{D2}
\ee
which means that the gravitational scale is much smaller then the
string scale. This explains why one does not see any backreaction
of the D-branes on the space-time in string perturbation theory.
D-branes have a finite tension and couple to gravity, but the
deviation from flat space caused by backreaction 
is only seen at scales of the order $r_S$. The only length
scale occuring in string perturbation theory is 
$\sqrt{\alpha'}$ and this is the minimal scale one can resolve when
probing D-branes with strings.

The R-R p-branes are solutions of the type II  effective actions. These
are valid at string tree level and therefore we need to be in the
perturbative regime (of the closed string sector), $g_S < 1$. 
Morover we have neglected $\alpha'$-corrections, which become relevant
when the curvature, mesured in string units, becomes large.
The condition for having small curvature is
\be
r_S \gg \sqrt{\alpha'} \;,
\ee
or, equivalently
\be
N_p g_S \gg 1
\ee
which is opposite to (\ref{D1}, \ref{D2}). The p-brane solution
is valid in the regime of the low energy effective field theory,
where stringy effects can be neglected.

Between the two regimes one can interpolate by changing the string 
coupling $g_S$, while keeping the charge $N_p$ fixed. In general it 
is not clear that one can believe in the results of such interpolations.
But in our case we know that the p-brane/D-brane is the object of 
minimal tension for the given charge. As a BPS state it sits in 
a special BPS multiplet. There is no mechanism compatible with supersymmetry
through which this state could decay or become a non-BPS state.
Besides these arguments, various quantities have been computed
in both regimes and agree with one another.

In string perturbation theory one also has D-branes with
$p>6$. Therefore one might wonder whether the corresponding
objects also exist as p-branes. The answer is yes, though these
so-called large branes have somewhat different properties than
the other branes. For example the seven- and eight-brane are
not flat in the transverse dimensions. The reason is that there are no
harmonic functions in transverse space that become constant at infinity
(this is similar to black holes
in $D<4$). The seven-brane carries magnetic charge under the IIB 
R-R scalar $A$. Its electric partner is a $(-1)$-brane, the D-instanton.
The eight-brane does not have a local source. It is a domain wall solution
which separates regions where the IIA mass parameter (which is
similar to a cosmological constant) takes different values.
The nine-brane is flat space.

\subsection{Further reading}

The type II effective actions and the corresponding p-brane
solutions can be found in the book \cite{Pol}. For extensive
reviews of BPS-branes in supergravity and string theory, 
see \cite{Stelle,West,Duff,Town}.

\section{Outlook}

In this final section we give an outlook on more recent developments.

\subsection{Eleven-dimensional M-theory}

Besides the R-R charged p-branes, type II string theories contain
various other BPS solutions. Since all these carry central charges
of the supersymmetry algebra,
they can be constructed systematically. The other string theories
also have their BPS solitons. Combining perturbative string theory
with the knowledge about the BPS states one can show that the 
strong coupling behaviour of any of the five string theories can
be described consistently by a dual theory. Moreover, one can
interrelate all five superstring theories by such string dualities.
These dualities have not been fully proven yet, but one has compared various 
accessible quantities and all these tests have
been successful. The dualities give a coherent picture where all
perturbative string theories are limits of one single underlying theory.

This is by now a huge subject, which deserves a separate set of
lectures. Here we will only illustrate it by reviewing Witten's 
analysis \cite{Wit:95}
of the strong coupling behaviour of type IIA string theory.
Consider the spectrum of finite mass objects in IIA string theory.
It starts with the massless IIA supergravity multiplet, then comes
an infinite series of excited string states with masses
(\ref{MassII})
\be
\alpha' M^2 \sim N  \;,
\ee
where $N=1,2,\ldots$. As further finite mass objects the 
theory contains states with $N_0$ D-0-branes, with masses (\ref{T-p})
\be
\alpha' M^2 \sim \frac{N_0}{g_S} \;.
\label{MassD0}
\ee

(One can show that there are no bound states at threshold, so the
states with $N_0>1$ are $N_0$-particle states.)
In the perturbative regime, $g_S \ll 1$, the D-0-branes are very heavy.
But when extrapolating to strong coupling, $g_S \rightarrow \infty$,
they become much lighter than any perturbative excitation. Since
the D-0-branes are BPS-states, we know that the mass formula (\ref{MassD0})
is not modified at strong coupling. For very large $g_S$
one gets a quasi-continuum of D-0-brane states above the massless
supergravity multiplet. The collective modes of a 
D-0-brane sit in a so-called short multiplet of
the IIA supersymmetry algebra. Short multiplets are special massive 
multiplets which saturate the BPS bound. They have 
less components than generic massive multiplets. The multiplet
of the D-0-brane is a massive
version of the supergravity multiplet: it has the same
number of states and the same spin content. Thus the low energy,
strong coupling spectrum looks like the Kaluza-Klein spectrum obtained
by dimensional reduction of an eleven-dimensional theory.
The only candidate is eleven-dimensional supergravity, the unique
supersymmetric theory in eleven dimensions. When comparing the
low energy, strong coupling spectrum of IIA string theory to the 
Kaluza-Klein spectrum of eleven-dimensional supergravity one finds
that both agree, provided
one relates  the string coupling to the radius $R_{11}$ 
of the additional
space dimension according to,
\be
g_S^2 = \left( \frac{R_{11}}{L_{\mscr{Pl}}} \right)^3 
\ee
and the string scale $\alpha'$ to the eleven-dimensional Planck length
$L_{\mscr{Pl}}$ according to:
\be
\alpha' = \frac{L_{\mscr{Pl}}^3}{R_{11}}  \;.
\ee
The eleven-dimensional Planck length is defined through the 
eleven-dimensional gravitational
coupling by: $\kappa_{(11)}^2 = L_{\mscr{Pl}}^9$.
The relation between the eleven-dimensional metric and the
IIA string frame metric is:
\be
ds^2_{11} = e^{2 \Phi/3} \left(
ds^2_{\mscr{IIA, Str}} + ( dx^{11} - A_{\mu} dx^{\mu} )^2 \right) \;,
\ee
where $\Phi$ is the IIA dilaton and the Kaluza-Klein gauge field
$A_{\mu}$ becomes the R-R one form.

This indicates that the strong coupling limit of IIA string theory 
is an eleven-dimensional theory, called M-theory. We do not have
enough information to give a complete definition, but we know that
M-theory has eleven-dimensional 
supergravity as its low energy limit. There must be additional
degrees of freedom, because
eleven-dimensional supergravity
is not consistent as a quantum theory. 
Even without a complete definition of M-theory,
one can find more evidence for the duality. 
Eleven-dimensional supergravity has BPS solitons, which properly reduce 
under dimensional reduction to the solitons of IIA string theory. In
particular it has a supersymmetric membrane solution, called the 
M-2-brane, which reduces to the fundamental IIA string.

\subsection{String dualities}

Let us now consider the other string theories. 
What about type IIB? The theory
has maximal supersymmetry, and its massless spectrum cannot
be obtained by dimensional reduction from a higher dimensional 
supersymmetric theory. The only obvious possibility is that it is selfdual,
which means that the strong and weak coupling limits take the same
form. One can show that inverting the coupling, 
$g_S \rightarrow g_S^{-1}$, preserves the form of the action
and is a symmetry of the BPS spectrum, 
if one simultanously interchanges the
fundamental IIB string with the D-1-brane. The strong coupling
limit is again a IIB string theory, with solitonic 
strings (D-1-branes) now playing the role of the fundamental 
objects. The transformation relating weak and strong coupling
is called S-duality and works the same way as the Montonen-Olive 
duality in four-dimensional $N=4$ Super-Yang-Mills theory.
It has also been verified that S-duality is respected by
instanton corrections to string amplitudes.

In a similar way, the strong coupling limit of the type I
string is the heterotic string with gauge group $SO(32)$,
and vice versa. We already saw that both theories have the 
same massless spectra, while the perturbative massive spectra
and interactions were different. Both theories cannot be
selfdual (for example, inverting the string coupling does not
preserve the form of the effective action). But once solitonic
BPS states are included, the BPS spectra are equal and reversing
the coupling relates the two effective actions. The heterotic
$SO(32)$ string is identified with the D-1-brane of type I.

What is left is to determine the strong coupling behaviour
of the $E_8 \times E_8$ heterotic string. This turns out to be
again eleven-dimensional M-theory  but this time compactified on an
interval instead of a circle. The interval has two ten-dimensional
boundaries, on which ten-dimensional vector multiplets with gauge 
group $E_8$ are located. This is also known as Horava-Witten theory.

Let us summarize the strong-coupling limits of the five
supersymmetric string theories:
\be
\begin{array}{|l|l|} \hline
\mbox{String theory} &  \mbox{Strong coupling dual} \\ \hline \hline
\mbox{IIA} & \mbox{M-theory on circle} \\ \hline
\mbox{IIB} & \mbox{IIB} \\ \hline 
\mbox{I} & \mbox{Heterotic } SO(32) \\ \hline
\mbox{Heterotic }  SO(32) & \mbox{I} \\ \hline
\mbox{Heterotic } E_8 \times E_8 & \mbox{M-theory on intervall} \\ \hline
\end{array}
\ee
These dualities fall into two classes: either one has a relation 
between strong and weak coupling. This is called S-duality.
Or the coupling is mapped to a geometric datum, the radius of
an additional dimension. There is a third type of string duality,
which leads to further relations between string theories. 
It is called T-duality and relates weak coupling to weak coupling,
while acting non-trivially on the geometry. Since weak coupling
is preserved, one can check that T-duality 
is preserved in perturbation theory.
By T-duality, the IIA string theory compactified on a circle
of radius $R$ is equivalent to IIB string theory compactified
on a circle of inverse radius in string length units, $\tilde{R}=
\frac{\alpha'}{R}$. One can take the decompactification
limit and obtain ten-dimensional IIB theory as the zero radius limit
of compactified IIA theory and vice versa. In the same way one
can relate the two heterotic string theories. 
When acting on open strings, T-duality exchanges Neumann boundary 
conditions with Dirichlet boundary conditions. Therefore the 
T-dual of type I string theory is a theory containing open strings
which are coupled to D-branes. Though one might consider this
as a solitonic sector of type I theory, it is sometimes called
type I' theory. Let us summarize the T-duals of the five
supersymmetric string theories:
\be
\begin{array}{|l|l|} \hline
\mbox{String theory} & \mbox{T-dual theory} \\ \hline \hline
\mbox{IIA} & \mbox{IIB} \\ \hline 
\mbox{IIB} & \mbox{IIA} \\ \hline
\mbox{I} & \mbox{I'} \\ \hline
\mbox{Heterotic } SO(32) & \mbox{Heterotic } E_8 \times E_8 \\ \hline
\mbox{Heterotic } E_8 \times E_8 & \mbox{Heterotic } SO(32) \\ \hline
\end{array}
\ee

Finally there is yet another relation between IIB string theory and
type I. Type IIB has supersymmetric D-9-branes. These D-branes are
space-filling, they correspond to adding open strings with Neumann
boundary conditions in all directions. From our earlier discussion 
we know that the only consistent coupling between open and closed 
superstrings is a non-oriented theory with gauge group $SO(32)$, namely
type I. This can be realized as a configuration in IIB string theory,
where one adds 32 D-9-branes together with additional non-dynamical
objects, so-called orientifold planes,
which reverse world-sheet parity. Type I string theory is
an \lq orientifold' of type IIB. More generally, after introducing
D-branes and orientifolds, the
type I, IIA and IIB string theories
can be considered as one theory in different backgrounds,
which can be transformed into another by T-duality and orientifolding.
The type I' theory, which we introduced above as the T-dual of 
type I,  can also be obtained as an orientifold
of type IIA. Therefore type I' and type I theory are also called type 
IA and type IB. 

Thus we see a bigger picture emerging once we include 
the BPS solitons of the five supersymmetric string theories.
All theories are related to one another and to eleven-dimensional
M-theory, and all strong couplings limits can be consistently
described. Therefore one believes today that the different string 
theories are perturbative limits of one single underlying
theory. Due to the role of D-branes and since there
is an eleven-dimensional limit, which cannot be described by 
perturbative string theory, one prefers to call it M-theory.

\subsection{Further reading.}

String dualities and how they relate the five supersymmetric
string theories to one another are discussed in the 
book \cite{Pol} and in various lectures notes. The paper
\cite{HaackKoersLuest} gives a nice overview of the various
dualities that we mentioned above. The lectures
\cite{deWLou} approache the subject from the side of effective
supergravity theories and string compactifications, whereas
\cite{West} is an introduction to supergravity which also covers
branes and string dualities. Other lecture notes on 
string theory and string dualities are \cite{Sen,Kiritsis,Sza}.

T-duality,
which we only mentioned briefly in these lectures is reviewed at
length in \cite{GPR}. The role of combined T- and S-dualities,
then called U-dualities, in string and M-theory compactifications
is reviewed in \cite{ObePio}.
D-branes and their
applications are discussed in \cite{Bachas,Johnson}. 
For a recent reviews
of open strings, see \cite{Sagnotti}.
The lectures \cite{Moh1} are devoted to the description
of BPS black holes in string theory. They also cover the 
ten-dimensional brane solutions of type II string theories and
how they are related by T- and S-duality. 
BPS solutions of eleven-dimensional
supergravity (M-branes) and their relation to the brane solutions of
type II string theory are explained in \cite{Town}.

\subsection{Lightning review of further topics}

Let us finally mention areas of active research 
together with some references, which might be useful
for the interested reader. 

\subsubsection{What is M-theory?}

So far M-theory was characterized by its relation to 
various perturbative string theories and through its
eleven-dimensional low energy limit, supergravity.
The fundamental open question is how to define M-theory
without recourse to a particular background, perturbation
theory or particular limits. The recent developments show
that besides strings also various branes have to be taken
into account as dynamical objects. The question which remains open 
is which
of these objects are truly fundamental. When considering all
p-branes as equally fundamental as strings, one immediately
faces the problems of how to quantize higher-dimensional 
objects. Among p-branes, 
strings ($p=1$) and particles particles ($p=0$) 
are singled out, because
their world-volume theories are free as long as the background
geometry is flat. This underlies the power of string perturbation
theory. The situation is completely different for higher-dimensional
branes ($p>1$), where the world-volume theory is a complicated
interacting theory, even in a flat background. Therefore no
analogon of string perturbation theory for these objects 
has been developed so far.
Alternatively, one particular kind of brane
might be the fundamental object, whereas all others are obtained
by dimensional reduction or as solitons. There are two candidates
for which concrete proposals have been made: the supermembrane and 
the D-0-brane.

\subsubsection{The supermembrane.}

Eleven-dimensional supergravity has a solitonic two-brane
solution, called the supermembrane or the M-2-brane. The 
three-dimension\-al action
for the collective modes of this solution contains a 
Nambu-Goto term and Wess-Zumino term, which describe the
coupling to gravity and to the three-form gauge field
of eleven-dimensional supergravity. One can then try
to treat this membrane as a fundamental object in an analogous
way to the fundamental string in string theory. Moreover one
can get back the IIA string by dimensional reduction.
Supermembrane 
theory is much more complicated then string theory, because
the world-volume theory does not become free in a flat 
background, as discussed above. Also note that there is no 
local Weyl invariance for p-branes with $p\not=1$. Therefore
there is no conformal world volume action and no analogon of
the Polyakov formulation.

At the WE-Haereus-Seminar, supermembrane theory was the subject
of lectures by Hermann Nicolai. A pedagogical introduction to the 
subject, which
also covers the relation to other approaches to M-theory
is provided by his Trieste lectures \cite{NicHel}.

\subsubsection{Matrix theory.}

In the matrix theory formulation of M-theory, also called
M(atrix) theory, the D-0-brane is the fundamental object.
More precisely, there is a conjecture due to Banks, Fischler,
Shenker and Susskind \cite{BFSS}, which claims that eleven-dimensional
M-theory in the infinite momentum frame is given exactly
by the limit $N\rightarrow \infty$ of the supersymmetric
$U(N)$ quantum mechanics describing a system of $N$ D-0-branes.

M(atrix) theory can be viewed as an alternative formulation
of supermembrane theory, since the finite--$N$--M(atrix) model
Hamiltonian is an approximation of the supermembrane 
Hamiltonian. In M(atrix) theory multi-membrane states are 
described by clusters of D-0-branes. Conversely D-0-branes
are contained in supermembrane theory as Kaluza-Klein modes
of the eleven-dimensional supergravity multiplet, which
consists of the zero mass states of the supermembrane.
Besides \cite{NicHel}, lectures on M(atrix) theory are
\cite{Bilal,Banks,Bigatti}.

\subsubsection{Black holes.}

While the fundamental definition of M-theory remains to be found,
string theory and D-branes have been applied to a variety of
problems in gravity, field theory and particle physics. One of
the most prominent applications is the description of black 
holes through D-branes, which elaborates on the relation
between D-branes and p-brane solutions discussed in section
5. Starting from p-branes in ten dimensions one can obtain four-dimensional
black holes by dimensional reduction. Performing the same reduction
with the corresponding D-brane configuration, one gets a description of the 
system where the
microscopic degrees of freedom are known. This can be used to 
compute the entropy of the black hole: one counts the number $N$
of microstates, i.e., excitations of the system, which belong to
the same macrostate, i.e., the same total energy, charge and
angular momementum:
\be
S = \log N \;.
\ee
In practice the statistical entropy $N$ is evaluated asymptotically
for very large black hole mass.

The result can be compared to the Bekenstein-Hawking entropy 
of the black hole, which is given in terms of the area $A$ of the
event horizon,
\be
S_{\mscr{BH}} = \frac{A}{4} \;.
\label{macro}
\ee
One finds that the two entropies agree, $S=S_{\mscr{BH}}$, which 
confirms that the D-brane picture correctly captures the 
microscopic degrees of freedom of the black hole \cite{StrVaf}. As mentioned
above $S$ is evaluated asymptotically, but we would like to stress
that the resulting $S$ matches exactly with the Bekenstein-Hawking
entropy. This is in contrast to other approaches, where 
both entropies have the same dependence on paramaters,  
while the numerical prefactor of the statistical  
cannot be determined precisely.

One can also compute and compare sub-leading contributions to both
entropies. Corrections to the statistical entropy have been computed
for Calabi-Yau compactifications of type II string theory and
eleven-dimensional M-theory \cite{MalStrWit,Vafa}, (see also 
\cite{deWCarMoh:99}). 
These match precisely with corrections
to the macroscopic black hole entropy, which are due to 
higher curvature terms in the effective 
action \cite{deWCarMoh:98,deWCarKaeMoh}. These higher curvature
terms modify the entropy in two ways. The first is an explicit
modification of the black hole solution and, hence, of the area $A$.
The second is a modification of the area law (\ref{macro}). As pointed
out by R. Wald \cite{Wald}, the validity of the first law of black hole
mechanics in presence of
higher curvature terms requires a modified definition of black hole
entropy. (The first law of black hole mechanics formulates the
conservation of energy. It expresses adiabatic changes of the mass
to changes in terms of parameters of the black hole solution.)
Both effects, the explicit change of the solution and the
modified definition of the entropy, 
change the entropy in a complicated way, but the
combined correction is relatively simple and precisely matches 
the statistical entropy. This is reviewed in \cite{Moh2}.

Besides entropy, the D-brane picture has been used to compute
Hawking radiation and greybody factors 
(see \cite{Mal,AharonyEtAl} for review and references). 
This is possible for branes
which are close to the BPS limit. In the D-brane picture one can compute
the emission, absorption and scattering of closed string states
by a D-brane. Again one finds agreement 
with a semiclassical treatment of the corresponding black hole solutions. 
Note, however,  that the method only applies to 
D-branes and p-branes which are close to the BPS limit.
The generalization to generic black holes remains an open problem,
though various proposals have been made. One idea, which applies 
to black holes without R-R charge is a correspondence principle between
black holes and fundamental strings \cite{Sus:93,HorPol:96}. 
The idea is that a black hole
evaporates through Hawking radiation until its size reaches the string
scale where it converts into a highly excited fundamental string.
This is supported by the observation that the entropies of black holes
and fundamental strings of equal mass match precisely when the 
Schwarzschild radius equals the string length. Another idea 
\cite{SfeSke}
is to
use string dualities to map four- (and five-)dimensional black 
holes to three-dimensional black holes (BTZ black holes \cite{BTZ}). 
Three-dimensional
gravity does not have local degrees of freedom, because the action is
a total derivative. If the space-time has boundaries one gets 
boundary degrees
of freedom which can be described by a two-dimensional conformal field
theory. Treating the horizon as a boundary, this  
can be used to compute the statistical entropy of 
three-dimensional black holes \cite{Carlip}. 
The dualities that one needs to connect
these three-dimensional to four-dimensional black holes are slightly
more general then those mentioned so far. In particular they change
the asymptotic geometry of space-time, so that one can map a 
higher-dimensional black hole
to a lower-dimensional one (times an internal, compact
space). One can argue that these transformations do not change the 
thermodynamic properties. Moreover one finds 
explicitly that the Bekenstein-Hawking 
entropy of the four-dimensional Schwarzschild black hole is matched
by the state counting of the dual three-dimensional black hole. 
A related approach is to use dualities to map Schwarzschild black holes
to brane configurations \cite{Englert}. Finally, the microscopic entropy of
Schwarzschild black holes has also been computed using Matrix theory,
see \cite{Bigatti} for review and references.

%%%

The most general and most promising approach to generic black holes 
is the AdS-CFT correspondence \cite{Mal:97,GubKlePol,Wit:98}. 
This correspondence and its generalizations
relate $D$-dimensional gravitational backgrounds to $(D-1)$--dimensional field
theories. One of the roots of this idea is the so-called holographic
principle \cite{tHo,Sus:94},
which claims that the physics beyond the horizon of 
a black hole can be described in terms of a field theory associated
with its horizon. The D-brane picture of black holes can  be viewed
as a realization of this idea, because here the 
interior region of the black hole has disappeared, while interactions
of the exterior region with the black hole are described as interactions
between closed strings in the bulk with open strings on the brane.
A more general version of the holographic principle is that gravity can
always be described in terms of a lower dimensional field theory.
The AdS-CFT correspondence, which we briefly describe below, can
be viewed as an attempt to realize this idea.

More about black holes in string theory can be found in 
\cite{Moh1} and in other reviews of the topic
including \cite{DauFre,You,Mal,Man,AharonyEtAl,Ske,Moh2,Bigatti}
and section 14.8 of \cite{Pol}.

\subsubsection{The AdS-CFT correspondence and its generalizations.}

The AdS-CFT correspondence is another consequence of the relation
between D-branes and p-brane solutions. Its most simple version
is obtained by considering a system of $N$ D-3-branes and
taking the limit $\alpha' \rightarrow 0$, while $N g_S$ and
$R/ \alpha'$ are kept fixed. Here $g_S$ is the string coupling
and $R$ the characteristic scale of separation between the branes.
In the D-brane picture gravity and massive string excitations
decouple and one is left with the effective theory of the 
massless open string modes, which is a four-dimensional
${\cal N}=4$ supersymmetric $U(N)$ gauge theory in the 
large $N$ limit. The corresponding limit in the p-brane
regime is the near horizon limit, where the geometry
takes the form $AdS^5 \times S^5$. The low energy excitations 
are described by supergravity on $AdS^5$. This observation
motivated Maldacena's conjecture \cite{Mal:97}:
five-dimensional supergravity on $AdS^5$
is a dual description of four-dimensional 
${\cal N}=4$ supersymmetric $U(N)$ gauge theory, the later being
a conformal field theory.
$AdS^5$ has an asymptotic
region which can be identified with (the conformal compactification of)
four-dimensional Minkowski space. This is called the boundary, and 
the conformal field theory is located there. One
finds a correspondence between fields $\phi(x_{(5)})$ 
of the bulk supergravity theory 
and operators ${\cal O}(x_{(4)})$ of the Yang-Mills theory on the boundary.
(Here $x_{(5)}$ are coordinates on the five-dimensional bulk and
$x_{(4)}$ are coordinates on the four-dimensional boundary.) 
A quantitative version
of the conjecture, due to Gubser, Klebanov, Polyakov 
\cite{GubKlePol}
and Witten \cite{Wit:98},
states that the generating functional for the correlators of
operators ${\cal O}(x_{(4)})$ with sources $\phi_0(x_{(4)})$
is given by the partition function
of the supergravity theory, evaluated in the background
$\phi(x_{(5)})$ with boundary values 
$\left. \phi(x_{5}) \right|_{\mscr{Boundary}} = \phi_0(x_{(4)})$,
according to:
\be
\left \langle e^{ \int d^4 x \phi_0(x_{(4)}) {\cal O}(x_{(4)}) }
\right \rangle = Z \left( \phi(x_{(5)}) \right) \;.
\ee

There are various generalizations of this basic form of the
correspondence, which relate other gravitational backgrounds to 
other gauge theories. One particular extension of the AdS-CFT
correspondence relates five-dimensional domain wall geometries to
renormalization group flows in non-conformal gauge theories.
In this setup the coordinate transverse to the domain wall
corresponds to the energy scale of the gauge theory 
\cite{ItzMalSonYan,BooSkeTow}. More recently, maximally 
supersymmetric gravitational 
wave backgrounds have moved to the center of interest
\cite{BerMalNas}.

Extensive reviews of the AdS-CFT correspondence can be found in
\cite{AharonyEtAl} and \cite{DHoFre}.

\subsubsection{Brane worlds.}

D-branes provide a new option for model building in particle 
physics. One can localize some or all matter and gauge fields
of the standard model on a three-brane, while gravity propagates
in the higher-dimensional bulk. Such models have the interesting
feature that the size of the extra dimensions can be quite large,
even in the sub-mm range. Moreover one can have
a fundamental (higher-dimensional) Planck scale of 1 TeV,
which provides a new approach to the gauge hierarchy problem.
A low gravitational scale of 1 TeV leads to 
spectacular predictions, like the mass production of
black holes at the LHC. Therefore brane worlds have been
a main activity in the string and particle physics community
over the last years. One should stress here that though 
TeV-scale gravity is possible within string theory, it is not
predicted.

There is a huge variety of brane world models, which
range from phenomenological models to models with explicit 
realization in string or M-theory, see for example 
\cite{RubSha,ArkH,AntArkH,RS1,RS2,Ovr}. 
In one variant, the so-called
Randall-Sundrum model (RS II model \cite{RS2}, to be precise), 
the extra dimensions are curved in such a way
that gravity is confined on the brane in a similar way as matter fields.
This opens the possibility of extra dimensions which a arbitrarily 
large, though invisible at low energies.

At the WE-Haereus Seminar brane worlds were the subject of
the lectures given by I. Antoniadis and A. Barvinsky, while
J. Gundlach reviewed tests of Newton's law at short distances.  
A nice review of mass scales and the possible sizes of extra 
dimensions in string theory can be found in \cite{Antoniadis}.
Experimental signatures of large extra dimensions are 
discussed in \cite{AntBen}.
One particular type of brane worlds, which occure in Calabi-Yau 
compactifications of Horava-Witten theory, is reviewed in \cite{Ovrut}. 
The lectures \cite{Kachru} give an introductions
brane worlds and warped compactifications.

\subsubsection{Compactifications and phenomenology.}

D-branes and p-branes have considerably extended the framework
of string compactifications, which aim to explaine how 
our four-dimensional world is embedded into the fundamental
ten- or eleven-dimensional theory. Whereas ten years ago
string phenomenology was synonymous with the study of the
heterotic $E_8 \times E_8$ string, compactified on 
complex three-dimensional Calabi-Yau manifolds, one now has
various other options to consider. Besides brane worlds one
can study compactifications where part of the standard model
particles are not string modes but descend from p-branes wrapped
on internal p-cycles. Switching on background fluxes 
of antisymmetric tensor fields, one obtains 
warped compactifications, where the characteristic length scale
of four-dimensional space-time becomes dependent on the position
in the internal space. A particular class of non-perturbative
IIB backgrounds can be described purely geometrically in terms of
so-called F-Theory.

The central problem of string compactifications is still the
problem of vacuum degeneracy. As we have seen, the vacuum expectation
value of the dilaton is not fixed at string tree level. In supersymmetric
theories this holds to all orders in perturbation theory. Similarly,
string compactifications in general have several scalar fields, called
moduli, which parametrize the shape and size of the internal manifold
and enter into the couplings of the effective field theory. 
The vacuum expectation values of these fields are not fixed, as long as
supersymmetry is unbroken. This ruins the predictive power that the
theory has in principle, and leads to continuous families of degenerate
vacua. Once supersymmetry is broken the moduli get fixed, but there
is a number of issues to be addressed: one needs to understand the
dynamical mechanism behind supersymmetry breaking, which requires
to understand the theory non-perturbatively. The potential generated
for the dilaton and for the moduli should have stable vacua and no runaway
behaviour. One needs sufficently large masses or sufficiently small
couplings for the moduli to avoid contradiction with empirical data.
Moreover, in string theory supersymmetry is closely related to
the absence of the tachyon, which one does not want to reintroduce.
One also wants that supersymmetry breaking occurs at a specific
scale, the most popular scenario being low energy supersymmetry where
the supersymmetric partners have masses of about 1 TeV. 
D-branes, p-branes and other new developments have added a variety
of new ways to address these problems, but a definite solution remains
to be found.

String compactifications on Calabi-Yau manifolds are reviewed in
\cite{Greene}. Lectures on warped compactifications and brane worlds
can be found in \cite{Kachru}. F-theory is for example explained in
\cite{Sen}. For an introduction to string and M-theory particle
phenomenology, see for example \cite{Lou:98,Dine}.

\subsubsection{Geometric and D-brane engineering, D-branes and
non-commutative field theory.}

In addition to the AdS-CFT correspondence, string theory
has lead to other new approaches to gauge theories and other 
field theories. In geometric engineering \cite{G-Engin} 
one starts 
from branes wrapped on cycles in an internal space, which 
typically is a Calabi-Yau manifold, whereas in D-brane engineering
\cite{D-Engin}
one studies D-brane configurations in a non-compact space-time.
In both cases one takes a low energy limit (similar to the one
discussed above in the context of the AdS-CFT correspondence) to
decouple gravity.

Another direction stimulated by 
string theory and D-branes is gauge theory on non-commutative
space-times. As mentioned in the lectures, the effective action
for a D-brane is of Born-Infeld type. It has been argued that
this can be reformulated as a Yang-Mills theory on a non-commutative
world volume, with a deformation parameter which is determined by
the bulk $B_{\mu \nu}$ field of the closed string sector \cite{SeiWit}.

Geometric engineering is reviewed in \cite{Klemm,Mayr}, while
gauge theory on non-commutative space-times is reviewed in
\cite{NC-rev}. For extensive lectures on D-branes, see
\cite{Bachas,Johnson}.

\subsubsection{Cosmology.}

Whereas string compactifications ususally aim at finding
four-dimensional Minkowski space with a realistic particle 
spectrum from string theory, one should of course try to
do better. Cosmological solutions of string theory should
shed light on the issue of the initial singularity, 
describe an inflationary phase (or an alternative 
mechanism which takes care of the problems of the old
hot big bang model), further describe the post-inflationary
phase and explain the smallness of the cosmological constant.
These problems have been mostly neglected by string theorists
for a long time, but nowadays they find increasing interest,
due to both new cosmological data and new theoretical
developments. In particular branes have been invoked for
either providing the mechanism for inflation or for providing
an alternative to inflation. 

Reviews of string cosmology can be found in \cite{LidWanCop,Eas,GasVen}.

\subsubsection{The challenge from de Sitter space.}

Since there is
empirical evidence in favour of a small, positive cosmological
constant, there has been a considerable interest in string
theory in de Sitter space over the last few years. De Sitter
space is a challenge for several reasons. First, most successful
applications of string theory to gravity depend on supersymmetry,
but supersymmetry is completely broken in presence of a positive
cosmological constant. Second, de Sitter space has cosmological
horizons, and the perturbative formalism which works for
Minkowski space as explained in section 3 cannot be applied.
Therefore de Sitter space requires a significant step beyond
that framework. For a review see \cite{SprStrVol}.

\subsubsection{Tachyon condensation and string field theory.}

As observed several times in these lectures, the appearence of 
tachyons is a generic feature of string theories when there
is no supersymmetry. Since the mass squared of a scalar particle
is given by the curvature of its potential at the stationary point
one is expanding around, this shows that one tries to expand the theory 
around a local maximum of the potential. Depending on the 
global form of the potential, the theory might be unstable,
or it might be that the scalar field rolls to a minimum.
This is referred to as tachyon condensation. 

Tachyons do not only occure in the groundstate of bosonic string 
theories, but also in D-branes configurations which are not BPS
states (non-BPS D-branes systems are reviewed in \cite{Schwarz}). 
Work starting with a paper 
by A. Sen \cite{Sen-Tachy}
provided strong evidence that tachyon condensation occurs 
in unstable non-BPS configurations of D-branes. Such systems have 
tachyonic open string states which condense.
The resulting stable vacuum is the closed string vacuum,
whereas the D-branes have decayed and therefore open strings are
absent. This work makes use of string field theory, which for
a long time was mostly neglected, because it is very complicated
and was believed of little practical use. The renewed interest in
string field theory might bring us one step forward 
towards a non-perturbative and background-independent formulation
of M-theory.

\section*{Acknowledgements}

First of all I would like to thank the organisers of the
271-th WE-Haereus seminar for organizing this very lively and 
stimulating school. I would also like to thank the participants for their
numerous questions and remarks, which I have tried to take
into account when writing these lecture notes. The same applies to
all the participants of my string theory course at the university of
Jena. My special thanks goes to Frank Saueressig and Christoph 
Mayer for their very helpful comments on the manuscript.

%INDEX%%%%%%%%%%%%%%%%%%%%%%%%%%%%%%%%%%%%%%%%%%%%%%%%%%%%%%%%%%%%%%%
% Please check with the editor of your book whether he plans to
% include a "mutual" subject index - if so, please code your entries
% in the standard syntax. For your own purposes you may print your
% "personal" index by using the following commands:
%
%\clearpage
%\addcontentsline{toc}{section}{Index}
%\flushbottom
%\printindex
%%%%%%%%%%%%%%%%%%%%%%%%%%%%%%%%%%%%%%%%%%%%%%%%%%%%%%%%%%%%%%%%%%%%%

\end{document}